\documentclass[prb,showpacs,twocolumn,aps,superscriptaddress,a4paper]{revtex4-1}
\usepackage{dcolumn,amssymb,amsmath,amsfonts,graphicx,latexsym,color,braket}

\definecolor{myblue}{rgb}{0,0,0.75}

\begin{document}

\title{Relativistic quantum field theory of stochastic dynamics in the Hilbert space}

\author{Pei Wang}
\email{wangpei@zjnu.cn}
\affiliation{Department of Physics, Zhejiang Normal University, Jinhua 321004, China}

\date{\today}

\begin{abstract}
In this paper, we develop an action formulation of stochastic dynamics
in the Hilbert space. In this formulation, the quantum theory of random unitary evolution is
easily reconciled with special relativity.
We generalize the Wiener process into 1+3-dimensional spacetime,
and then define a scalar random field which keeps invariant under Lorentz transformations.
By adding to the action of quantum field theory a coupling term between random and quantum fields, we obtain a
random-number action which has the statistical spacetime translation and Lorentz symmetries.
The canonical quantization of the theory results in a Lorentz-invariant
equation of motion for the state vector or density matrix. We derive the path integral formula
of $S$-matrix, based on which we develop a diagrammatic technique for doing the calculation.
We find the diagrammatic
rules for both the stochastic free field theory and stochastic $\phi^4$-theory.
The Lorentz invariance of the random $S$-matrix is strictly proved by using the diagrammatic
technique. We then develop a diagrammatic technique for calculating the density matrix of
final quantum states after scattering. In the absence of interaction, we obtain the exact expressions of
both $S$-matrix and density matrix. In the presence of interaction, we prove a simple relation
between the density matrices of stochastic and conventional $\phi^4$-theory. Our formalism
leads to an ultraviolet divergence which has the similar origin as that in quantum field theory.
The divergence can be canceled by renormalizing the coupling strength to random field.
We prove that the stochastic quantum field theory is renormalizable even in the presence of interaction.
In the models with a linear coupling between random and quantum fields, the random field
excites particles out of the vacuum, driving the universe towards an infinite-temperature state.
The number of excited particles follows the Poisson distribution. The collision between particles
is not affected by the random field. But the signals of colliding particles are gradually covered by
the background excitations caused by random field.
\end{abstract}

\maketitle

\tableofcontents

\section{Introduction}

In the past decades, the stochastic processes in the Hilbert space
have been under active investigation. They are widely employed to simulate
the evolution of quantum states in open systems~\cite{Breuer02}, and also employed
at a more fundamental level, i.e. in the attempts to solve the
quantum measurement problem~\cite{Bassi03,Bassi13,Tumulka09}. Various stochastic processes display
interesting properties which are distinguished from those of deterministic evolutions in the Hilbert space.

The application of stochastic processes in quantum physics has a long history.
In the 1990s, it was shown that there exists a stochastic-process representation
for the dynamics of an open quantum system which used to be represented
by a master equation of density matrix~\cite{Dalibard92,Dum92,Gisin92}. Transforming a master equation
into an equivalent stochastic process, i.e. the so-called unravelling of
the master equation, quickly became a popular numerical simulation method for open systems.
The efficiency of this method (sometimes called the Monte Carlo
wave function) comes from the fact that the dimension of Hilbert space is
much smaller than the dimension of the vector space of density matrix~\cite{Plenio98,Daley14}.
In quantum optics, the Monte Carlo wave function has found its application in the studies
of laser cooling~\cite{Castin95}, quantum Zeno effects~\cite{Power96}, Rydberg atomic gases~\cite{Ates12,Hu13} or
dissipative phase transitions~\cite{Raghunandan18}.

Almost at the same time, the stochastic processes were employed
in the attempts to solve the quantum measurement problem.
In the 1980s and 1990s, a family of spontaneous collapse models were proposed
for explaining the randomness of a measurement outcome
and also the lack of quantum superposition in the macroscopic world. In these models,
the wave function collapse (sometimes called state-vector reduction)
is seen as an objective stochastic process with no conscious observer.
Examples of the collapse models include the Ghirardi-Rimini-Weber
model~\cite{GRW} (GRW), quantum mechanics with universal position localization~\cite{Diosi89,Bassi05} (QMUPL),
continuous spontaneous localization~\cite{Pearle89,Ghirardi90} (CSL),
gravity-induced collapse or Di\'{o}si-Penrose model~\cite{Penrose96},
energy-eigenstate collapse~\cite{Hughston96,Adler00},
CSL model with non-white Gaussian noise~\cite{Bassi02,Pearle98,Adler07nonwhite,Adler08,Bassi09},
completely quantized collapse models~\cite{Pearle05,Pearle08},
dissipative generalization of Di\'{o}si-Penrose model~\cite{Bahrami14},
and models driven by complex stochastic fluctuations of the spacetime metric~\cite{Gasbarri17}.
These models have different predictions from quantum mechanics
and the difference is measurable. Recently, there are increasing number of
experimental proposals for testing these collapse models~\cite{Ghirardi99,Marshall03,Bassi05EXP,Adler05,Arndt14,Nimmrichter11,Pontin19,Vinante16,Vinante17,Adler07,Lochan12,Donadi21,Bahrami14EXP,Bahrami13,Bedingham14,Bera14,Li17,Bahrami18,Tilloy19,Bilardello17,Leontica21,Gasbarri21,Kaltenbaek21}.

Currently, the application of stochastic processes in quantum theories
is based on the stochastic differential equation (SDE).
In quantum mechanics or quantum field theory (QFT), the state vector
follows a deterministic unitary evolution in the Hilbert space, governed by
the Schr\"{o}dinger equation. For using stochastic processes in a quantum theory,
one assumes a random unitary evolution of state vector. This can be achieved
by introducing a random term into the Schr\"{o}dinger equation,
which then turns into a SDE whose
solution is a stochastic process in the Hilbert space. Within such a process,
the initial state vector does not uniquely determine the trajectory
of the state vector at the following time. Instead, the solution
is an ensemble of trajectories with some specific probability distribution.

Despite the success of SDE approach in the simulation of open systems
or in the spontaneous collapse models, this approach has a disadvantage ---
it is hard to incorporate the Lorentz symmetry into this approach.
The Lorentz symmetry is a fundamental symmetry in high energy physics,
it is then necessary to study the stochastic processes that respect Lorentz symmetry,
if one hopes to make use of these processes for describing the collision of elementary particles.
In the context of spontaneous collapse models, an intense effort has been made to
create a SDE of state vector with Lorentz symmetry~\cite{Gisin89,Pearle90Book,Ghirardi90r,Diosi90,Pearle98,Bedingham11,Pearle05rela,Myrvold17,Pearle15,Tumulka06,Tumulka20,Jones20,Jones21}. Unfortunately, no satisfactory SDE has been found up to now.
One reason is that it is difficult to show the Lorentz symmetry of a model
through the differential-equation (Hamiltonian) approach.
In fact, a Lorentz-invariant QFT is always built by starting from
a Lorentz-invariant action, because the latter is much easier to obtain.
It is common sense in QFTs that the Lorentz invariance becomes clear only if one
chooses the action formulation and path-integral approach, instead of the Hamiltonian approach~\cite{Weinberg}.
But, up to today, the action formulation and path-integral approach
are still absent for stochastic processes. For an easy incorporation of Lorentz symmetry
in stochastic processes, it is necessary to develop these tools. This is the purpose of present paper.

In this paper, we develop the action formulation and path-integral approach
for generic random unitary evolutions in the Hilbert space. A great advantage
of our approach is that the construction of relativistic QFTs with
stochastic dynamics becomes easy through our approach. Simply speaking,
our approach is as follows. We start from a random-number action that
has statistical symmetries (e.g., the Lorentz symmetry), go through the standard canonical
quantization, and finally reach a path-integral formula of the random $S$-matrix.
We develop a general diagrammatic method for calculating the random
$S$-matrix and then the density matrix. For canceling the divergence in
ultraviolet limit, we find a way to renormalize the coupling strength
between noise field and quantum field. After the renormalization, the
physical quantities like $S$-matrix and density matrix acquire finite values.

To demonstrate our method, we study three specific models. The first model is a
non-relativistic one, i.e. a harmonic oscillator with a white-noise force acting on it.
This model is mainly used to explain the quantization process of a random action
and establishment of the path integral formalism. We find that the effect of random force is to randomize
the position of a wave packet's center, and the model can be used to simulate
the decoherence of wave function in an environment in which
the pointer observable is position. The main part of this paper is devoted to the second
and third models which both have Lorentz symmetry. While the second model describes
a massive spin-zero scalar field coupled linearly to a white-noise field,
the third model additionally includes the $\phi^4$-interaction between particles.
In the absence of interaction, exact expressions of scattering matrix
and density matrix are obtained. After the interaction is considered, we derive all the
diagrammatic rules for perturbative calculations. In the two relativistic
models, we find that the effect of white-noise field is to thermalize the universe towards an
infinite-temperature state by exciting particles with Poisson distribution from the vacuum.
Our findings show that the action formulation succeeds in describing
the Lorentz-invariant stochastic processes in the Hilbert space,
even the calculation is more complicated than that of conventional QFTs.

The paper is organized as follows. In Sec.~\ref{sec:harmonic}, we introduce
the action formulation of stochastic dynamics using the example of nonrelativistic harmonic oscillator.
This example is a good starting point for readers who are not familiar with stochastic calculus.
Our formulation is then applied to relativistic QFT in Sec.~\ref{sec:phi},
in which the random-number action, canonical quantization,
path integral, diagrammatic rules and renormalization are explained
step by step. The expressions of state vector and density matrix are obtained for
arbitrary initial states including the vacuum and single-particle states.
In Sec.~\ref{sec:phi4}, we quantize the stochastic $\phi^4$-theory,
deriving the diagrammatic rules, using them to solve the two-particle
collision problem and discussing the difference between the predictions
of our theory and conventional QFT. Finally, Sec.~\ref{sec:summary} summarizes our results.

\section{Action formulation of random unitary evolution in the Hilbert space\label{sec:harmonic}}

\subsection{The Wiener process and a random action\label{sec:ha:ac}}

Let us start our discussion from a simple physical system - one-dimensional harmonic oscillator
whose action reads
\begin{equation}\label{eq:ha:lag0}
I_0 = \int^{t'}_{t_0} dt \left[ \frac{1}{2} m \left( \frac{d x}{dt} \right)^2 - \frac{1}{2}m \omega^2
x^2 \right],
\end{equation}
where $m$ is the mass, $x(t)$ is the position of particle at time $t$, and $\omega$ is the
oscillating frequency. In classical mechanics, given the initial and final positions, e.g. $x(t_0)=x_0$
and $x(t')=x'$, one can find the path of particle by minimizing the action.

Suppose there is a random force acting on the particle. We hope that the
dynamics keeps Markovian, which means that
the particle's future state depends only upon its current state but is independent
of its past state. This requirement puts strong constraint on the random force which
must be white. A mathematically elegant way of
describing a white-noise force is to add a random term
into the action, which then becomes
\begin{equation}\label{eq:ha:lag}
I_W = I_0 + \lambda \int^{t'}_{t_0} dW_t \ x(t),
\end{equation}
where $\lambda$ is the strength of random coupling, and $W_t$ is the Wiener
process. The integral over Wiener process is the so-called It\^{o} integral, defined as
\begin{equation}\label{eq:ha:ito}
\int^{t'}_{t_0} dW_t \ x(t) = \lim_{\Delta t\to 0} \ \sum_{j=0}^{N-1} \Delta W_{t_j} x(t_j),
\end{equation}
where $t_0 < t_1 < \cdots < t_{N-1} < t_N = t'$ with $t_{j+1} - t_j = \Delta t$ is a partition
of the interval $[t_0, t']$, and $\Delta W_{t_j} = W_{t_{j}+\Delta t}- W_{t_j}$ is an infinitesimal
increment. Note that the Wiener process is non-differentiable everywhere, hence,
we cannot express $dW_t $ as $\displaystyle \frac{dW_t }{dt}dt $ because $\displaystyle\frac{dW_t}{dt}$
does not exist. As a consequence,
we cannot see the action as an integral of some Lagrangian, because the latter is not well-defined.

A defining property of the Wiener process is that the increments $\Delta W_{t_j}$ are all independent
Gaussian random variables with zero mean and $\Delta t$ variance. Therefore,
when doing the calculations involving an action like Eq.~\eqref{eq:ha:lag},
one can start from its time-discretization form in which
$\left\{ \Delta W_{t_0}, \Delta W_{t_1},\cdots, \Delta W_{t_{N-1}}\right\}$
is a set of independent Gaussians, and only take the limit $\Delta t\to 0$
in the final results.

Next we minimize the action~\eqref{eq:ha:lag} to find the classical equation of motion.
Bearing in mind that $I_W$ is a random number whose randomness comes
from $W_t$, we define the minimization of $I_W$
in a pathwise way. For each path of $W_t$, or in the discretized version, for each
value of the vector $\left(W_{t_0}, W_{t_1}, \cdots, W_{t_{N-1}}\right)$, we minimize
$I_W$ to obtain the path $x(t)$ which then becomes a functional of $W_t$. Since $W_t$ is a random path,
$x(t)$ is also a random path (or stochastic process). In probability theory, one usually says
that the set of paths of $W_t$ forms a sample space, and the probability
of a path $x(t)$ or other quantities can be defined by seeing them as a functional of $W_t$.

The minimization of $I_W$ results in a SDE
\begin{equation}\label{eq:ha:Lang}
m d v = -m\omega^2 x dt + \lambda dW_t,
\end{equation}
where $v = dx/dt$ is the velocity. Note that without the random force ($\lambda=0$),
the minimization of action gives the Euler-Lagrangian equation.
But in the presence of randomness, Eq.~\eqref{eq:ha:Lang} is recognized as the
Langevin equation. Next we focus on how to quantize this theory.

\subsection{Canonical quantization \label{sec:ha:ca}}

We start from the action~\eqref{eq:ha:lag} and go through the canonical
quantization process. Notice that we keep viewing $W_t$ as a classical noise during
the quantization, thereafter, our formalism is different from the quantum Langevin equation~\cite{Breuer02}.
First, we perform the Legendre transformation to
obtain a Hamilton equation that is equivalent
to the equation of motion~\eqref{eq:ha:Lang}. Since the action is in a quadratic
form of velocity and the velocity is not coupled to $W_t$,
we then define the canonical momentum as
$ p(t) = {\delta I_W}/{\delta v(t)} = m v(t)$,
where $\delta/ \delta v$ is the functional derivative. $p(t)$ is understood as a functional of
$W_t$ and then a random variable. In the same reason as Lagrangian, the Hamiltonian
is not well-defined, instead, we can only define something like its time integral,
namely the Hamiltonian integral reading
\begin{equation}\label{eq:ha:ca:orH}
\begin{split}
\mathcal{H} & = \int_{t_0}^{t'} dt \ v(t) p(t) - I_W \\ & = 
\int_{t_0}^{t'}dt \  \left[\frac{p(t)^2}{2m}+ \frac{1}{2} m\omega^2 x(t)^2\right] - \lambda
\int_{t_0}^{t'} dW_t \ x(t) .
\end{split}
\end{equation}
Again, $\mathcal{H}$ being a random variable must be seen as a functional of $W_t$.

Now taking the variation of $\mathcal{H}$ with respect to $x(t)$ and $p(t)$, we obtain
\begin{equation}\label{eq:ha:ca:H}
\begin{split}
\delta \mathcal{H} = &
\int_{t_0}^{t'} \left( dt \frac{p(t)}{m}\right) \  \delta p(t) \\ & + \int_{t_0}^{t'} \left(dt \ m\omega^2 x(t) - \lambda
dW_t  \right)\ \delta x(t) .
\end{split}
\end{equation}
In Eq.~\eqref{eq:ha:ca:H}, the bracketed terms
describe the change of $\mathcal{H}$ caused by $\delta p(t)$ and $\delta x(t)$, respectively.
They play the roles of $ \frac{\displaystyle\delta \mathcal{H}}{\displaystyle\delta p(t)}dt$ and
$\frac{\displaystyle\delta \mathcal{H} }{\displaystyle\delta x(t)}dt$. The Hamilton equation
is traditionally built by letting $\delta \mathcal{H} /\delta p(t)$ and $\delta \mathcal{H} /\delta x(t)$
equal $dx(t)/dt$ and $-d p(t)/dt$, respectively.
But in the presence of randomness, we have to avoid using
$\delta \mathcal{H} /\delta p(t)$ or $\delta \mathcal{H} /\delta x(t)$
which are not well-defined. Instead, we let the bracketed terms equal $dx(t)$
and $-dp(t)$, respectively, and then obtain
\begin{equation}\label{eq:ha:ca:Heq}
\begin{split}
& dx(t) = \frac{p(t)}{m} dt , \\
&  -dp(t) = m\omega^2 x(t) dt - \lambda dW_t ,
\end{split}
\end{equation}
where $dx(t) = x(t+dt)-x(t)$ and $dp(t)= p(t+dt)-p(t)$
are the differentials of canonical coordinate and momentum, respectively.
It is easy to verify that Eq.~\eqref{eq:ha:ca:Heq} is equivalent to the Hamilton equation
of a harmonic oscillator in the case $\lambda=0$. As $\lambda$ is finite,
Eq.~\eqref{eq:ha:ca:Heq} is equivalent to the
equation of motion~\eqref{eq:ha:Lang}. Therefore, Eq.~\eqref{eq:ha:ca:Heq} is the replacement
of the Hamilton equation in the presence of random forces, just as the Langevin equation~\eqref{eq:ha:Lang}
is the replacement of the Euler-Lagrangian equation.

The quantization is to simply replace the canonical coordinate $x(t)$ and momentum $p(t)$
by the operators $\hat x(t)$ and $\hat p(t)$, respectively. We need these operators
satisfy exactly the same equation:
\begin{equation}\label{eq:ha:ca:Heqop}
\begin{split}
& d \hat x(t) = \frac{\hat p(t)}{m} dt, \\
& -d\hat p(t) = m\omega^2 \hat x(t) dt - \lambda dW_t.
\end{split}
\end{equation}
This can be done by defining the commutator $\left[ \hat x(t), \hat p(t) \right] = i$.
We choose the unit $\hbar = 1$ throughout this paper. And the infinitesimal
transformation from $\hat x(t)$ or $\hat p(t)$ to $\hat x(t+dt)$ or $\hat p(t+dt)$, respectively,
now becomes a unitary transformation expressed as
\begin{equation}\label{eq:ha:ca:uni}
\begin{split}
& \hat x(t+dt) = e^{i d\hat{\mathcal{H}}_t} \hat x(t) e^{-i d\hat{\mathcal{H}}_t }, \\
& \hat p(t+dt) = e^{i d\hat{\mathcal{H}}_t} \hat p(t) e^{-i d\hat{\mathcal{H}}_t },
\end{split}
\end{equation}
where $d\hat{\mathcal{H}}_t$ is the differential of Hamiltonian integral which equals
the Hamiltonian multiplied by $dt$ as $\lambda=0$. By reexpressing
the right-hand side of Eq.~\eqref{eq:ha:ca:orH} in a time-discretization form, we easily find
\begin{equation}\label{eq:ha:ca:infH}
d \hat{\mathcal{H}}_t =  \left[\frac{\hat p(t)^2}{2m}+ \frac{1}{2} m\omega^2 \hat x(t)^2\right]dt -
 \lambda \hat x(t) dW_t .
\end{equation}
Substituting Eq.~\eqref{eq:ha:ca:infH} into Eq.~\eqref{eq:ha:ca:uni} and using the
commutator relation, we recover Eq.~\eqref{eq:ha:ca:Heqop}.

Therefore, the canonical quantization still works even in the presence of random forces.
But it must be adapted. Due to the non-differentiability of Wiener process, some physical quantities become
non-differentiable so that we cannot use their
derivatives, instead, we use their differentials.
This explains why we do not use the Lagrangian and Hamiltonian which
are the derivatives of $I_W$ and $\mathcal{H}$, respectively, because $I_W$ and $\mathcal{H}$
are non-differentiable as $\lambda \neq 0$.

In quantum mechanics, Eq.~\eqref{eq:ha:ca:Heqop} or the equivalent Eq.~\eqref{eq:ha:ca:uni}
are understood as the equations of operators in the Heisenberg picture.
In this picture, the wave function does not change with time.
The next step is to choose a reference time $t_0$ and change into the Schr\"{o}dinger picture.
We use $\hat x(t_0)\equiv \hat x $ and $\hat p(t_0)\equiv \hat p$ as the
coordinate and momentum operators, respectively, in the Schr\"{o}dinger picture.
They are connected to the operators in the Heisenberg picture by
a unitary evolution reading
\begin{equation}\label{eq:ha:ca:HtoS}
\begin{split}
& \hat x(t') = \hat U^\dag(t',t_0) \hat x(t_0) \hat U(t',t_0), \\
& \hat p(t') = \hat U^\dag(t',t_0) \hat p(t_0) \hat U(t',t_0),
\end{split}
\end{equation}
where, by iteratively using Eq.~\eqref{eq:ha:ca:uni}, we find
\begin{equation}\label{eq:ha:ca:unisch}
\hat U(t', t_0) = \lim_{\Delta t\to0} e^{- i \Delta \hat{\mathcal{H}}_{t_0} }
e^{- i \Delta\hat{\mathcal{H}}_{t_1} } \cdots
e^{- i \Delta\hat{\mathcal{H}}_{t_{N-1}} }
\end{equation}
with $t_j = t_0+ j \Delta t$ and $t' = t_N$. 
Note that $\Delta\hat{\mathcal{H}}_{t_{j}} $ is the change of Hamiltonian
integral over the interval $\left[t_j,t_j+\Delta t\right]$. In this paper, we use the differential symbol $d$ as we
want to emphasize an infinitesimal change, but the difference symbol $\Delta $
for a finite change.

The wave function in the Schr\"{o}dinger picture
evolves unitarily according to
\begin{equation}\label{eq:ha:ca:wave}
\ket {\psi(t')} = \hat U(t',t_0) \ket {\psi(t_0)},
\end{equation}
where $\ket {\psi(t_0)}$ is the wave function at the reference time
which equals the wave function in the Heisenberg picture. As $\lambda\neq 0$, the
infinitesimal unitary transformation $e^{- i d\hat{\mathcal{H}}_{t} }$ depends on $dW_{t}$,
being then time-dependent. We are dealing with an evolution similar to that governed by
a time-dependent Hamiltonian.
According to the definition~\eqref{eq:ha:ca:unisch}, an infinitesimal evolution from $t$
to $t+dt$ is $\hat U(t+dt,t) = e^{- i d\hat{\mathcal{H}}_{t} }$, where
$d\hat{\mathcal{H}}_{t}$ in Eq.~\eqref{eq:ha:ca:infH} is expressed in terms of $\hat x(t)$ and $\hat p(t)$.
Using Eq.~\eqref{eq:ha:ca:HtoS},
we replace $\hat x(t)$ and $\hat p(t)$ by the corresponding operators
in the Schr\"{o}dinger picture, respectively, and then find
$\hat U(t+dt,t) = \hat U^\dag(t,t_0)
e^{-i d\hat{\mathcal{H}}^S_{t} }
 \hat U(t,t_0)$, where
\begin{equation}\label{eq:ha:ca:infHS}
d\hat{\mathcal{H}}^S_{t} = \left(\frac{\hat p^2}{2m}+ \frac{1}{2} m\omega^2 \hat x^2\right)dt -
 \lambda \hat x dW_t 
\end{equation}
is the differential of Hamiltonian integral~\eqref{eq:ha:ca:infH} but with
all the operators replaced by their time-independent versions
in the Schr\"{o}dinger picture. We call $d\hat{\mathcal{H}}^S_{t}$
the Hamiltonian differential in the Schr\"{o}dinger picture.
Using the unitarity of $\hat U$ and noticing
$\hat U(t+dt,t_0) = \hat U (t,t_0) \hat U(t+dt,t)$, we obtain
\begin{equation}\label{eq:ha:ca:schop}
\hat U(t+dt,t_0) = e^{-i d\hat{\mathcal{H}}^S_{t}} \hat U(t,t_0),
\end{equation}
By iteratively using Eq.~\eqref{eq:ha:ca:schop}, we find
\begin{equation}\label{eq:ha:ca:schev}
\hat U(t',t_0) = \lim_{\Delta t\to 0} e^{-i \Delta\hat{\mathcal{H}}^S_{t_{N-1}}} \cdots 
e^{-i \Delta\hat{\mathcal{H}}^S_{t_{1}}}  e^{-i \Delta\hat{\mathcal{H}}^S_{t_{0}}}.
\end{equation}
The evolution operator is now reexpressed in a form that we are familiar with.
In this expression, the time dependence of $\Delta\hat{\mathcal{H}}^S_{t} $
comes only from $\Delta W_t$ but not from the operators.

From Eq.~\eqref{eq:ha:ca:schop} and~\eqref{eq:ha:ca:wave},
we easily derive
\begin{equation}\label{eq:ha:ca:scheq}
\ket{\psi(t+dt)} = e^{-i d\hat{\mathcal{H}}^S_{t}} \ket{\psi(t)}.
\end{equation}
This equation tells us how the wave function evolves. The unitarity of
$e^{-i d\hat{\mathcal{H}}^S_{t}}$ guarantees the normalization of wave function
in a pathwise way, that is $\braket{\psi(t)|\psi(t)}= 1$ for arbitrary $t$ and
path of $W_t$. We calculate
the differential of wave function. Notice that one must keep the second order terms
in the Taylor series of $e^{-i d\hat{\mathcal{H}}^S_{t}}$, because the exponent contains
not only $dt$ but also $dW_t$ and $\left(dW_t\right)^2 \sim dt$ is in the first order
of $dt$ according to the It\^{o} calculus. The result is
\begin{equation}\label{eq:ha:ca:dsch}
\begin{split}
d \ket{\psi(t)} =& -i dt \left( \frac{\hat p^2}{2m}+ \frac{1}{2} m\omega^2 \hat x^2 \right)
\ket{\psi(t)} \\ & + i\lambda dW_t \ \hat x \ket{\psi(t)} -\frac{\lambda^2}{2} dt \ \hat x^2 \ket{\psi(t)}.
\end{split}
\end{equation}
In the case $\lambda=0$, Eq.~\eqref{eq:ha:ca:dsch} reduces to
the Schr\"{o}dinger equation. As $\lambda \neq 0$, Eq.~\eqref{eq:ha:ca:dsch} becomes
a SDE which describes a random unitary evolution in
the Hilbert space. Notice that $\ket{\psi(t)}$ is a functional of $W_t$, being
then a random vector.

To see how the random term in Eq.~\eqref{eq:ha:ca:dsch} works,
we temporarily neglect the Hamiltonian term (the first term in the right-hand side
of the equal sign), and then the solution becomes
$\psi(t,x)=e^{i\lambda x \left(W_t-W_{t_0}\right)} \psi(t_0,x)$, where $\psi(t,x)
= \braket{x|\psi(t)}$ is the wave function in real space. This simple analysis
indicates that the random force after quantization tends to contribute a random
phase to the wave function.

It is worth comparing Eq.~\eqref{eq:ha:ca:dsch} with the collapse model - QMUPL.
The difference is that the coupling between $\hat x$ and the
Wiener process in Eq.~\eqref{eq:ha:ca:dsch} is
purely imaginary but it is real in the QMUPL model, and there is a
lack of the nonlinear term $\bra{\psi(t)}\hat x \ket{\psi(t)}dW_t $
in Eq.~\eqref{eq:ha:ca:dsch}. Such a difference leads to a different fate of
the wave packet, as will be further discussed in next.

Let us shortly discuss the master equation of density matrix. The density
matrix is defined as $\hat \rho(t) = \text{E} \left(\ket{\psi(t)}\bra{\psi(t)}\right)$
where $\text{E}\left(\cdot\right)$ denotes the mean over all the paths of
$W_t$. From Eq.~\eqref{eq:ha:ca:dsch}, it is straightforward to derive
\begin{equation}\label{eq:ha:ca:lind}
\begin{split}
\frac{d\hat \rho(t)}{dt}=& -i \left[\frac{\hat p^2}{2m}+ \frac{1}{2} m\omega^2 \hat x^2,
\hat \rho(t) \right] \\ & + \lambda^2 \left(\hat x \hat \rho(t)\hat x - \frac{1}{2} \hat x^2 \hat
\rho(t) - \frac{1}{2} \hat \rho(t) \hat x^2\right).
\end{split}
\end{equation}
This is a typical Lindblad equation with the first term in the right-hand side
describing the unitary evolution governed by the Hamiltonian and the second
term describing the decoherence. Indeed, Eq.~\eqref{eq:ha:ca:lind} is the
master equation for single-particle Brownian motion in the macroscopic
and high-temperature limit~\cite{Zurek03}. It is typically employed to describe an environment-induced
decoherence when position is the instantaneous pointer observable. To see the decoherence, we
temporarily neglect the first term, and then the solution of Eq.~\eqref{eq:ha:ca:lind}
becomes
\begin{equation}\label{eq:ha:ca:rhoxy}
\rho_{x,y}(t) =e^{-\frac{\lambda^2}{2} \left(x-y\right)^2 t} \rho_{x,y}(0),
\end{equation}
where $\rho_{x,y}(t) = \bra{x} \hat \rho(t)\ket{y}$ is the density matrix element
in the coordinate space. It is clear that the diagonal elements with $x=y$ keep invariant
but the off-diagonal elements decay exponentially with a rate proportional to the squared distance.
The model~\eqref{eq:ha:lag}, which is originally proposed for demonstrating our approach,
can also be used to simulate a master equation which describes the rapid destruction of
nonlocal superposition and emergent classicality of the eigenstates of position.

\subsection{Path integral approach \label{sec:ha:path}}

The density matrix gives only a statistical description of the quantum state.
To see how a wave packet evolves, we need solve the SDE~\eqref{eq:ha:ca:dsch}.
A possible method is to assume a form of the wave function (such as a Gaussian function),
obtain the SDE of undetermined parameters, and write down the solution in terms of
Wiener process~\cite{Bassi05}. This method is
difficult to be generalized to QFT. Here we choose a different way - the path integral approach.

Let us calculate the propagator
\begin{equation}
\braket{t'x'|t_0x_0}= \bra{x'} \hat U(t',t_0) \ket{x_0},
\end{equation}
where $t'>t_0$ is an arbitrary ending time. Expressing $\hat U(t',t_0)$
in terms of infinitesimal unitary evolutions and utilizing the complete relations
in the coordinate and momentum spaces, we obtain
\begin{equation}
\begin{split}
\braket{t'x'|t_0x_0}= & \lim_{\Delta t\to0} \int \left(\prod_{j=1}^{N-1} dx_{j}\right)\left(
\prod_{j=0}^{N-1} dp_{j}\right) \\
& \times \prod_{j=0}^{N-1} \left(\braket{x_{j+1}|p_j} \bra{p_j}
e^{-i \Delta\hat{\mathcal{H}}^S_{t_{j}}}\ket{x_{j}}\right),
\end{split}
\end{equation}
where $x'= x_N$, $t'= t_N$ and $\braket{x_{j+1}|p_j}=e^{ip_jx_{j+1}}/\sqrt{2\pi}$.
The Baker-Campbell-Hausdorff formula tells us
\begin{equation}
e^{-i \Delta\hat{\mathcal{H}}^S_{t_{j}}}  = e^{-i \frac{\hat p^2}{2m}\Delta t }
e^{-i \left( \frac{1}{2} m\omega^2 \hat x^2\Delta t-
 \lambda \hat x \Delta W_t\right) } e^{\hat X},
\end{equation}
where $\hat X $ is a series of operators with the lowest order terms proportional
to $\Delta t\Delta W_t$ or $\Delta t^2$, therefore, $e^{\hat X}$
can be dropped in the limit $\Delta t\to 0$ according to the It\^{o} calculus.
Indeed, in the It\^{o} calculus, except for the linear terms,
only $\Delta W_t^2$ need to be kept in the second order terms, and
all the third and higher order terms can be neglected. Integrating out
the momentum $p_0, \cdots, p_{N-1}$, we obtain
\begin{equation}\label{eq:ha:path:pa}
\begin{split}
&\braket{t'x'|t_0x_0} \\ = & \lim_{\Delta t\to0} \left(\frac{2\pi i \Delta t}{m}\right)^{-N/2} 
\int \left(\prod_{j=1}^{N-1} dx_{j}\right) \\
& exp \bigg\{ i \displaystyle\sum_{j=0}^{N-1} \bigg[\left( \frac{1}{2}m \left(
\frac{x_{j+1}-x_j}{\Delta t} \right)^2- \frac{1}{2} m\omega^2 x_j^2  
\right)\Delta t \\ & +\lambda x_j \Delta W_{t_j} \bigg] \bigg\}.
\end{split}
\end{equation}
The exponent in Eq.~\eqref{eq:ha:path:pa} is recognized as
$i I_W$ in the limit $\Delta t \to 0$. We can then rewrite it
by employing the path integral notation:
\begin{equation}
\braket{t'x'|t_0x_0} = \int D x(t) \ e^{iI_W}.
\end{equation}

It is encouraging to see that the path integral approach is working and leads to
the same Feynman integral formula even in the presence of white noise.
In general, if the Wiener process is coupled to the coordinate but not to
the velocity, then the definition of momentum keeps the same and the random term
can be treated as an additional potential, thereafter, the Feynman's formula
keeps the same.

\subsection{Stochastic dynamics of wave packet \label{sec:ha:omega}}

Next we work out the propagator for the special case $\omega =0$.
In Eq.~\eqref{eq:ha:path:pa}, we can see $i I_W$ as a quadratic
function of the vector $x = (x_1,\cdots,x_{N-1})^T$ and reexpress it as
\begin{equation}\label{eq:ha:om:IW}
iI_W = -\displaystyle\frac{1}{2}\frac{m}{i \Delta t} \ x^T A x + i B^T x + C ,
\end{equation}
where
\begin{equation}
A = \left( \begin{array}{cccc}
2 & -1 & 0 & \cdots \\
-1 & 2 & -1 & \cdots \\
0 & -1 & 2 & \cdots \\
\cdots & \cdots & \cdots & \cdots 
\end{array}\right)
\end{equation}
is a $(N-1)$-dimensional tridiagonal matrix,
\begin{equation}\label{eq:ha:om:B}
B = \left( \begin{array}{c}
\lambda \Delta W_{t_1} - \displaystyle \frac{m}{\Delta t} x_0 \\
\lambda \Delta W_{t_2} \\
\cdots \\
\cdots \\
\lambda \Delta W_{t_{N-2}} \\
\lambda \Delta W_{t_{N-1}} - \displaystyle \frac{m}{\Delta t} x'
\end{array}\right)
\end{equation}
is a $\left(N-1\right)$-dimensional vector, and $A$, $B$ and $C$
are all independent of $x_j$ for $j=1,\cdots N-1$. To calculate
the multi-variant Gaussian integral in Eq.~\eqref{eq:ha:path:pa}, we use the formula
\begin{equation}\label{eq:ha:path:oz}
\begin{split}
& \int \left(\prod_{j=1}^{N-1} dx_{j}\right) e^{iI_W(x_1, x_2,\cdots, x_{N-1})} \\
& = \left(\frac{m}{i 2\pi \Delta t }\right)^{-\frac{N-1}{2}} \left( \text{Det}A\right)^{-1/2}
 e^{iI_W(\bar{x}_1, \bar{x}_2,\cdots, \bar{x}_{N-1})},
\end{split}
\end{equation}
where $\text{Det}A$ is the determinant of $A$, and
 $\left(\bar{x}_1,\cdots,\bar{x}_{N-1}\right)$
is the stationary point of $I_W$ defined by $\partial I_W/\partial x_j=0$ for each $j$.
$\bar{x}(t_j)=\bar{x}_j$ can be also seen as the classical path which minimizes
$I_W$, and $\bar{I}_W=I_W(\bar{x})$ is the classical action.

Notice that the Wiener process is linearly coupled to the canonical coordinate
in our model. This is a very useful property in the calculation. Because the coefficient
matrix $A$ contains no $\Delta W_t$, its determinant or inverse are exactly
the same as those in the absence of randomness, hence, the familar
path integral formula can be directly applied. Usually, the path integral
results in a simple expression only if the random term is linear.

By using mathematical induction, we easily prove $\text{Det}A = N$.
And according to Eq.~\eqref{eq:ha:om:IW}, the stationary point of $I_W$ is
$\bar{x} = -\left({\Delta t}/{m} \right) A^{-1} B$.
By some tidy calculation, we find that the matrix elements of
the inverse of $A$ can be expressed as
$A^{-1}_{j,j'} = \frac{1}{N} \left[N \text{min}(j,j')  - jj'\right]$
with $j,j'=1,\cdots,N-1$ and $\text{min}(j,j')$ denoting the smaller one
between $j$ and $j'$. The stationary point is found to be
\begin{equation}\label{eq:ha:om:xj}
\begin{split}
\bar{x}_j = & \frac{(N-j)x_0+jx'}{N} \\ & -\frac{\lambda \Delta t}{m} \sum_{j'=1}^{N-1}\Delta W_{t_{j'}}
\left(\text{min}(j,j') -\frac{jj'}{N}\right).
\end{split}
\end{equation}
The first term is recognized as the position of the particle at time $t_j$ if
it is moving at a constant velocity from the coordinates $(t_0,x_0)$ to $(t',x')$,
as being expected since the classical path without random forces is
a straight line in the spacetime.
And the second term gives the contribution of the random force to the path.
By using Eq.~\eqref{eq:ha:path:oz} and~\eqref{eq:ha:om:xj}, we finally evaluate
Eq.~\eqref{eq:ha:path:pa} to be
\begin{equation}
\braket{t'x'|t_0x_0} = \sqrt{\frac{m}{2\pi i (t'-t_0)}} e^{i \bar{I}_W},
\end{equation}
where $\bar{I}_W$ reads
\begin{widetext}
\begin{equation}\label{eq:ha:om:ibarw}
\begin{split}
\bar{I}_W = & \lim_{\Delta t\to 0} \bigg\{ \frac{1}{2}m\frac{\left(x'-x_0\right)^2}{t'-t_0} 
+ \sum_{j=0}^{N-1} \lambda \Delta W_{t_j} \frac{x_0(t'-t_j)+x'(t_j-t_0)}{t'-t_0} \\ &
- \frac{\lambda^2}{2m} \sum_{j,j'=0}^{N-1} \Delta W_{t_j} \Delta W_{t_{j'}}
\frac{\left[t'-\text{max}(t_j,t_{j'})\right]\left[\text{min}(t_j,t_{j'})-t_0\right]}{t'-t_0} \bigg\} \\
= & \frac{1}{2}m\frac{\left(x'-x_0\right)^2}{t'-t_0} 
+ \lambda \int^{t'}_{t_0} d W_{t} \frac{x_0(t'-t)+x'(t-t_0)}{t'-t_0} 
- \frac{\lambda^2}{2m} \int^{t'}_{t_0} d W_{t_1} \int^{t'}_{t_0} d W_{t_2} 
\frac{\left[t'-\text{max}(t_1,t_2)\right]\left[\text{min}(t_1,t_2)-t_0\right]}{t'-t_0}.
\end{split}
\end{equation}
\end{widetext}
Using the independent-increment property of $W_t$, it is easy to see that $\bar{I}_W$
depends only upon the difference $t'-t_0$ but is independent of the initial time.
As $\lambda=0$, only the first term of $\bar{I}_W$ survives and our result
repeats the well-known propagator of a free particle. As $\lambda \neq 0$, the
propagator is different from the free-particle one by a random phase that is
the combination of second and third terms of $\bar{I}_W$.

Using the propagator, we can study the evolution of a general wave packet.
Suppose the quantum state is pure with a wave function ${\psi}(t_0,x_0)$ at the initial time $t_0$,
the wave function at an arbitrary later time $t'$ is then
\begin{equation}\label{eq:ha:in:ptpxp}
\psi(t', x') = \int dx_0 \braket{t'x'|t_0x_0} \psi(t_0, x_0).
\end{equation}
In the study of the decoherence of a single particle, the initial state is usually assumed to
be a Gaussian wave packet centered at $q_0$ with an averaged momentum $p_0$, reading
$\psi(t_0, x_0)=\left(\pi \sigma_0^2\right)^{-1/4} 
exp\left[ -\frac{\left(x_0-q_0\right)^2}{2\sigma_0^2} + ip_0 x_0 \right]$
with $\sigma_0$ denoting the initial packet width. Let us
see how this wave packet evolves in course of time. The propagator being independent of the
choice of $t_0$ makes the calculation easier (we can simply set $t_0=0$). By using Eq.~\eqref{eq:ha:in:ptpxp},
we obtain the expression of $\psi(t', x')$. We are only interested in the probability distribution of
the particle's position, which is
\begin{equation}
\left| \psi(t', x')\right|^2 = \left[\pi \sigma^2(t')\right]^{-1/2} 
e^{-\frac{\left(x'-q(t')\right)^2}{\sigma^2(t')}}.
\end{equation}
The packet width $\sigma(t')= \sigma_0\sqrt{1+\frac{\left(t'-t_0\right)^2}{m^2\sigma_0^4}}$
is increasing linearly with time as the evolution time is much larger than $m\sigma_0^2$.
While the center of packet is at
\begin{equation}\label{eq:ha:path:qt}
q(t') = q_0 + \frac{p_0+p_1}{m}(t'-t_0).
\end{equation}
with $p_1=\lambda \displaystyle\int_{t_0}^{t'} dW_t \left[1- (t-t_0)/(t'-t_0)\right]$ being a functional
of $W_t$. Eq.~\eqref{eq:ha:path:qt} tells us that the packet center is changing at a velocity
${(p_0+p_1)}/{m}$ which contains a constant part $p_0/m$ and
a random part $p_1/m$ with $p_1$ being just the accumulated change of momentum
caused by the random force.

It is now clear that the packet is widening at a speed independent of
the random force. While the Hamiltonian controls the widening of the wave packet,
the effect of the random force is to randomize the position of the wave packet's center.

The action~\eqref{eq:ha:lag} is the simplest action with Markovian property that
after quantization can describe a random unitary evolution in Hilbert space.
The quantization techniques developed in this section can be easily generalized to a
field theory.

\subsection{Statistical symmetry \label{sec:ha:sym}}

Finally, we discuss the symmetry of model~\eqref{eq:ha:lag}.
We focus on the time translational symmetry.
In general, the path of $W_t$ is time-dependent, therefore, the
model~\eqref{eq:ha:lag} does not have the time translational
symmetry in a pathwise way. For a specific path of $W_t$,
if we perform the coordinate transformations $\tilde{t} = t+\tau$ and $\tilde{x} =x$,
then in the new coordinates $(\tilde{t},\tilde{x})$, the particle's path becomes
$\tilde{x}(\tilde{t}) = \tilde{x}(t+\tau) = x(t)$ and the action becomes
\begin{equation}
\begin{split}
\tilde{I}_W = & \int^{\tilde{t}'}_{\tilde{t}_0}d \tilde{t}
\left[ \frac{1}{2} m \left( \frac{d \tilde{x}}{d\tilde{t}} \right)^2 - \frac{1}{2}m \omega^2
\tilde{x}^2 \right] \\ & + \lambda \int^{\tilde{t}'}_{\tilde{t}_0} dW_{\tilde{t}} \ \tilde{x}(\tilde{t})
\end{split}
\end{equation}
with $\tilde{t}_0 = t_0+\tau$ and $\tilde{t}'={t}'+\tau$. The first term of $\tilde{I}_W$
is equal to that of $I_W$ in Eq.~\eqref{eq:ha:lag}, indicating that
the action keeps invariant under time translation as $\lambda=0$.
But due to $dW_t \neq dW_{\tilde{t}}$,
the second term of $\tilde{I}_W$ is different from that of $I_W$, therefore,
the time translational symmetry is explicitly broken by the random force.

On the other hand, we should not forget that $dW_t$ and $dW_{\tilde{t}}$
are both random numbers with exactly the same distribution, because the Wiener
process has stationary independent increments. Indeed, the vector
of increments from $t_0$ to $t'$, i.e. $\left(\Delta W_{t_0}, \Delta W_{t_1}, \cdots \Delta W_{t_{N-1}}\right)$
has exactly the same probability distribution as
$\left(\Delta W_{\tilde{t}_0}, \Delta W_{\tilde{t}_1}, \cdots \Delta W_{\tilde{t}_{N-1}}\right)$
that is the increments from $\tilde{t}_0$ to $\tilde{t}'$. In probability theory, we say that
the two vectors are equal in distribution to each other.
As emphasized above, $I_W$ is a functional of $\Delta W_t$ within
the interval $(t_0,t')$, and $\tilde{I}_W$ is the same functional but of $\Delta W_{\tilde{t}}$
in the interval $(\tilde{t}_0,\tilde{t}')$.
As a consequence, $I_W$ and $\tilde{I}_W$
are equal in distribution to each other, written as
\begin{equation}
I_W \stackrel{d}{=} \tilde{I}_W.
\end{equation}
The action keeps invariant in the meaning of probability distribution.
We say that the action has a statistical symmetry.

The statistical symmetry of action has important consequences.
First, since the Hamiltonian integral is the Legendre transformation of
action, $\hat{\mathcal{H}}$ and then $d \hat{\mathcal{H}}$ are both
statistically invariant under time translation. Then the Heisenberg
equation~\eqref{eq:ha:ca:Heqop} must be also statistically invariant,
so is the unitary evolution operator $\hat U(t',t_0)$. In the Schr\"{o}dinger
picture, the evolution operator $e^{- i d\mathcal{H}^S_t}$ and then
the SDE of wave function~\eqref{eq:ha:ca:dsch} have the statistical symmetry.
And because the density matrix is the expectation value
of $\ket{\psi(t)}\bra{\psi(t)}$, its dynamical equation~\eqref{eq:ha:ca:lind}
must have an explicit time translational symmetry, as been easily seen.
Therefore, a statistical symmetry of action indicates an explicit symmetry of master equation.
This conclusion stands for the other symmetries.

By using the path integral method, we have expressed the
propagator $\braket{t'x'|t_0x_0}$ as the integral of the exponential of action.
It is straightforward to see that the propagator has a statistical time
translational symmetry. In other words, the probability distribution of $\braket{t'x'|t_0x_0}$ depends
only upon the time difference $t'-t_0$, being independent of the initial time $t_0$.
This can be seen in the expression~\eqref{eq:ha:om:ibarw}.

In a model of random unitary evolution, the
conventional symmetry in quantum mechanics
is replaced by the statistical symmetry. Once if the action
has some statistical symmetry, by using the aforementioned quantization technique,
the resulting equation of motion or propagator will have the same
symmetry. This theorem can help us to construct a stochastic quantum theory
with more complicated symmetries, e.g. the Lorentz symmetry.

\section{Relativistic quantum field theory of random unitary evolution\label{sec:phi}}

Using the action approach developed above, we can now study
a quantum field theory in which the state vector experiences a random unitary evolution.
It is natural to put next two constraints on the theory.
First, the random evolution of state vector should be Markovian. In other words,
once if we know the current state vector, the future
state vector should be independent of the past one. After all,
if the information in the distant past is necessary for predicting
the future of universe, any theoretical prediction would be impossible.
Second, the Lorentz symmetry and spacetime translational symmetry
must be preserved, at least in a statistical way.
But we do not expect an explicit symmetry and its properties in the theory.

\subsection{Random scalar field and statistical Lorentz symmetry\label{sec:phi:ran}}

Let us start from the simplest Lorentz-invariant action
\begin{equation}
I_0 = \int d^4 x \left(-\frac{1}{2}\partial_\mu \phi(x) \partial^\mu \phi(x) - \frac{1}{2}
m^2 \phi^2(x) \right),
\end{equation}
where $x=(x^0,x^1,x^2,x^3)$ is the 1+3-dimensional spacetime coordinates,
$\phi(x)$ is a real scalar field and $\partial_\mu = {\partial}/{\partial x^\mu}$.
The sign of metric is chosen to be $\left(-, +, + ,+\right)$. And we use
the units $c=\hbar =1$. In QFT, $I_0$
is the action of a free boson of spin-zero and mass $m$.

According to the aforementioned action approach, we need to add to $I_0$
a random term without breaking the Lorentz symmetry. And
it is reasonable to first try a linear coupling between $\phi$ and some random field.
Following Eq.~\eqref{eq:ha:lag}, we formally write down the new action as
\begin{equation}\label{eq:phi:ra:IW}
I_W = I_0 + \lambda \int dW(x) \ \phi(x),
\end{equation}
where $dW(x)$ is the generalization of $dW_t$ in the 1+3-dimensional spacetime
and $\lambda$ is the coupling constant.
However, the mathematically precise definition of $dW(x)$ is not easy to see.
In above, we see $dW_t$ as the differential of the Wiener process.
One might naively think $dW(x)$ to be also the differential of some
stochastic process $W(x)$.
Unfortunately, it is difficult if not impossible to find such a $W(x)$. Because there are infinite
paths connecting two different points (say $x_1$ and $x_2$) in a multi-dimensional spacetime.
If we see $W(x)$ as a function of $x$, then $W(x_2)-W(x_1)$
must be an unambiguously defined random number, and can be obtained by accumulating
the infinitesimal increments along arbitrary path connecting $x_1$ and $x_2$.
But there is no way to guarantee that the sum of increments along
two different paths are the same, because these increments are random
numbers and they should be independent of each other (the Wiener process
has independent increments).
To further clarify the above idea, let us consider
$W(x^0+\Delta x^0, x^1+\Delta x^1)-W(x)$. Here we omit the coordinates $x^2$ and $x^3$
by simply assuming them to be constants. There are two paths connecting
$x$ and $(x^0+\Delta x^0, x^1+\Delta x^1)$ with the intermediate point
being $(x^0, x^1+\Delta x^1)$ and $(x^0+\Delta x^0, x^1)$, respectively.
It is obvious to see
\begin{equation}\label{eq:phi:ra:in}
\begin{split}
&\left[W(x^0+\Delta x^0, x^1+\Delta x^1)-
W(x^0, x^1+\Delta x^1)\right] \\ & + \left[W(x^0, x^1+\Delta x^1)- W(x)\right] \\
= & \left[W(x^0+\Delta x^0, x^1+\Delta x^1)-
W(x^0+\Delta x^0, x^1)\right] \\ & + \left[W(x^0+\Delta x^0, x^1)- W(x)\right].
\end{split}
\end{equation}
But this is impossible if the four increments in the brackets are independent random numbers.
Therefore, it is unreasonable to see $dW(x)$ as a differential.

On the other hand, in Eq.~\eqref{eq:ha:ito} we have defined $\displaystyle\int dW_t x(t)$
as the sum of $\Delta W_{t_j} x(t_j)$ with $\left(\Delta W_{t_0},
\Delta W_{t_1}, \cdots\right)$ being a sequence of independent random numbers.
In this definition and the following calculations, we never used
the existence of $W_t$ itself, even $W_t$ can be redefined as
the sum of $\Delta W_{t}$ over the interval $(0,t)$. In other words,
we can see $dW_t $ as an independent random number,
and whether $dW_t$ is the differential
of some function or not is unimportant. This observation inspires us
to see $dW(x)$ also as a random number but not the differential of $W(x)$.
Since $I_0$ in Eq.~\eqref{eq:phi:ra:IW} is a four-dimensional integral,
by comparing Eq.~\eqref{eq:phi:ra:IW} with Eq.~\eqref{eq:ha:lag}, we
define the variance of $dW(x)$ to be $d^4 x$, i.e. an infinitesimal
volume in the spacetime.

\begin{figure}[tbp]
\vspace{0.5cm}
\includegraphics[width=0.9\linewidth]{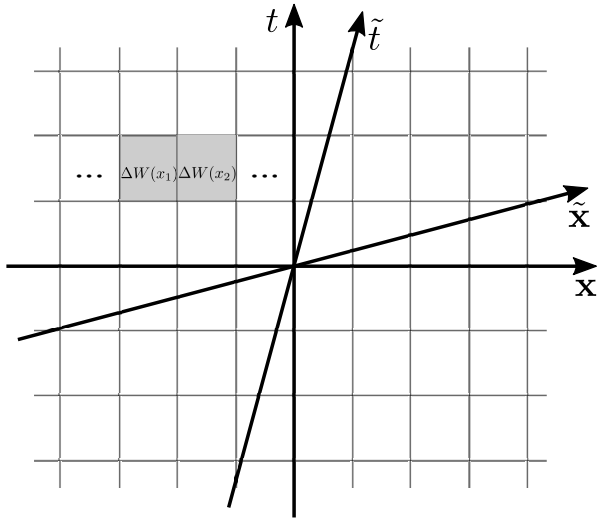}
\caption{A schematic illustration of the partition of spacetime
and the definition of random field. $\Delta W(x_1)$ and
$\Delta W(x_2)$, being assigned to different spacetime elements, are two
independent Gaussians of the same distribution. The coordinates $\left(\tilde t,\tilde{\bold{x}}\right)$
is a Lorentz transformation of $\left(t,\bold{x}\right)$.}\label{fig:spacetime}
\end{figure}
Now let us make a precise definition of $\displaystyle\int dW(x) \phi(x)$.
Partitioning the spacetime into a set of small elements of volume $\Delta^4 x$
(see Fig.~\ref{fig:spacetime} for a schematic illustration),
we then assign to each element at coordinate $x$
a Gaussian random number $\Delta W(x)$ which has zero mean and the variance $\Delta^4 x$,
and suppose $\left(\Delta W(x_1), \Delta W(x_2),\cdots\right)$ with $x_1\neq x_2\neq \cdots$
to be independent random numbers. We then define
\begin{equation}\label{eq:phi:ra:Wint}
\int dW(x) \ \phi(x) = \lim_{\Delta^4 x \to 0} \sum_{x} \Delta W(x) \phi(x),
\end{equation}
where the sum is over all the spacetime elements.

It is necessary to explain why the limit in Eq.~\eqref{eq:phi:ra:Wint} exists. Choose
$\Delta^4 x$ to be small enough so that the change of $\phi(x)$
within each element is negligible. Let us further partition
the element at $x_j$ into $M$ smaller pieces centered at $y^{(j)}_1, y^{(j)}_2,\cdots,y^{(j)}_M$,
respectively, with a volume $\Delta^4 y = \Delta^4 x/M$.
According to the definition of the random field, we assign to each piece at $y^{(j)}_i$
an independent Gaussian random number $\Delta W(y^{(j)}_i)$
of mean zero and variance $\Delta^4 y$.
Now the sum over the smaller elements at $y^{(j)}_i$ becomes
$\sum_{i,j} \Delta W(y^{(j)}_i) \phi(y^{(j)}_i) \approx \sum_j \phi(x_j)
\left(\sum_i\Delta W(y^{(j)}_i) \right)$.
But according to the properties of independent Gaussians,
$\sum_i\Delta W(y^{(j)}_i)$ is indeed a Gaussian of mean zero and variance $\Delta^4 x$
and then it has the same probability distribution as $\Delta W(x_j)$, furthermore,
$\sum_i\Delta W(y^{(j)}_i)$ at different $x_j$ are independent of each other. Therefore,
the vector $\left(\sum_i\Delta W(y^{(1)}_i), \sum_i\Delta W(y^{(2)}_i),\cdots\right)$
has exactly the same probability distribution as
$\left(\Delta W(x_1), \Delta W(x_2),\cdots\right)$.
As a consequence, $\sum_{i,j} \Delta W(y^{(j)}_i) \phi(y^{(j)}_i)$ has
the same distribution as $\sum_j \Delta W(x_j)\phi(x_j)$, or
they are equal in distribution to each other.
If $\Delta^4 x$ is small enough, further partitioning the spacetime
results in no change in the sum.
We say that the limit in Eq.~\eqref{eq:phi:ra:Wint} is well-defined
in the meaning of convergence in distribution.

It is well known that the action $I_0$ has the explicit spacetime translation
and Lorentz symmetries. We then check the symmetries of $\displaystyle\int dW(x) \phi(x)$.
First, suppose $\tilde x = x+a$ to be a spacetime translation with $a$
being a constant vector. In the new coordinate system, the integral becomes
$\displaystyle\int dW(\tilde{x}) \tilde{\phi}(\tilde{x})$, where the scalar field transforms as
$\tilde{\phi}(\tilde{x}) =\phi(x)$. Since an arbitrary vector
$\left(\Delta W(\tilde{x}_1), \Delta W(\tilde{x}_2),\cdots\right)$
equals in distribution to $\left(\Delta W({x}_1), \Delta W({x}_2),\cdots\right)$,
according to the definition~\eqref{eq:phi:ra:Wint}, we then have
\begin{equation}
\int dW(x) \phi(x)  \stackrel{d}{=} \int dW(\tilde{x}) \tilde{\phi}(\tilde{x}).
\end{equation}
The action~\eqref{eq:phi:ra:IW} has the statistical spacetime translational symmetry.

Next, suppose $\tilde{ x} = L x $ to be a Lorentz transformation. Under this
transformation, the scalar field still keeps invariant, i.e. $\tilde{\phi}(\tilde{x}) =\phi(x)$.
But we have to consider the possibility of $\Delta W(x)$ changing under $L$.
Without loss of generality, we use $\Delta\tilde{W}(\tilde x)$ to denote
the random field in the new coordinate system. For a specific partition
of the spacetime, $\Delta \tilde{W}(\tilde x)$ is a Gaussian of mean zero and
variance $\Delta^4 \tilde{x}$, according to definition. Recall that the Lorentz transformations do not
change the volume of spacetime (the determinant of metric tensor
is unity), hence, we have $\Delta^4 \tilde{x}= \Delta^4 {x}$, and then
$\Delta \tilde{W}(\tilde x)\stackrel{d}{=} \Delta W(x) $. Moreover,
the vector $\left(\Delta \tilde{W}(\tilde{x}_1), \Delta \tilde{W}(\tilde{x}_2),\cdots\right)$
equals in distribution to $\left(\Delta W({x}_1), \Delta W({x}_2),\cdots\right)$.
Due to this invariant property of $\Delta W(x)$, we call it a random scalar field.
Following the definition~\eqref{eq:phi:ra:Wint}, we obtain
\begin{equation}
\int dW(x) \phi(x)  \stackrel{d}{=} \int d\tilde{W}(\tilde{x}) \tilde{\phi}(\tilde{x}).
\end{equation}
The action~\eqref{eq:phi:ra:IW} has the statistical Lorentz symmetry.

It is now clear that the random scalar field in a stochastic QFT plays the role
of scalar field in QFT. By coupling $dW(x)$ to arbitrary field or combination
of fields that transform as a scalar, we can obtain a Lorentz-invariant random action.
The physical meaning of $dW(x)$ is clear. Its independence property
indicates that $dW(x)$ is a spacetime white noise, being both temporally and
spatially local. And $I_W$ describes scalar bosons driven by a background
white noise.

It is worth emphasizing that the action $I_W$ is
understood as a functional of the random field or
in the discretized version, the random vector $\left(\Delta W(x_1), \Delta W(x_2),\cdots\right)$.
All the other physical quantities derived from the action should
be similarly treated as the functionals. And then they are also random numbers.

\subsection{Canonical quantization\label{sec:phi:quan}}

Following the approach in Sec~\ref{sec:ha:ac}, we minimize the
action~\eqref{eq:phi:ra:IW}
and find
\begin{equation}\label{eq:phi:ca:EL}
\left[ - \partial_t^2 \phi(x)+ \nabla^2 \phi(x) -m^2 \phi(x) \right] d^4 x + \lambda dW(x)=0.
\end{equation}
This is a stochastic version of the Euler-Lagrangian equation.
One must be careful when using Eq.~\eqref{eq:phi:ca:EL}. Indeed,
nothing guarantees that the partial derivative of $\phi$ exists. Eq.~\eqref{eq:phi:ca:EL}
can only be treated as a difference equation based on a partition of
spacetime with finite $\Delta^4 x$, and $\partial_t$ or $\nabla$ should be
seen as quotients between finite differences. The limit $\Delta^4 x \to 0$ is only taken
after one obtains the solution $\phi$.
Solving the difference equation is
a difficult if not impossible task. Fortunately, we never need Eq.~\eqref{eq:phi:ca:EL}
or its solution in the quantization process.

One must bear in mind that the spacetime has already
been partitioned (see Fig.~\ref{fig:spacetime}) in the beginning, and
the action and all the following equations are defined
in a discretized spacetime with the derivative symbol
being a convenient abbreviation for quotient.
From now on, unless otherwise noted,
all the equations are based on a finite partition of spacetime. The limit
$\Delta^4 x \to 0$ is only taken after we obtain the final results.
But for convenience, we will also frequently use the integral symbol as an abbreviation
of summation when there is no ambiguity.

After a Legendre transformation of action, we find the Hamiltonian integral to be
\begin{equation}
\begin{split}
\mathcal{H} = & \int d^4 x \left[\frac{1}{2}\pi^2(x) + \frac{1}{2}
\left(\nabla\phi(x)\right)^2 +\frac{1}{2}m^2 \phi^2(x) \right] \\ &
-\lambda \int dW(x) \phi(x),
\end{split}
\end{equation}
where $\phi(x)$ and $\pi(x)$ are the canonical coordinate and momentum, respectively.
As we did in Sec.~\ref{sec:ha:ca}, we study the variation of $\mathcal{H}$ with respect to $\phi(x)$
and $\pi(x)$ and then obtain the Hamilton equation which reads
\begin{equation}\label{eq:phi:qu:ha}
\begin{split}
& \pi(x) \Delta^4 x = \Delta^3 \bold{x} 
\left[\phi(t+\Delta t, \bold{x})-\phi(t,\bold{x})\right], \\
& \left[ -\nabla^2 \phi(x) + m^2\phi(x) \right]\Delta^4 x - \lambda \Delta W(x)\\ & = -\Delta^3\bold{x}
\left[\pi(t+\Delta t,\bold{x})-\pi(t,\bold{x})\right],
\end{split}
\end{equation}
where we have used the relation $\Delta^4x= \Delta t \Delta^3\bold{x}$
with $\bold{x}$ denoting the spatial coordinate. We use $\Delta$ in place of $d$
to emphasize that Eq.~\eqref{eq:phi:qu:ha} is a finite difference equation.
It is easy to verify that Eq.~\eqref{eq:phi:qu:ha} is equivalent to Eq.~\eqref{eq:phi:ca:EL}.

The quantization process is to replace the canonical coordinate and momentum
by the operators $\hat\phi(x)$ and $\hat\pi(x)$, respectively. As in a conventional QFT,
they satisfy the commutator
$\left[\hat\phi(t,\bold{x}), \hat\pi(t,\bold{x}')\right]= i\delta^3(\bold{x}-\bold{x}')$,
or in a discretized version, $\left[\hat\phi(t,\bold{x}), \hat\pi(t,\bold{x}')\right]
= i\delta_{\bold{x},\bold{x}'} /\Delta^3 \bold{x}$, where $\delta^3(\bold{x}-\bold{x}')$
and $\delta_{\bold{x},\bold{x}'}$ are the Dirac and Kronecker $\delta$-functions, respectively.
With this commutation relation,
the Hamilton equation can be rewritten in terms of a unitary transformation, reading
\begin{equation}\label{eq:phi:qu:uni}
\begin{split}
& \hat{\phi}(t+\Delta t,\bold{x})= e^{i \sum_{\bold{x}}
\Delta \hat{\mathcal{H}}(t,\bold{x})}\hat{\phi}(t,\bold{x})
e^{-i \sum_{\bold{x}}\Delta \hat{\mathcal{H}}(t,\bold{x})}, \\
& \hat{\pi}(t+\Delta t,\bold{x})= e^{i \sum_{\bold{x}}
\Delta \hat{\mathcal{H}}(t,\bold{x})}\hat{\pi}(t,\bold{x})
e^{-i \sum_{\bold{x}}\Delta \hat{\mathcal{H}}(t,\bold{x})},
\end{split}
\end{equation}
where
\begin{equation}\label{eq:phi:qu:dH}
\begin{split}
\Delta \hat{\mathcal{H}}(x)=&\Delta^4 x \left[\frac{1}{2}\hat{\pi}^2(x) + \frac{1}{2}
\left(\nabla\hat{\phi}(x)\right)^2 +\frac{1}{2}m^2 \hat{\phi}^2(x) \right] \\
& -\lambda \Delta W(x)\hat{ \phi}(x)
\end{split}
\end{equation}
is the Hamiltonian integral within a single spacetime element.
Note a difference from the single-particle case. The exponent
of the unitary operator now becomes a sum over the simultaneous spacetime elements,
i.e. $\sum_{\bold{x}}
\Delta \hat{\mathcal{H}}(t,\bold{x})$. In the continuous limit, this sum
changes into an integral over the space. As $\lambda=0$,
the exponent becomes the Hamiltonian times $\Delta t$, repeating
what is well-known in conventional QFTs. But as $\lambda \neq 0$,
since $\Delta W(x)/\Delta^4 x$ is not well-defined in the limit $\Delta^4 x\to 0$,
we have to be satisfied with the lengthy expression $\sum_{\bold{x}}
\Delta \hat{\mathcal{H}}(t,\bold{x})$.
It is straightforward to verify that
the Hamilton equation~\eqref{eq:phi:qu:ha} can be
derived from Eq.~\eqref{eq:phi:qu:uni} and~\eqref{eq:phi:qu:dH}.

Next we change from the Heisenberg picture to the Schr\"{o}dinger picture.
We follow the process explained in detail in Sec.~\ref{sec:ha:ca}. Given
a reference time $t_0$, the operators $\hat \phi(t',\bold{x})$ and
$\hat\pi(t',\bold{x})$ in the Heisenberg picture are connected to
those in the Schr\"{o}dinger picture, i.e. $\hat \phi(\bold{x})\equiv\hat \phi(t_0,\bold{x})$ and
$\hat \pi(\bold{x})\equiv\hat\pi(t_0,\bold{x})$ by a unitary transformation reading
$\hat \phi(t',\bold{x}) = \hat U^\dag (t',t_0)\hat \phi(\bold{x})\hat U(t',t_0)$ and
$\hat \pi(t',\bold{x}) = \hat U^\dag (t',t_0)\hat \pi(\bold{x})\hat U(t',t_0)$.
The unitary operator is expressed as
\begin{equation}\label{eq:phi:ca:Ut0}
\begin{split}
\hat U(t',t_0) =& e^{-i \sum_{\bold{x}_{N-1}}\Delta \hat{\mathcal{H}}^S(t_{N-1},\bold{x}_{N-1})}
 \cdots e^{-i \sum_{\bold{x}_1}\Delta \hat{\mathcal{H}}^S(t_1,\bold{x}_1)}\\ & 
\times e^{-i \sum_{\bold{x}_0}\Delta \hat{\mathcal{H}}^S(t_0,\bold{x}_0)},
\end{split}
\end{equation}
where $t_j = t_0 +j \Delta t$, $t'= t_N$ and
\begin{equation}
\begin{split}
\Delta \hat{\mathcal{H}}^S(t,\bold{x}) = & \Delta^4 x \bigg[\frac{1}{2}
\hat{\pi}^2(\bold{x}) + \frac{1}{2}
\left(\nabla\hat{\phi}(\bold{x})\right)^2 \\ &+\frac{1}{2}m^2 \hat{\phi}^2(\bold{x}) \bigg] 
-\lambda \Delta W(t,\bold{x}) \hat{\phi}(\bold{x})
\end{split}
\end{equation}
is the infinitesimal Hamiltonian integral in the Schr\"{o}dinger picture.
Note that $\Delta \hat{\mathcal{H}}^S(x)$ depending on
$\Delta W(x)$ is a random operator and $\hat U(t',t_0)$ is then a random unitary operator.
The wave function experiences a random unitary evolution, which reads
$\ket{\psi(t')}= \hat U(t',t_0) \ket{\psi(t_0)}$.

Let us write down an infinitesimal form of the unitary evolution, even
we will not use it in the path integral formalism. We use $\displaystyle\int_{R^3}
dW(x) = \displaystyle\lim_{\Delta^3\bold{x}\to 0}\sum_{\bold{x}} 
\Delta W(t,\bold{x})$ to denote an integral
over the 3-dimensional space with $t$ fixed (notice its difference
from $\int dW(x)$). The change of wave function
over an infinitesimal time interval is then expressed as
\begin{equation}\label{eq:phi:ca:sch}
\begin{split}
d\ket{\psi(t)} = & -i dt \hat H_0 \ket{\psi(t)} + i\lambda \int\displaylimits_{R^3}
dW(t,\bold{x}) \hat \phi(\bold{x}) \ket{\psi(t)} \\
& -\frac{\lambda^2}{2} \int\displaylimits_{R^3} 
dW(t,\bold{x}_1) \int\displaylimits_{R^3}dW(t,\bold{x}_2)
\hat\phi(\bold{x}_1)\hat\phi(\bold{x}_2) \ket{\psi(t)},
\end{split}
\end{equation}
where
\begin{equation}\label{eq:phi:ca:H0}
\hat H_0= \int d^3\bold{x} \left[\frac{1}{2}
\hat{\pi}^2(\bold{x}) + \frac{1}{2}
\left(\nabla\hat{\phi}(\bold{x})\right)^2 +\frac{1}{2}m^2 \hat{\phi}^2(\bold{x}) \right]
\end{equation}
is the well-known Hamiltonian of a free massive boson. It is worth
emphasizing that in the Schr\"{o}dinger picture, $\hat{\phi}(\bold{x})$
and $\hat{\pi}(\bold{x})$ are independent of $dW(x)$, being simply
the field operators of free bosons. By using the
independence property of $dW(x)$, we further find that the density matrix satisfies
a Lindblad equation which reads
\begin{equation}\label{eq:phi:ca:rho}
\begin{split}
\frac{d\hat \rho(t)}{dt} = & -i \left[\hat H_0, \hat \rho(t) \right] + \lambda^2
\bigg(\int d^3\bold{x} \hat \phi(\bold{x}) \hat \rho(t)\hat \phi(\bold{x}) \\ & 
-\frac{1}{2} \int d^3\bold{x}  \hat \phi^2(\bold{x}) \hat \rho(t) -
\frac{1}{2}   \hat \rho(t) \int d^3\bold{x}\hat \phi^2(\bold{x}) \bigg).
\end{split}
\end{equation}

The equations of motion keep invariant under both the spatial and temporal
translations. Under the transformation $\tilde{t}=t+\tau$, $\tilde{\bold{x}}=\bold{x}
+ \bold{\xi}$, $\hat{ \tilde{\rho}}(\tilde t)=\hat \rho(t)$, $\hat{\tilde{\phi}}(\tilde{\bold{x}})
=\hat{{\phi}}(\bold{x})$ and $\hat{\tilde{\pi}}(\tilde{\bold{x}})=\hat{{\pi}}(\bold{x})$,
Eq.~\eqref{eq:phi:ca:rho} does keep invariant. Moreover, by using the
relation $dW(t,\bold{x})\stackrel{d}{=} dW(\tilde{t},\tilde{\bold{x}})$, we also
find Eq.~\eqref{eq:phi:ca:sch} to keep invariant in the statistical meaning.
The invariance of equations of motion is a direct consequence of the invariance
of action~\eqref{eq:phi:ra:IW}.

Note that similar equations of motion have been studied in the SDE approach
to relativistic spontaneous collapse models (CSL-type models).
But, in this paper, we derive these equations from a Lorentz-invariant action.
In the context of collapse models, it was shown that Eq.~\eqref{eq:phi:ca:rho} leads
to an infinite rate of particle number production.
Here, we will not solve Eq.~\eqref{eq:phi:ca:sch} or~\eqref{eq:phi:ca:rho}.
Instead, we turn to the path integral formalism in which the
wave function and density matrix can be obtained and also the
Lorentz invariance manifests naturally. In the path-integral
formalism, one can even remove the infinity by renormalizing $\lambda$,
as will be shown next. After we develop the renormalization technique,
we will revisit Eq.~\eqref{eq:phi:ca:sch} and~\eqref{eq:phi:ca:rho} in Sec.~\ref{sec:phi:revisit}.

\subsection{Path integral approach and $S$-matrix\label{sec:phi:scatter}}

In QFTs, one usually supposes the system to be initially
prepared at a state vector $\ket{\alpha}$,
and is interested in its final state after a long evolution. To avoid
a divergent phase, we turn to the interaction picture in which
the evolution operator becomes $\hat U_I(t',t_0)=e^{i t'\hat H_0}
\hat U(t',t_0) e^{-i t_0 \hat H_0}$ with $\hat H_0$ being the Hamiltonian of free
particle (see Eq.~\eqref{eq:phi:ca:H0}). The $S$-matrix is then defined
as $S_{\beta,\alpha} = \bra{\beta}\hat U_I(T,-T)\ket{\alpha}$ with $T$ being
much larger than the interacting time among particles. The limit $T\to\infty$
is usually taken, but let us consider a finite $T$ at current stage. In a stochastic QFT,
$S_{\beta,\alpha} $ depends on $dW(x)$, being then a random number.
Next we study how to calculate $S_{\beta,\alpha} $ in the path integral approach.

For convenience, we use $\ket{\phi}$ with $\phi$ being a
function of $\bold{x}$ to denote an eigenstate of $\hat \phi(\bold{x})$
which satisfies $\hat \phi(\bold{x}) \ket{\phi} = \phi(\bold{x}) \ket{\phi} $.
The states $\left\{ \ket{\phi}\right\}$ form an orthonormal basis of
the Hilbert space with the complete relation reading $
\displaystyle\int  \prod_{\bold{x}} d\phi(\bold{x}) \ket{\phi} \bra{\phi} =1$.
Let us first calculate the propagator $\bra{\phi'} \hat U(t',t_0)\ket{\phi_0}$
with $\phi_0$ and $\phi'$ being two arbitrary functions of $\bold{x}$.

In Sec.~\ref{sec:ha:path}, we proved that the path integral approach
can be employed to calculate the propagator of a harmonic oscillator.
A real scalar field can be seen as a set of harmonic oscillators
located at different positions in the three-dimensional space.
In the action~\eqref{eq:phi:ra:IW}, the random field is linearly coupled to the canonical
coordinate, as same as in the action of harmonic oscillator.
Therefore, the formalism in Sec.~\ref{sec:ha:path}
can be applied here. By inserting a sequence of
$\displaystyle\int  \prod_{\bold{x}} d\phi_j(\bold{x}) \ket{\phi_j} \bra{\phi_j}$ with
$j=1,2,\cdots N-1$ into Eq.~\eqref{eq:phi:ca:Ut0}, we find
\begin{equation}\label{eq:phi:path:p0}
\begin{split}
& \bra{\phi'} \hat U(t',t_0)\ket{\phi_0} = \mathcal{N}_{\pi} \int \prod_{j=1}^{N-1}\prod_{\bold{x}}
d \phi_j(\bold{x}) \ \times \\& exp\bigg\{i \Delta t \sum_{j=0}^{N-1}\sum_{\bold{x}} \Delta^3 \bold{x} 
\left[\frac{1}{2} \bigg(\frac{\phi_{j+1}(\bold{x})- \phi_j(\bold{x})}{\Delta t}\right)^2 \\ & -\frac{1}{2}
\left(\nabla \phi_j(\bold{x})\right)^2-\frac{1}{2}m^2 \phi_j^2(\bold{x}) \bigg]+
i \lambda \sum_{\bold{x}} \Delta W(t_j,\bold{x}) \phi_j(\bold{x}) \bigg\},
\end{split}
\end{equation}
where $ \mathcal{N}_{\pi}$ is a field-independent factor originated from the integral
over canonical momentum, $\phi_0$ and $\phi'= \phi_N$ are the initial and final configurations
of field, respectively, and $\phi_j(\bold{x}) = \phi(t_j,\bold{x})$ with $j=1,2,\cdots
N-1$ is the intermediate-time configuration of the quantum field.
In the continuous limit, the exponent in Eq.~\eqref{eq:phi:path:p0} becomes $iI_W$.

Let us use $\hat U^{(0)}(t',t_0)= e^{-i\hat H_0(t'-t_0)}$ to denote the evolution
operator of the free-particle Hamiltonian, and in the interaction picture
it becomes $\hat U^{(0)}_I(t',t_0) = 1$. We use $\ket{0}$ to denote the free-particle vacuum.
The $S$-matrix is then expressed as
\begin{widetext}
\begin{equation}\label{eq:phi:path:Sm}
\begin{split}
S_{\beta,\alpha}= & \frac{\bra{\beta}\hat U_I(T,-T)\ket{\alpha}}{\bra{0} \hat U^{(0)}_I(T,-T)\ket{0}}\\
= & \frac{ \displaystyle\int \displaystyle\prod_{\bold{x}} d\phi_T(\bold{x}) d\phi_{-T}(\bold{x})
\bra{\beta}e^{iT\hat H_0} \ket{\phi_T} \bra{\phi_T} \hat U(T,-T)
 \ket{\phi_{-T}} \bra{\phi_{-T}}e^{iT\hat H_0} \ket{\alpha} }
 {\displaystyle\int \displaystyle\prod_{\bold{x}} d\phi_T(\bold{x}) d\phi_{-T}(\bold{x})
\bra{0}e^{iT\hat H_0} \ket{\phi_T} \bra{\phi_T} \hat U^{(0)}(T,-T)
 \ket{\phi_{-T}} \bra{\phi_{-T}}e^{iT\hat H_0} \ket{0}}.
\end{split}
\end{equation}
\end{widetext}
In the numerator, $\bra{\phi_T} \hat U(T,-T) \ket{\phi_{-T}} $
can be written as a path integral (see Eq.~\eqref{eq:phi:path:p0}) in which we set
$t_j= -T+j2T/N$, and $\phi_T= \phi_N$ and $\phi_{-T}= \phi_0$ are
the final and initial configurations, respectively.
Similarly, $\bra{\phi_T} \hat U^{(0)}(T,-T) \ket{\phi_{-T}}$ in the denominator is a path
integral but with $\lambda=0$, i.e. in the absence
of random field. The field-independent factors such as $\mathcal{N}_{\pi}$
in the denominator cancel those in the numerator (the denominator is
introduced for this purpose). We then express the $S$-matrix as
\begin{equation}\label{eq:phi:path:Sba}
\begin{split}
S_{\beta,\alpha}=  \frac{ \displaystyle\int D\phi 
\bra{\beta}e^{iT\hat H_0} \ket{\phi_T} \bra{\phi_{-T}}e^{iT\hat H_0} \ket{\alpha} e^{iI_W} }
 {\displaystyle\int D\phi
\bra{0}e^{iT\hat H_0} \ket{\phi_T}  \bra{\phi_{-T}}e^{iT\hat H_0} \ket{0} e^{iI_0} },
\end{split}
\end{equation}
where $\displaystyle\int D\phi$ is the abbreviation of $\displaystyle\int \displaystyle\prod_{\bold{x}}
\prod_{j=0}^N d\phi_j(\bold{x})$. Eq.~\eqref{eq:phi:path:Sba} is as same
as the expression of $S$-matrix in a conventional QFT.
Indeed, the action~\eqref{eq:phi:ra:IW} is similar to
that of bosons in an external potential, but with the
deterministic potential (say $V(x)$) being replaced by a random one.
And $dW(x)$ plays the role of $V(x)d^4 x$. The difference between $dW(x)$ and $V(x)d^4 x$
is that $\left[d W(x)\right]^2\sim d^4 x$ cannot be neglected but $\left[V(x)d^4 x\right]^2$ can be.
However, in the path integral approach, $dW(x)$ stays in the exponent of $e^{iI_W}$ and no
expansion of $e^{iI_W}$ is needed, hence, $\left[d W(x)\right]^2$
does not appear in the calculation. This explains why the $S$-matrix
has the same form.

In the QFTs, $\ket{\alpha}$ or $\ket{\beta}$ are usually chosen to the free-particle states
with specific momentum, e.g. $\ket{\bold{p}_1\bold{p}_2\cdots\bold{p}_n}
=\hat a_{\bold{p}_1}^\dag \hat a_{\bold{p}_2}^\dag\cdots\hat a_{\bold{p}_n}^\dag \ket{0}$
where $ \hat a_{\bold{p}}^\dag$ denotes the creation operator of a particle of momentum $\bold{p}$.
It is well known that the field operators of free bosons are associated to
the creation and annihilation operators by
$\hat \phi(\bold{x})=\displaystyle\frac{1}{\sqrt{(2\pi)^3}}
\displaystyle\int \displaystyle\frac{d^3\bold{p}}{\sqrt{2E_{\bold{p}}}}
\left(e^{i\bold{p}\cdot\bold{x}} \hat a_{\bold{p}}+
e^{-i\bold{p}\cdot\bold{x}} \hat a_{\bold{p}}^\dag \right) $ and
$\hat \pi(\bold{x})=\displaystyle\frac{-i}{\sqrt{(2\pi)^3}}
\displaystyle\int d^3\bold{p} \displaystyle\sqrt{\frac{E_{\bold{p}}}{{2}}}
\left(e^{i\bold{p}\cdot\bold{x}} \hat a_{\bold{p}}-
e^{-i\bold{p}\cdot\bold{x}} \hat a_{\bold{p}}^\dag \right) $, where $E_{\bold{p}}=
\sqrt{m^2+\bold{p}^2}$ is the dispersion relation. These two equations
are sufficient for deriving an expression of $\bra{\phi_t}e^{-it\hat H_0}
\ket{\bold{p}_1\cdots\bold{p}_n} $ with arbitrary $n$ and $t$
(see e.g. Ref.~[\onlinecite{Weinberg}]). Especially for $n=0$, the result is
\begin{equation}\label{eq:phi:path:pp}
\bra{\phi_t}e^{-it\hat H_0}
\ket{0} = \mathcal{N}_{\epsilon} e^{-\frac{1}{2}\int d^3\bold{x}d^3\bold{x}' \epsilon(\bold{x},
\bold{x}') \phi_t(\bold{x})\phi_t(\bold{x}')},
\end{equation}
where $\mathcal{N}_{\epsilon}$ is an unimportant field-independent constant
and $\epsilon(\bold{x},\bold{x}')= \displaystyle\frac{1}{(2\pi)^3}\displaystyle\int
d^3\bold{p}e^{i\bold{p}\cdot(\bold{x}-\bold{x}')} E_{\bold{p}}$ is the Fourier transformation
of dispersion relation, being real and symmetric with respect to $\bold{x}$
and $\bold{x}'$. For $n>0$, the result is a functional
derivative of~\eqref{eq:phi:path:pp}, which can be generally expressed as
\begin{equation}\label{eq:phi:path:phit}
\begin{split}
&\bra{\phi_t}e^{-it\hat H_0}
\ket{\bold{p}_1\cdots\bold{p}_n} =\mathcal{N}_{\epsilon} 
e^{-\frac{1}{2}\int d^3\bold{x}d^3\bold{x}' \epsilon(\bold{x},
\bold{x}') \phi_t(\bold{x})\phi_t(\bold{x}')} \\&
\times \int \prod_{j=1}^n d^3 \bold{x}_j
\ e^{i \sum_{j=1}^n \left(\bold{p}_j \cdot \bold{x}_j -t E_{\bold{p}_j}\right)}
\bigg\{ \prod_{j=1}^n \sqrt{\frac{2E_{\bold{p}_j}}{(2\pi)^3}} \phi_t(\bold{x}_j) \\
& - \frac{\delta^3(\bold{x}_1-\bold{x}_2)}{(2\pi)^3}  
\prod_{j\neq 1,2} \sqrt{\frac{2E_{\bold{p}_j}}{(2\pi)^3}} \phi_t(\bold{x}_j) - \cdots \\&
- \frac{\delta^3(\bold{x}_1-\bold{x}_2)}{(2\pi)^3}  \frac{\delta^3(\bold{x}_3-\bold{x}_4)}{(2\pi)^3}  
\prod_{j\neq 1,2,3,4} \sqrt{\frac{2E_{\bold{p}_j}}{(2\pi)^3}} \phi_t(\bold{x}_j) - \cdots \\ & -\cdots\bigg\}.
\end{split}
\end{equation}
Inside the curly brackets, the first term is a product of
$\phi_t(\bold{x}_1), \phi_t(\bold{x}_2),\cdots,$ and $\phi_t(\bold{x}_n)$ with
each $\phi_t(\bold{x}_j)$ accompanied by a factor
$\sqrt{2E_{\bold{p}_j}/(2\pi)^3}$. The other terms have a minus sign.
To obtain them, we take account of all ways of pairing the fields in the set
$\left\{\phi_t(\bold{x}_1), \phi_t(\bold{x}_2),\cdots,\phi_t(\bold{x}_n)\right\}$,
and if $\phi_t(\bold{x}_i)$ and $\phi_t(\bold{x}_j)$ are paired, we then
replace their products by $\delta^3(\bold{x}_i-\bold{x}_j)/(2\pi)^3$.

Setting $t=\mp T$ in Eq.~\eqref{eq:phi:path:phit}, we obtain the
expressions of $\bra{\phi_{-T}}e^{iT\hat H_0}\ket{\alpha}$ and
$\bra{\beta}e^{iT\hat H_0}\ket{\phi_T}= \bra{\phi_T}e^{-iT\hat H_0}\ket{\beta}^*$.
Substituting these expressions into Eq.~\eqref{eq:phi:path:Sm}, we are then able to
calculate the $S$-matrix. Notice that the factor $\mathcal{N}_\epsilon$ in the
denominator of $S_{\beta,\alpha}$
cancels that in the numerator. What is left in the numerator or denominator
is a path integral of an exponential function multiplied by the product
of fields at $t=\pm T$.

Let us consider $S_{0,0}$ (the initial and final states
are both vacuum). According to QFT, the exponential functions in the expressions of
$\bra{\phi_{-T}}e^{iT\hat H_0}\ket{0}$ and $\bra{0}e^{iT\hat H_0}\ket{\phi_T}$ need to be
replaced by using
\begin{equation}\label{eq:phi:path:eta}
\begin{split}
& \int d^3\bold{x}d^3\bold{x}' 
\epsilon(\bold{x},\bold{x}') \left[\phi_T(\bold{x})\phi_T(\bold{x}')+
\phi_{-T}(\bold{x})\phi_{-T}(\bold{x}')\right] \\ &
= \eta \int^T_{-T} dt \ e^{-\eta\left| t\right|} \epsilon(\bold{x},\bold{x}')\phi(t,\bold{x})\phi(t,\bold{x}'),
\end{split}
\end{equation}
where $\phi(t,\bold{x})=\phi_t(\bold{x})$ is the configuration of field at time $t$,
and $\eta>0$ is an infinitesimal. The relation~\eqref{eq:phi:path:eta} is strict in
the limit $T\to\infty$ and $\eta \to 0$. But in QFTs, it is applicable once if $T$
is much larger than the other time scales and $\eta$ is much smaller
than the other energy scales. Now the $S$-matrix element becomes
\begin{equation}\label{eq:phi:path:S00}
\begin{split}
S_{0,0} =  \frac{ \displaystyle\int D\phi \ e^{iI_W^{(\eta)} }}
 {\displaystyle\int D\phi \ e^{iI_0^{(\eta)}} },
\end{split}
\end{equation}
where
\begin{equation}
\begin{split}
& I_0^{(\eta)} = I_0+ \frac{i}{2}\eta \int dt d^3\bold{x}d^3\bold{x}' \
\epsilon(\bold{x},\bold{x}') \phi(t,\bold{x}) \phi(t,\bold{x}') e^{-\eta\left|t\right|},\\
& I_W^{(\eta)}=I_W + \frac{i}{2}\eta \int dt d^3\bold{x}d^3\bold{x}' \
\epsilon(\bold{x},\bold{x}') \phi(t,\bold{x}) \phi(t,\bold{x}') e^{-\eta\left|t\right|}
\end{split}
\end{equation}
are the actions with the so-called $i\eta$ terms.

Both $I_0^{(\eta)}$ and $I_W^{(\eta)}$ are quadratic forms of $\phi$, hence,
the integrals in Eq.~\eqref{eq:phi:path:S00} are the Gaussian integrals
which can be done by using the formula~\eqref{eq:ha:path:oz}. To use this formula,
we need in principle to know the determinant of the quadratic term's coefficient matrix and also
the stationary point of the action. Fortunately, the determinants for $I_0^{(\eta)}$ and $I_W^{(\eta)}$
are exactly the same, because $I_W^{(\eta)}$ is only different from $I_0^{(\eta)}$
by a linear term of $\phi$. As a consequence, the determinant in the denominator of
Eq.~\eqref{eq:phi:path:S00} cancels that in the numerator. Furthermore, $I_0^{(\eta)}$
contains only quadratic terms so that its value at the stationary point is zero, and then
the denominator becomes unity after the cancellation. We obtain
$S_{0,0} =e^{i \bar{I}_W^{(\eta)}}$ where $\bar{I}_W^{(\eta)}= {I}_W^{(\eta)}(\bar{\phi})$
is the value of ${I}_W^{(\eta)}$ at the stationary-point. Studying the variation of ${I}_W^{(\eta)}$
with respect to $\phi$, we find the stationary point to satisfy
\begin{equation}\label{eq:phi:path:pbar}
\begin{split}
& \Delta^4 x\bigg[\left(-\partial_\mu\partial^\mu+m^2\right) \bar{\phi}(x) \\ & - i\eta e^{-\eta\left|t\right|}
\int d^3\bold{x}' \epsilon(\bold{x},\bold{x}')  \bar{\phi}
(t,\bold{x}') \bigg] = \lambda \Delta W(x),
\end{split}
\end{equation}
where we use the notation $x=(t,\bold{x})$. Eq.~\eqref{eq:phi:path:pbar}
is a linear equation of $\bar{\phi}$, which can be solved by using the Green's function
(or Feynman propagator) defined by
\begin{equation}\label{eq:phi:path:Ge}
\begin{split}
& \left(-\partial_\mu\partial^\mu+m^2\right)G(x,x') \\ & - i\eta e^{-\eta\left|t\right|}\int d^3\bold{x}''
\epsilon(\bold{x},\bold{x}'') G(t\bold{x}'',x') = \delta^4(x-x').
\end{split}
\end{equation}
The solution of Eq.~\eqref{eq:phi:path:Ge} is
\begin{equation}\label{eq:phi:path:Gxxx}
\begin{split}
G(x,x') = & \frac{1}{(2\pi)^4} \int d^4 p \frac{e^{ip(x-x')} }{p^2+m^2-i\eta} \\
= & \theta(t-t') \frac{i}{(2\pi)^3}\int \frac{d^3\bold{p}}{2E_\bold{p}}e^{i\bold{p}\cdot
(\bold{x}-\bold{x}')- iE_\bold{p}(t-t')} \\ &
+ \theta(t'-t) \frac{i}{(2\pi)^3}\int \frac{d^3\bold{p}}{2E_\bold{p}}e^{i\bold{p}\cdot
(\bold{x}'-\bold{x})- iE_\bold{p}(t'-t)},
\end{split}
\end{equation}
where we have used the fact that $\eta$ is infinitesimal. In Eq.~\eqref{eq:phi:path:Gxxx},
$G(x,x')$ is expressed in terms of the four-dimensional and also
three-dimensional integrals. Two different expressions
are useful in different contexts. With the help of Green's function, the solution of
Eq.~\eqref{eq:phi:path:pbar} can be written as $\bar{\phi}(x) = \lambda \displaystyle\int dW(x') G(x,x')$.
Then the $S$-matrix element is found to be
\begin{equation}\label{eq:phi:path:S00r}
S_{0,0} = exp \left\{\frac{i}{2} \lambda^2 \int dW(x) dW(x') G(x,x')\right\}.
\end{equation}
The expression of $S_{0,0}$ is surprisingly simple.
As $\lambda=0$, we obtain $S_{0,0}=1$, recovering
the well-known results in QFTs. But as $\lambda \neq 0$,
$S_{0,0}$ is a functional of $dW(x)$,
being a random number. More properties of $S_{0,0}$ will
be discussed in next sections.

The Lorentz invariance of $S_{0,0}$
is straightforward to see. Suppose $\tilde{x}=Lx$ to be a Lorentz transformation. The Feynman propagator
expressed in terms of four-dimensional notation is clearly Lorentz invariant, i.e.
$G(x,x') = G(\tilde{x},\tilde{x}')$. In Sec.~\ref{sec:phi:ran}, we already show $dW(x)
\stackrel{d}{=} d\tilde{W}(\tilde{x})$, and more generally,
$\left(dW(x_1),dW(x_2),\cdots\right) \stackrel{d}{=} \left(d\tilde{W}(\tilde{x}_1),
d\tilde{W}(\tilde{x}_2),\cdots\right)$. As a consequence, we have
\begin{equation}
\int dW(x) dW(x') G(x,x') \stackrel{d}{=}\int d\tilde{W}(\tilde{x})
d\tilde{W}(\tilde{x}') G(\tilde{x},\tilde{x}').
\end{equation}
Therefore, under a Lorentz transformation, $S_{0,0}$ is statistically invariant
as what we expect, since the vacuum state itself keeps invariant.

\subsection{Diagrammatic rules for $S$-matrix\label{sec:phi:dia}}

Next, we show how to calculate an arbitrary $S$-matrix
element, say $S_{\bold{p}'_1\cdots\bold{p}'_m,
\bold{p}_1\cdots\bold{p}_n}$ where $\ket{\bold{p}_1\cdots\bold{p}_n}$ and
$\ket{\bold{p}'_1\cdots\bold{p}'_m}$ are the initial and final state vectors, respectively.

For easy to understand, we use $S_{\bold{p}',\bold{p}}$ as an example.
We choose $n=1$ in Eq.~\eqref{eq:phi:path:phit} and
substitute it into Eq.~\eqref{eq:phi:path:Sba}, obtaining
\begin{equation}\label{eq:phi:dia:Spp}
\begin{split}
S_{\bold{p}',\bold{p}} = &
\sqrt{\frac{2E_\bold{p}}{(2\pi)^3}}e^{iTE_\bold{p}}
\sqrt{\frac{2E_{\bold{p}'}}{(2\pi)^3}} e^{iTE_{\bold{p}'}}
\int d^3 \bold{x}d^3\bold{x}' e^{i\bold{p}\cdot \bold{x}-i\bold{p}'\cdot \bold{x}'} \\
& \times \left(\frac{\displaystyle\int D\phi \ \phi(-T,\bold{x}) \phi(T,\bold{x}') e^{iI_W^{(\eta)}}}
{\displaystyle\int D\phi \ e^{iI_0^{(\eta)}}}\right).
\end{split}
\end{equation}
The bracketed term in Eq.~\eqref{eq:phi:dia:Spp} includes a path integral
of $e^{iI_W^{(\eta)}}\phi(-T,\bold{x}) \phi(T,\bold{x}')$.
In the calculation of $S_{\bold{p}'_1\cdots\bold{p}'_m,
\bold{p}_1\cdots\bold{p}_n}$ with generic $n$ and $m$, we repeatedly run
into path integrals of this type. To calculate this integral,
we turn it into a functional derivative of $\displaystyle\int D\phi \ e^{i I_W^{(J)}}$ in which $I_W^{(J)}$
reads
\begin{equation}
I_W^{(J)} = I_W^{(\eta)} + \int d^4 x J(x)\phi(x)
\end{equation}
with $J(x)$ being a source. For example, the bracketed term in Eq.~\eqref{eq:phi:dia:Spp}
is turned into
\begin{equation}\label{eq:phi:dia:dpdp}
\begin{split}
\left(-i\right)^2\frac{\delta^2}{\delta J(-T,\bold{x})\delta J(T,\bold{x}')}
\left.\left(\frac{\displaystyle\int D\phi \ e^{iI_W^{(J)}}}
{\displaystyle\int D\phi \ e^{iI_0^{(\eta)}}}\right) \right|_{J=0}.
\end{split}
\end{equation}

The bracketed term in Eq.~\eqref{eq:phi:dia:dpdp} is evaluated by using the same method
as we evaluate Eq.~\eqref{eq:phi:path:S00}.
We notice that the quadratic
terms of $I_W^{(J)}$ and $I_0^{(\eta)}$ are the same.
As a consequence, we have $\int D\phi e^{iI_W^{(J)}}/
\int D\phi e^{iI_0^{(\eta)}} =  e^{iI_W^{(J)}(\bar{\phi})}$ where $\bar{\phi}$
is the stationary point of $I_W^{(J)}$, satisfying Eq.~\eqref{eq:phi:path:pbar} with an additional $J(x) \Delta^4 x$
added to the right-hand side. It is straightforward to find
$\bar{\phi}(x)= \displaystyle\int d^4x' G(x,x')J(x') + \lambda \displaystyle\int dW(x') G(x,x')$,
and then we obtain
\begin{equation}\label{eq:phi:dia:DD}
\begin{split}
& \frac{\displaystyle\int D\phi \ e^{iI_W^{(J)}}}
{\displaystyle\int D\phi \ e^{iI_0^{(\eta)}}} =  e^{i\mathcal{F}\left[J\right]}, \\
& \mathcal{F}\left[J\right] = \frac{1}{2}
\left(\int d^4 xJ(x)+\lambda\int dW(x)\right) \\ & \ \ \ \  \ \ \ \ \ \ \times
\left(\int d^4 x'J(x')+\lambda\int dW(x')\right) G(x,x') .
\end{split}
\end{equation}
As $J(x)=0$, $e^{i\mathcal{F}[J]}$ reduces to $S_{0,0}$.

By substituting Eq.~\eqref{eq:phi:dia:DD} into Eq.~\eqref{eq:phi:dia:dpdp}
and then into Eq.~\eqref{eq:phi:dia:Spp}, we find
\begin{equation}\label{eq:phi:dia:Sppres}
\begin{split}
S_{\bold{p}',\bold{p}} = & S_{0,0} \bigg\{\delta^3(\bold{p}-\bold{p}') - \lambda^2
\frac{ 1}{\sqrt{(2\pi)^32E_{\bold{p}}}}\frac{ 1}{\sqrt{(2\pi)^32E_{\bold{p}'}}}\times
\\ & \int dW(x) e^{i\left(\bold{p}\cdot\bold{x}-E_\bold{p} t\right)}
\int dW(x') e^{-i\left(\bold{p}'\cdot\bold{x}'-E_{\bold{p}'} t'\right)}\bigg\}.
\end{split}
\end{equation}
Here we have used the property $G(x,x')=G(x',x)$ and Eq.~\eqref{eq:phi:path:Gxxx}.
As immediately seen, $S_{0,0}$ is a factor of $S_{\bold{p}',\bold{p}}$.
This is a common feature of $S$-matrix elements. Since $S_{0,0}$ is a random number, all the
$S$-matrix elements must be random numbers.
Besides the factor $S_{0,0}$, $S_{\bold{p}',\bold{p}}$ contains two terms
with the first one being equal to $\braket{\bold{p}'|\bold{p}}$ which is the $S$-matrix without random field.
The second term of $S_{\bold{p}',\bold{p}}$ is
the random-field modification, describing how
a random field drives the particle away from its initial state.
The second term is proportional to $\lambda^2$ and disappears as $\lambda$ goes to zero.
It is clear that the momentum is not conserved in a stochastic QFT.
As we have proved in Sec.~\ref{sec:phi:ran},
the space translational symmetry is only preserved statistically.
It is then not surprising that the conservation of momentum is absent.

Similarly, by substituting Eq.~\eqref{eq:phi:path:phit} into
Eq.~\eqref{eq:phi:path:Sba} and then changing the path integrals into
the functional derivatives of $e^{i\mathcal{F}\left[J\right]}$, we obtain
$S_{\bold{p}'_1\cdots\bold{p}'_m,
\bold{p}_1\cdots\bold{p}_n}$ for generic $n$ and $m$, which reads
\begin{widetext}
\begin{equation}\label{eq:phi:dia:sexp}
\begin{split}
S_{\bold{p}'_1\cdots\bold{p}'_m,
\bold{p}_1\cdots\bold{p}_n}= &
(-i)^{n+m} \int \left(\prod_{j=1}^n d^3 \bold{x}_j\right) \left(\prod_{l=1}^m d^3 \bold{x}'_l\right) 
\prod_{j=1}^n \left(\sqrt{\frac{2E_{\bold{p}_j}}{(2\pi)^3}} 
e^{i \left( T E_{\bold{p}_j}+ \bold{p}_j \cdot \bold{x}_j \right)} \right)
\prod_{l=1}^m \left(\sqrt{\frac{2E_{\bold{p}'_l}}{(2\pi)^3}}
e^{i\left(T E_{\bold{p}'_l}-\bold{p}'_l \cdot \bold{x}'_l\right)} \right) 
 \\ & \times 
 \left. \frac{\displaystyle\delta^{n+m} }{\delta 
J(-T,\bold{x}_1) \cdots \delta J(-T,\bold{x}_n)\delta J(T,\bold{x}'_1)\cdots
\delta J(T,\bold{x}'_m)}  e^{i\mathcal{F}\left[J\right]} \right|_{J=0} \ \
-\cdots .
\end{split}
\end{equation}
\end{widetext}
Here we only show part of $S_{\bold{p}'_1\cdots\bold{p}'_m,
\bold{p}_1\cdots\bold{p}_n}$ that comes from the first term inside
the brackets of Eq.~\eqref{eq:phi:path:phit}, and use
the ellipsis to denote the other parts of $S_{\bold{p}'_1\cdots\bold{p}'_m,
\bold{p}_1\cdots\bold{p}_n}$ that come from the minus-sign terms inside
the brackets of Eq.~\eqref{eq:phi:path:phit}.

Because the third-order derivative of
$\mathcal{F}$ with respect to $J$ is always zero,
the $(n+m)$th derivative of $e^{i\mathcal{F}}$
at $J=0$ must be a sum of items with each one being
the product of a sequence of functions that are
either $\delta \mathcal{F}/\delta J(x)$ or
$\delta^2 \mathcal{F}/\delta J(x)\delta J(x') $.
For example, one of these items can be written as
\begin{equation}
\left.\frac{\delta^2 \mathcal{F}}{\delta J(-T,\bold{x}_1)\delta J(T,\bold{x}'_1)} \right|_{J=0}
\left.\frac{\delta \mathcal{F}}{\delta J(-T,\bold{x}_2)}\right|_{J=0} \cdots.
\end{equation}
To obtain the $(n+m)$th derivative of $e^{i\mathcal{F}}$, we must consider all the ways
of partitioning the set $\{J(-T,\bold{x}_1)$, $\cdots ,J(-T,\bold{x}_n)$,
$J(T,\bold{x}'_1)$, $\cdots$, $J(T,\bold{x}'_m)\}$ with each subset consisting of either
a pair of $J$s or a single $J$. We should not forget the ellipsis in Eq.~\eqref{eq:phi:dia:sexp}.
By using the identity
\begin{equation}
\begin{split}
& i\delta^3
\left(\bold{p}_1+\bold{p}_2\right) = 
\sqrt{\frac{2E_{\bold{p}_1}}{(2\pi)^3}}\sqrt{\frac{2E_{\bold{p}_2}}{(2\pi)^3}} \times \\ &
\int d^3 \bold{x}_1 d^3 \bold{x}_2 e^{\mp i\left(\bold{p}_1\cdot\bold{x}_1+
\bold{p}_2\cdot\bold{x}_2\right)} \left.\frac{\delta^2 \mathcal{F}}{\delta J(\pm T, \bold{x}_1)
\delta J(\pm T,\bold{x}_2)}\right|_{J=0},
\end{split}
\end{equation}
we find that what the ellipsis represents partly cancel
the first term in the right-hand side of Eq.~\eqref{eq:phi:dia:sexp}. As a consequence of
the cancellation, all the partitions with a pair of $J$s at $T$ or a pair of $J$s at $-T$ are removed.
Therefore, we only need to consider the partition in which
one $J$ at $T$ is paired with one $J$ at $-T$.

\begin{figure}[tbp]
\vspace{0.2cm}
\includegraphics[width=0.9\linewidth]{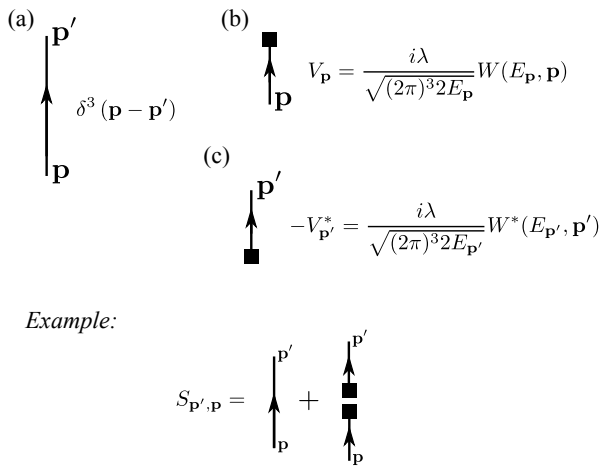}
\caption{Graphical representation in the evaluation of $S$-matrix. The bottom
panel shows the graphical expression of $S_{\bold{p}',\bold{p}}$ as an example.}\label{fig:FM}
\end{figure}
A diagrammatic formalism is usually adopted to keep track of all the ways
of grouping $J$s. A diagram consists of solid lines, each representing a pair of $J$s,
and square dots, each representing an unpaired $J$.
We integrate out the variables $\bold{x}_1$, $\cdots$, $\bold{x}_n$
and $\bold{x}'_1$, $\cdots$, $\bold{x}'_m$ in Eq.~\eqref{eq:phi:dia:sexp}.
After the integration, each line running from below into the diagram is labelled an initial momentum $\bold{p}$,
and each line running upwards out of the diagram is labelled a final momentum $\bold{p}'$.
The rules for calculating $S$-matrix is summarized in Fig.~\ref{fig:FM}.
More specifically:

\noindent (a) The pairing of one $J$ at $T$ with one $J$ at $-T$ gives
$\delta^2 \mathcal{F}/\delta J(T,\bold{x}')\delta J(-T,\bold{x}) \sim 
G(T\bold{x}',-T\bold{x})$. After integrating out $\bold{x}$ and $\bold{x}'$, we obtain
$\delta^3\left(\bold{p}'-\bold{p}\right)$. This is represented
by a solid line carrying an arrow pointed upwards from the initial
momentum $\bold{p}$ to the final momentum $\bold{p}'$ (see Fig.~\ref{fig:FM}(a)).

\noindent (b) An unpaired $J$ at $-T$ gives
$\delta \mathcal{F}/\delta J(-T,\bold{x}) \sim \displaystyle\int dW(x') G({x}',-T\bold{x})$.
After substituting the expression of $G$ in and integrating out $\bold{x}$ and $\bold{x}'$,
we obtain
\begin{equation}
V_\bold{p}=\frac{{i\lambda }}{\sqrt{(2\pi)^32E_\bold{p}}}W(E_\bold{p},\bold{p}),
\end{equation}
where $W(E_\bold{p},\bold{p})$ is defined by the Fourier transformation
\begin{equation}\label{eq:phi:dia:Wpd}
W(p)= \displaystyle \int dW(x) e^{ipx}
\end{equation}
and $(E_\bold{p},\bold{p})$ is an on-shell four-momentum. This is
represented by an arrowed line starting from an initial momentum $\bold{p}$
and ending at a square dot (see Fig.~\ref{fig:FM}(b)).

\noindent (c) The result is similar for an unpaired $J$ at $T$, which is
represented by an arrowed line starting from a square dot and ending at
a final momentum $\bold{p}'$ (see Fig.~\ref{fig:FM}(c)).

\noindent (d) Each $S$-matrix element has the factor $S_{0,0}$.

When drawing the diagrams of $S_{\bold{p}'_1\cdots\bold{p}'_m,
\bold{p}_1\cdots\bold{p}_n}$, we can
connect arbitrary initial $\bold{p}_j$ to arbitrary final $\bold{p}'_l$, or connect
arbitrary $\bold{p}_j$ (or $\bold{p}'_l$) to a square dot. But we cannot connect
an initial $\bold{p}_j$ to another initial $\bold{p}_l$, nor can we connect a final
$\bold{p}'_j$ to another final $\bold{p}'_l$. Bearing this mind, we find that
$S_{\bold{p}',\bold{p}}$ can be expressed as a sum of two diagrams (see Fig.~\ref{fig:FM}
the bottom panel). By using the diagrammatic technique, we repeat the result
in Eq.~\eqref{eq:phi:dia:Sppres}.

Since an external line (into or out of the diagram) can be connected to
a square dot, $S_{\bold{p}'_1\cdots\bold{p}'_m,
\bold{p}_1\cdots\bold{p}_n}$ is nonzero even for $n\neq m$.
In the presence a random driving field,
particles with finite energy can be generated from the vacuum, or annihilated into the vacuum.
The conservation of particle number is broken.

In conventional QFT, the Lorentz invariance of $S$-matrix manifests
as a relation between $S_{\bold{p}'_1\cdots\bold{p}'_m,
\bold{p}_1\cdots\bold{p}_n}$ and $S_{\tilde{\bold{p}}'_1\cdots\tilde{\bold{p}}'_m,
\tilde{\bold{p}}_1\cdots\tilde{\bold{p}}_n}$ where $\tilde{\bold{p}}=L\bold{p}$
is the Lorentz transformation of momentum $\bold{p}$. This relation holds for
the stochastic QFT. But since the $S$-matrix element of stochastic QFT is a random number,
the equality must be replaced by the equality in distribution. We have
\begin{equation}\label{eq:phi:dia:Lo}
S_{\bold{p}'_1\cdots\bold{p}'_m,
\bold{p}_1\cdots\bold{p}_n} \stackrel{d}{=}\prod_{j=1}^n \sqrt{\frac{E_{\tilde{\bold{p}}_j}}
{E_{\bold{p}_j}}} \prod_{l=1}^m \sqrt{\frac{E_{\tilde{\bold{p}}'_l}}
{E_{\bold{p}'_l}}} \ 
S_{\tilde{\bold{p}}'_1\cdots\tilde{\bold{p}}'_m,
\tilde{\bold{p}}_1\cdots\tilde{\bold{p}}_n}.
\end{equation}
Eq.~\eqref{eq:phi:dia:Lo} is easy to prove. As shown in Fig.~\ref{fig:FM},
besides some irrelevant constants,
the $S$-matrix consists of the factors $\delta^3(\bold{p'}-\bold{p})$, $W(E_{\bold{p}},\bold{p})/
\sqrt{E_\bold{p}}$ and $W^*(E_{\bold{p}'},\bold{p}')/
\sqrt{E_{\bold{p}'}}$. The $\delta$-function satisfies
${E_{\tilde{\bold{p}}}}\delta^3\left(\tilde{\bold{p}}'-\tilde{\bold{p}}\right)
={E_{{\bold{p}}}}\delta^3\left({\bold{p}}'-{\bold{p}}\right)$,
and then it does transform as what Eq.~\eqref{eq:phi:dia:Lo} shows.
$W(E_{\bold{p}},\bold{p})/\sqrt{E_\bold{p}}$ or $W^*(E_{\bold{p}'},\bold{p}')/
\sqrt{E_{\bold{p}'}}$ also transform in the same way, once if
$W(E_{\bold{p}},\bold{p}) \stackrel{d}{=}\tilde{W}(E_{\tilde{\bold{p}}},\tilde{\bold{p}})$,
which can be proved by using $dW(x)\stackrel{d}{=} d\tilde{W}(\tilde{x})$ and
$px=(Lp)(Lx)$ in Eq.~\eqref{eq:phi:dia:Wpd}.
In general, $W(p)$ and $\tilde{W}(\tilde{p})$ both have a Gaussian
distribution with zero mean and the same variance for arbitrary four-momentum $p$,
or we can say that $W\left(p\right)\stackrel{d}{=}\tilde{W}\left(\tilde{p}\right)$ is a Lorentz
scalar (its property will be discussed later).

The Lorentz invariance of $S$-matrix indicates that
the distribution of state vectors after particles interact with each other
in a collision experiment does not depend on which laboratory frame of reference we choose.
It is through Eq.~\eqref{eq:phi:dia:Lo} that the Lorentz
symmetry of a stochastic QFT can be tested in experiments.

On the other hand, in conventional QFTs, the spacetime translational symmetry
indicates that the $S$-matrix vanishes unless the
four-momentum is conserved. As mentioned above, this is not true for
the stochastic QFT. $S_{\bold{p}'_1\cdots\bold{p}'_m,
\bold{p}_1\cdots\bold{p}_n}$ as a random number is nonzero
even for $\bold{p}'_1+ \cdots + \bold{p}'_m \neq \bold{p}_1+ \cdots +\bold{p}_n$
or $E_{\bold{p}'_1}+ \cdots + E_{\bold{p}'_m} \neq E_{\bold{p}_1}
+ \cdots +E_{\bold{p}_n}$. The explicit translational symmetry is broken.
And the statistical translational symmetry is
hidden in the equations of motion (see Eq.~\eqref{eq:phi:ca:sch} and~\eqref{eq:phi:ca:rho}),
as we have analyzed.

\subsection{Excitation of the vacuum\label{sec:phi:sin}}

Assuming the universe to be initially ($t=-T$) at the vacuum state of free particles, we study
how the state vector evolves randomly in the Hilbert space.
As we already show, the energy does not conserve and particles can be generated
from the vacuum under the random driving. By using the $S$-matrix, we can obtain the
state vector at $t=T$, which is the state of universe after an evolution of period $2T$.

\begin{figure}[tbp]
\vspace{0.2cm}
\includegraphics[width=0.9\linewidth]{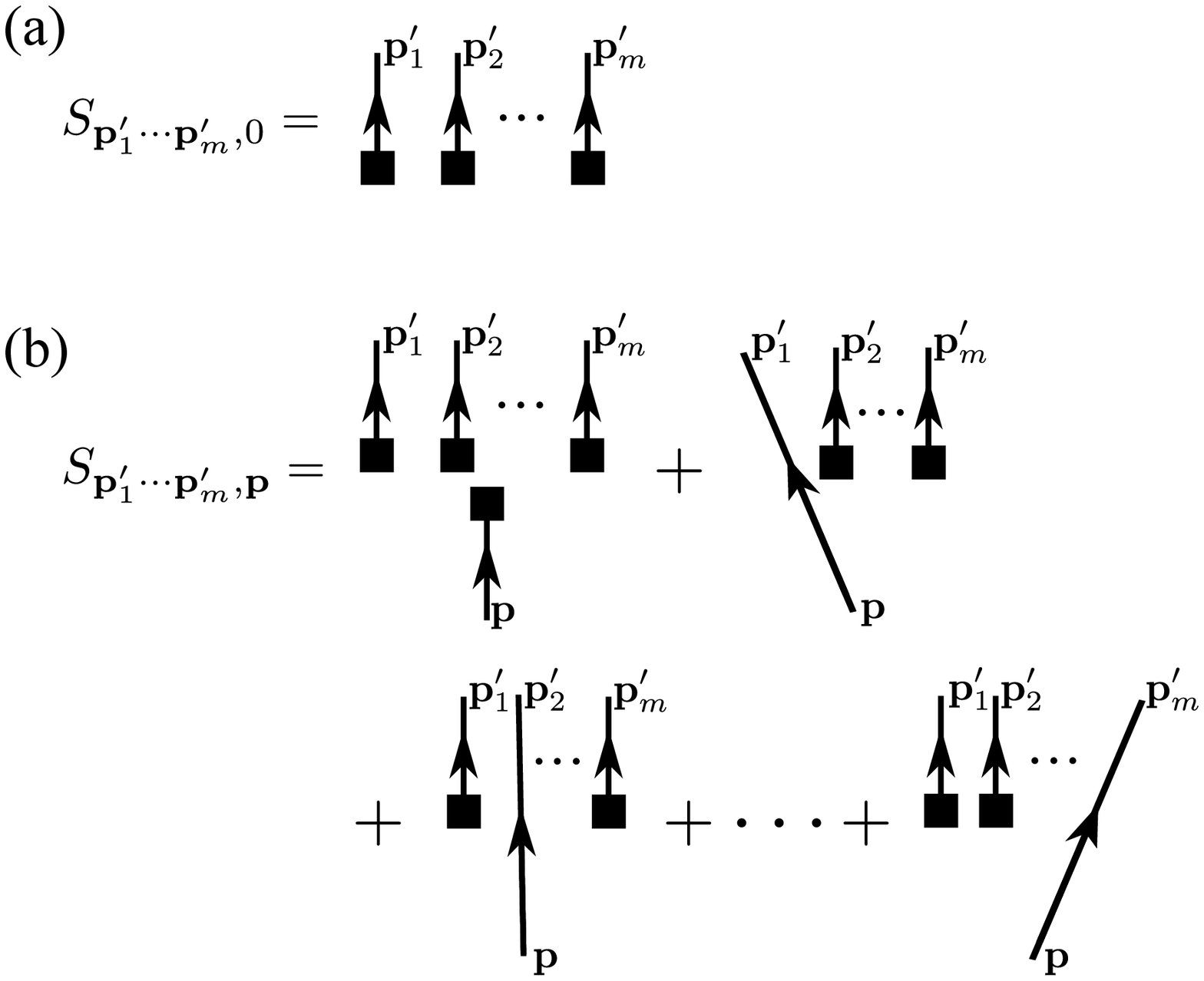}
\caption{The graphical expressions of (a) $S_{\bold{p}'_1\cdots\bold{p}'_m,0}$
and (b) $S_{\bold{p}'_1\cdots\bold{p}'_m,\bold{p}}$.}\label{fig:fate}
\end{figure}
Using the definition of $S$-matrix and the complete relation
$ \displaystyle\sum_{m=0}^\infty \displaystyle\frac{1}{m!}  \displaystyle\int 
 \displaystyle\prod_{j=1}^m
d^3 \bold{p}'_j \ket{\bold{p}'_1\cdots\bold{p}'_m}\bra{\bold{p}'_1\cdots\bold{p}'_m} =1$, we find
\begin{equation}\label{eq:phi:sin:0ra}
\hat U_I(T,-T) \ket{0} = \sum_{m=0}^\infty \frac{1}{m!} \int \prod_{j=1}^m
d^3 \bold{p}'_j \ S_{\bold{p}'_1\cdots\bold{p}'_m,0} \ket{\bold{p}'_1\cdots\bold{p}'_m}.
\end{equation}
The $S$-matrix is calculated by using the diagrammatic technique (see Fig.~\ref{fig:fate}(a)),
which reads
\begin{equation}
\begin{split}
S_{\bold{p}'_1\cdots\bold{p}'_m,0} = S_{0,0} \prod_{j=1}^m \left(-V^*_{\bold{p}'_j} \right).
\end{split}
\end{equation}
We then obtain
\begin{equation}\label{eq:phi:sin:U0}
\hat U_I(T,-T) \ket{0} = S_{0,0} \ exp\left\{-\int d^3\bold{p}
V^*_{\bold{p}} \hat a^\dag_{\bold{p}}\right\}\ket{0}.
\end{equation}

Due to the unitarity of $\hat U_I(T,-T)$, the vector
$\hat U_I(T,-T) \ket{0}$ should be normalized for an arbitrary
configuration of random field. This can be proved as follows.
We reexpress $S_{0,0}$ in terms of $W(p)$ or $V_\bold{p}$.
Substituting Eq.~\eqref{eq:phi:path:Gxxx} into Eq.~\eqref{eq:phi:path:S00r},
we find
\begin{equation}\label{eq:phi:sin:S2}
\begin{split}
& S_{0,0} = exp\left\{\frac{\lambda^2}{2}\frac{i}{(2\pi)^4}
\displaystyle \int \frac{d^4{p}}{p^2+m^2-i\eta} \left|W(p)\right|^2 \right\}, \\ &
 \left| S_{0,0} \right|^2 = exp\left\{- \displaystyle \int d^3\bold{p}
 \left| V_{\bold{p}} \right|^2\right\}.
 \end{split}
\end{equation}
By using $e^{\hat A} e^{\hat B}=
 e^{\hat B}e^{\hat A} e^{\left[\hat A,\hat B\right]}$ which stands for
arbitrary $\hat A$ and $\hat B$ once if $\left[\hat A,\hat B\right]$ is a $C$-number,
we immediately verify that the norm of $\hat U_I(T,-T) \ket{0}$ is unity.

In Eq.~\eqref{eq:phi:sin:U0}, $S_{0,0}$ and $V_{\bold{p}}$ are both a functionsal
of $W(p)$. We are now in a position to discuss the properties of $W(p)$.
According to the definition~\eqref{eq:phi:dia:Wpd}, there exists a one-to-one
map between $W(p)$ and $dW(x)$. A functional
of $dW(x)$ is also a functional of $W(p)$. Therefore, we can use
the set of configurations of $W(p)$ as the sample space,
instead of configurations of $dW(x)$. A random number (such as $S_{0,0}$)
is seen as a functional of $W(p)$.

The evolution we are interested in is during the time interval
$[-T,T]$, and we assume the volume of space to be finite (denoted by $V$),
hence, the total volume of spacetime is $2TV$.
Such a spacetime has already been discretized into elements of volume $\Delta^4x$.
There are then totally $\mathcal{M}=2TV/\Delta^4x$ elements. As a Fourier
transform, $W(p)$ is defined over the four-momentum
space which is reciprocal to the four-dimensional
spacetime. According to definition, the $p$-space
has a total volume $(2\pi)^4/\Delta^4x$ with each element having a volume $\Delta^4 p = (2\pi)^4/(2TV)$.
The total number of $p$-space elements is also $\mathcal{M}$.

Now, $W(p)=\sum_x \Delta W(x) e^{ipx}$
is a linear combination of independent Gaussian random numbers,
thereafter, it has a Gaussian distribution. $\Delta W(x)$ being real indicates $W(-p)=W^*(p)$,
we then only consider the set of $W(p)$s with $p^0 >0$.
There are $\mathcal{M}/{2}$ different $p$s with $p^0 >0$. But since each $p$ corresponds to two
random numbers which are $\text{Re}W(p)$ (real part of $W(p)$) and $\text{Im}W(p)$
(imaginary part of $W(p)$), there are totally $\mathcal{M}$ random numbers.
We put all these random numbers in a column vector, and then
the Fourier transformation can be expressed in a matrix form, which reads
\begin{widetext}
\begin{equation}\label{eq:phi:dwpana}
\begin{split}
\frac{1}{\sqrt{\mathcal{M}/2}} \left(\begin{array}{c}
{\text{Re} W(p_1)}\\ 
\text{Im} W(p_1) \\ 
\text{Re} W(p_2) \\ 
\text{Im} W(p_2) \\
\cdots
\end{array}\right) = \frac{1}{\sqrt{\mathcal{M}/2}} \left(\begin{array}{ccccc}
\cos\left(p_1x_1\right) & \cos\left(p_1x_2\right) 
& \cos\left(p_1x_3\right) & \cos\left(p_1x_4\right) & \cdots \\
\sin\left(p_1x_1\right) & \sin\left(p_1x_2\right) 
& \sin\left(p_1x_3\right) & \sin\left(p_1x_4\right) & \cdots \\
\cos\left(p_2x_1\right) & \cos\left(p_2x_2\right) 
& \cos\left(p_2x_3\right) & \cos\left(p_2x_4\right) & \cdots \\
\sin\left(p_2x_1\right) & \sin\left(p_2x_2\right) 
& \sin\left(p_2x_3\right) & \sin\left(p_2x_4\right) & \cdots \\
\cdots & \cdots & \cdots & \cdots & \cdots
\end{array}
 \right)
 \left(\begin{array}{c}
{\Delta W(x_1)}\\ 
\Delta W(x_2) \\ 
\Delta W(x_3) \\ 
\Delta W(x_4) \\
\cdots
\end{array}\right).
\end{split}
\end{equation}
\end{widetext}
It is easy to see that the transformation matrix is orthogonal.
If $\left(\Delta W(x_1) ,\Delta W(x_2),\cdots\right)^T$ is a vector
of independent Gaussian random numbers, so must be 
$\left(\text{Re} W(p_1), \text{Im} W(p_1),\cdots \right)^T$. 
The mean of $\text{Re} W(p)$ or $\text{Im} W(p)$ is zero,
and the variance is $TV = {(2\pi)^4}/{\left(2\Delta^4 p\right)}$.
$\Delta^4 p$ or $TV$ are scalars under Lorentz transformation, which proves
that $W(p)$ is a Lorentz scalar. Especially, if $p$ is on-shell ($p^0=E_\bold{p}>0$), the real and imaginary
parts of $W(E_\bold{p},\bold{p})$ are independent,
both having zero mean and variance ${(2\pi)^4}/(2{\Delta^4 p})$.
For $\bold{p}\neq \bold{p}'$, $W(E_\bold{p},\bold{p})$ and $W(E_{\bold{p}'},\bold{p}')$
are independent of each other.

Let us analyze the expression of state vector at $t=T$ (see Eq.~\eqref{eq:phi:sin:U0}).
Since we know the distribution of $W(p)$, the probability distribution of
$\hat U_I(T,-T) \ket{0}$ is also clear.
As $\lambda=0$, Eq.~\eqref{eq:phi:sin:U0} reduces to $\hat U_I(T,-T) \ket{0}
=\ket{0}$. The vacuum state keeps invariant in the absence of
random field. As $\lambda\neq 0$, we reexpress $\hat U_I(T,-T) \ket{0}$
as $S_{0,0} \displaystyle\prod_{\bold{p}} e^{-\gamma_\bold{p}
 \hat a^\dag_{\bold{p}}} \ket{0}$ with $\gamma_\bold{p}=\Delta^3\bold{p}
V^*_\bold{p}$. It is clear that the final state is a coherent state, i.e. the eigenstate of annihilation
operator $\hat a_\bold{p}$ with the eigenvalue
$-\gamma_\bold{p}$. In the final state, the number of particles is uncertain.
Moreover, $\gamma_\bold{p}$ at different $\bold{p}$s are
independent random numbers. The final state is then a product
of random coherent states in the momentum space. According to the above analysis
and the facts $\Delta^4p = \Delta^3\bold{p} \Delta p^0$ and
$\Delta p^0 = 2\pi/(2T)$, the mean of $\left| \gamma_\bold{p}\right|^2$ is $\Delta^3\bold{p}{\lambda^2 T}/{E_\bold{p}}$.
For a bigger $\lambda$, we have a bigger probability
of observing more excitations at $t=T$, and this probability increases with $T$.
The probability of observing a high-energy particle is smaller
than that of a low-energy one, and the probability gradually vanishes in the limit $E_\bold{p}\to \infty$
since $\gamma_\bold{p}$ is inversely proportional to $\sqrt{E_\bold{p}}$.
The random field generates particles from the vacuum.
As a consequence, the temperature of universe increases.
More properties of the final state will be discussed after we obtain the density matrix.

It is worth mentioning that what we obtain is $\hat U_I(T,-T)\ket{0}$,
while the state vector in the Schr\"{o}dinger picture
is in fact $\hat U(T,-T)\ket{0}$. But the relation between interaction and
Schr\"{o}dinger pictures is simple, which leads to $\hat U(T,-T)\ket{0}= e^{-iT\hat H_0}\hat U_I(T,-T)\ket{0}$.
Since we express $\hat U_I(T,-T)\ket{0}$ in the momentum
space, $e^{-iT\hat H_0}$ only results in a phase factor additional to $\gamma_\bold{p}$.

Beyond the vacuum, we consider the case of initial state being
$\ket{\bold{p}_1\cdots\bold{p}_n}$ for arbitrary $n$.
Using the diagrammatic technique, we calculate the $S$-matrix and
then $\hat U_I(T,-T)\ket{\bold{p}_1\cdots\bold{p}_n}$. For example,
Fig.~\ref{fig:fate}(b) displays the diagrams of $S_{\bold{p}'_1\cdots\bold{p}'_m,\bold{p}}$
as $n=1$. For arbitrary $n$, the final state vector is found to be
\begin{equation}\label{eq:phi:sin:Up}
\hat U_I(T,-T)\ket{\bold{p}_1\cdots\bold{p}_n} =
\prod_{j=1}^n \left(\hat a^\dag_{\bold{p}_j}+V_{\bold{p}_j}\right)
\hat U_I(T,-T)\ket{0}.
\end{equation}
If there are initially $n$ particles, after an evolution of period $2T$,
the universe is in a random coherent state, similarly as the initial state is a vacuum.
Eq.~\eqref{eq:phi:sin:Up} tells us that the excitation caused by random driving is a background effect,
which is not affected by whether there are particles or not at the initial time,
but depends only on the driving strength and period.

Since a state vector can always be expressed as a linear
combination of $\ket{\bold{p}_1\cdots\bold{p}_n}$, by using Eq.~\eqref{eq:phi:sin:Up}
one can study the evolution of an arbitrary initial state.
For example, if we use $\ket{\bold{x}}= \displaystyle\int d^3\bold{p} \
e^{-i\bold{p}\cdot \bold{x}} \ket{\bold{p}}$ as the initial state which
describes a particle localized at a specific position in the space,
the state vector at $t=T$ is then $\displaystyle\int d^3\bold{p} e^{-i\bold{p}\cdot\bold{x}}
\left(\hat a^\dag_\bold{p} + V_\bold{p}\right)  \hat U_I(T,-T)\ket{0}$.
Here we see $\hat U_I(T,-T)\ket{0}$ again, which describes the background excitation
caused by random field.

\subsection{Diagrammatic rules for density matrix\label{sec:phi:den}}

The final quantum state is a random vector in the Hilbert space.
To further understand its properties,
we calculate the corresponding density matrix.
The density matrix encodes less information than the random vector
(there exist different distributions of random vector
that correspond to the same density matrix), but it
is more transparent, providing a simple picture of what happened
during the evolution.

We use $\ket{\alpha}$ to denote the initial state vector.
In the interaction picture, the density matrix at $t=T$ is expressed as
\begin{equation}\label{eq:phi:den:rha}
\begin{split}
\hat \rho^I_\alpha(T) = & \text{E} \left( \hat U_I(T,-T) \ket{\alpha}\bra{\alpha}
\hat U_I^\dag(T,-T)\right) \\
= & \sum_{n,m=0}^\infty \frac{1}{n!\ m!}  \int \prod_{l=1}^m
d^3\bold{p}'_l \prod_{j=1}^n d^3 \bold{q}'_j  \times \\ & \rho^{(\alpha)}_{\bold{p}'_1\cdots
\bold{p}'_m, \bold{q}'_1\cdots\bold{q}'_n}  \ket{\bold{p}'_1\cdots\bold{p}'_m}
 \bra{\bold{q}'_1\cdots\bold{q}'_n},
\end{split}
\end{equation}
where
\begin{equation}\label{eq:phi:den:rpp}
 \rho^{(\alpha)}_{\bold{p}'_1\cdots
\bold{p}'_m, \bold{q}'_1\cdots\bold{q}'_n} = \text{E} \left( S_{\bold{p}'_1\cdots\bold{p}'_m,\alpha}
S^*_{\bold{q}'_1\cdots\bold{q}'_n,\alpha} \right)
\end{equation}
is the matrix element in the momentum basis.

Without loss of generality, we choose the initial state to be
$\ket{\alpha}=\ket{\bold{p}_1\cdots\bold{p}_a}$ (an arbitrary
state vector can be expressed as a linear combination of $\ket{\bold{p}_1\cdots\bold{p}_a}$).
The $S$-matrices in Eq.~\eqref{eq:phi:den:rpp} can be obtained
by using the diagrammatic technique. And $\rho^{(\alpha)}$
is the product of one $S$-matrix element and the complex conjugate of another averaged over the
configurations of random field or $W(p)$. Each $S$-matrix element has a factor $S_{0,0}$,
hence, the product has a factor $\left| S_{0,0}\right|^2$.
And according to Fig.~\ref{fig:FM}, the product has also
the factors $V_\bold{p}$ or $V^*_{\bold{p}'}$ if the diagrams of $S$-matrices contain square dots.
To obtain $\rho^{(\alpha)}$, we need to evaluate something like
$\text{E} \left(\left| S_{0,0}\right|^2 V_{\bold{p}_1} V^*_{\bold{p}'_1}\cdots\right)$.

It is necessary to derive a general formula for the expectation value.
Since $\text{Re} W(E_{\bold{p}},\bold{p})$ and
$\text{Im} W(E_{\bold{p}},\bold{p})$ are independent Gaussians of
zero mean and variance $(2\pi)^4/(2\Delta^4 p)$, $\left(\text{Re}V_{\bold{p}_1},
\text{Im}V_{\bold{p}_1}, \text{Re}V_{\bold{p}_2},\cdots\right)$ is then
a sequence of independent Gaussians with each having zero mean and
variance $\sigma^2_\bold{p}= \lambda^2 T/(2\Delta^3\bold{p} E_\bold{p})$.
Let us use $\chi_\bold{p}^+$ and $\chi_\bold{p}^-$ to denote
two independent functions of $\bold{p}$. We find
\begin{equation}\label{eq:phi:den:ES}
\begin{split}
& \text{E} \left(\left| S_{0,0}\right|^2 exp\left\{ \int d^3\bold{p} \ \chi^+_\bold{p}
V_\bold{p} +  \int d^3\bold{p} \ \chi^-_\bold{p}
V_\bold{p}^* \right\}\right) \\
= & \int \prod_{\bold{p}} \left(
\frac{d \text{Re}V_{\bold{p}} \ d\text{Im}V_{\bold{p}} }{2\pi\sigma^2_\bold{p}}\right)
exp\left\{ -\sum_\bold{p} \frac{ \left| V_{\bold{p}}\right|^2}{2\sigma^2_\bold{p}}\right\} \\
& \times exp \{ \sum_\bold{p} \Delta^3 \bold{p}\left(- \left| V_{\bold{p}}\right|^2
+ \chi^+_\bold{p}
V_\bold{p} +  \chi^-_\bold{p}
V_\bold{p}^*  \right) \} \\ 
= & \frac{1}{Z} exp \left\{ \sum_\bold{p} \Delta^3\bold{p} \ \frac{1}{1+\tilde{E}_\bold{p}}
 \chi^+_\bold{p}\chi^-_\bold{p} \right\},
\end{split}
\end{equation}
where $Z = \displaystyle\prod_\bold{p} \frac{1+\tilde{E}_\bold{p}}{\tilde{E}_\bold{p}}$
is called the partition function and $\tilde{E}_\bold{p}=
E_\bold{p}/\left(\lambda^2 T\right)$ is the dimensionless
dispersion relation. The expectation of $\left|S_{0,0}\right|^2$ multiplied by
a sequence of $V_\bold{p}$ or $V_{\bold{p}'}^*$ is equal to the functional derivative of Eq.~\eqref{eq:phi:den:ES}
at $\chi^+=\chi^-=0$. As easily seen, the expectation value is nonzero if and only if
each $V_\bold{p}$ ($V_{\bold{p}'}^*$) is paired with a $V^*_\bold{p}$
($V_{\bold{p}'}$) in the sequence, which can be expressed as
\begin{equation}\label{eq:phi:den:SV}
\begin{split}
& \text{E} \left(\left| S_{0,0}\right|^2\prod_{j=1}^n V_{\bold{p}_j} 
\prod_{l=1}^n V^*_{\bold{p}'_l} \right) \\ & = \frac{1}{Z} \left(\prod_{j=1}^n
\frac{1}{1+\tilde{E}_{\bold{p}_j}}\right)  \sum_{\pi} \left[ \prod_{j=1}^n \delta^3\left(\bold{p}_j-
\bold{p}'_{\pi_j}\right) \right], 
\end{split}
\end{equation}
where we have used $\delta^3(\bold{p}-\bold{p}')=\delta_{\bold{p},\bold{p}'} /\Delta^3\bold{p}$,
$\pi$ denotes the permutation of $\left\{1,2,\cdots,n\right\}$ and the
sum is over all the $n!$ permutations. As $n=0$,
we have $ \text{E} \left(\left| S_{0,0}\right|^2\right) = 1/Z$.
As $n>0$, Eq.~\eqref{eq:phi:den:SV} consists of $n!$ terms which
come from the $n!$ ways of pairing $V_{\bold{p}'}^*$s with $V_\bold{p}$s.

\begin{figure}[tbp]
\vspace{0.2cm}
\includegraphics[width=0.9\linewidth]{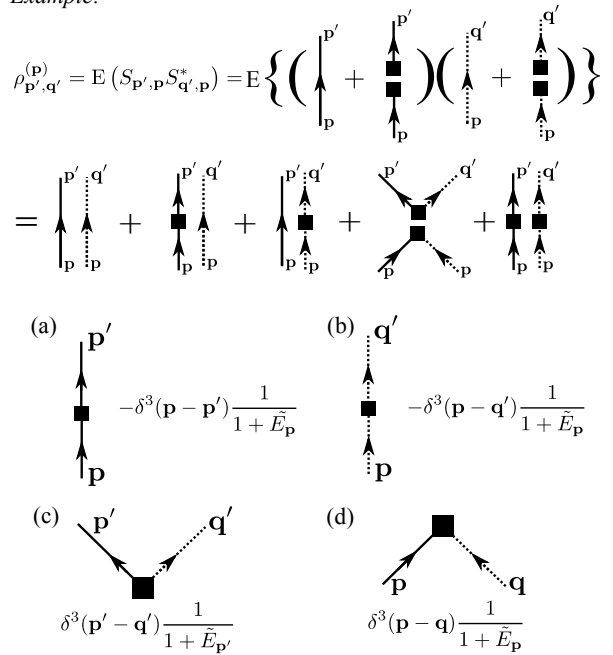}
\caption{The top panel is the graphical expression of
$\rho^{(\bold{p})}_{\bold{p}',\bold{q}'}$ which serves as an example
explaining how to calculate the density matrix by using the diagrammatic technique.
(a), (b), (c) and (d) explain the diagrammatic rules.}\label{fig:den}
\end{figure}
We are now prepared to calculate $\rho^{(\alpha)}_{\bold{p}'_1\cdots
\bold{p}'_m, \bold{q}'_1\cdots\bold{q}'_n}$. Each $S$-matrix element is
represented by a sum of Feynman diagrams in which a square dot represents $V_\bold{p}$
or $V_{\bold{p}'}^*$. Therefore, $\rho^{(\alpha)}$ can be represented by
a sum of paired-diagrams with each paired-diagram
consisting of one diagram from $S_{\bold{p}'_1\cdots\bold{p}'_m,\alpha}$
and the other from $S^*_{\bold{q}'_1\cdots\bold{q}'_n,\alpha}$ (see Fig.~\ref{fig:den}).
For easy to distinguish, we draw the diagram of
$S_{\bold{p}'_1\cdots\bold{p}'_m,\alpha}$ in solid lines and that of
$S^*_{\bold{q}'_1\cdots\bold{q}'_n,\alpha}$ in dotted lines. The
process of computing the expectation value is represented by pairing each dot
representing $V_\bold{p}$ with a dot representing $V_{\bold{p}'}^*$.
In the paired-diagram, we merge two paired dots into a single dot. Different ways of
pairing the dots result in different diagrams. A paired-diagram
is made of solid lines, dotted lines and square dots with each dot
connected to two lines. Fig.~\ref{fig:den} lists the possible
components of a paired-diagram. The rules for calculating $\rho^{(\alpha)}$ is summarized as follows:

\noindent (a) For a square dot with a leaving solid line of momentum
$\bold{p}'$ and an entering solid line of momentum $\bold{p}$,
include a factor $-\delta^3(\bold{p}-\bold{p}')/(1+\tilde{E}_\bold{p})$
(see Fig.~\ref{fig:den}(a)).

\noindent (b) For a dot with a leaving dotted line of momentum
$\bold{q}'$ and an entering dotted line of momentum $\bold{p}$, include a factor
$-\delta^3\left(\bold{p}-\bold{q}'\right)/(1+\tilde{E}_\bold{p})$
(see Fig.~\ref{fig:den}(b)).

\noindent (c) For a dot with a leaving solid line of momentum
$\bold{p}'$ and a leaving dotted line of momentum $\bold{q}'$,
include a factor $\delta^3\left(\bold{p}'-\bold{q}'\right)/(1+\tilde{E}_{\bold{p}'})$
(see Fig.~\ref{fig:den}(c)).

\noindent (d) For a dot with an entering solid line of momentum $\bold{p}$ and
an entering dotted line of momentum $\bold{q}$,
include a factor $\delta^3\left(\bold{p}-\bold{q}\right)/(1+\tilde{E}_{\bold{p}})$
(see Fig.~\ref{fig:den}(d)).

\noindent (e) For each paired-diagram, include a factor $1/Z$.

A square dot in the paired-diagram is similar to a vertex in the Feynman diagram. The $\delta$-function
ensures that the momentum is conserved at each dot, just
as it is conserved at each vertex. But a dot can be simultaneously connected to
one solid and one dotted lines, and in this case, to be consistent with
the momentum conservation, one needs to think of the momentum of dotted line as changing the sign.

\begin{figure}[tbp]
\vspace{0.2cm}
\includegraphics[width=0.9\linewidth]{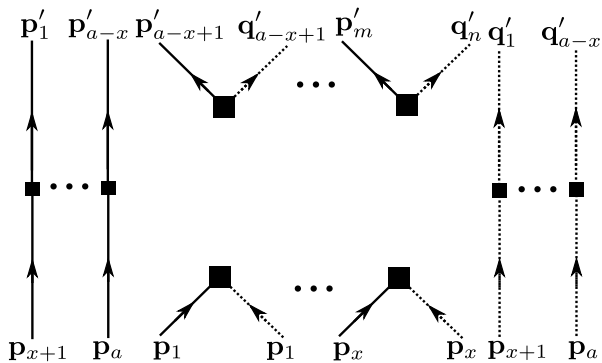}
\caption{Graphical representation of $\rho^{(\bold{p}_1\cdots\bold{p}_a)}_{\bold{p}'_1\cdots
\bold{p}'_m, \bold{q}'_1\cdots\bold{q}'_n}$. This figure explains how to prove
the equal-momentum condition.}\label{fig:equm}
\end{figure}
Let us use the diagrammatic rules to derive the general condition
of $\rho^{(\bold{p}_1\cdots\bold{p}_a)}_{\bold{p}'_1\cdots
\bold{p}'_m, \bold{q}'_1\cdots\bold{q}'_n}$ being nonzero. In a paired-diagram
that has nonzero contribution to $\rho^{(\bold{p}_1\cdots\bold{p}_a)}_{\bold{p}'_1\cdots
\bold{p}'_m, \bold{q}'_1\cdots\bold{q}'_n}$,
there are totally $a$ solid lines and the same number of dotted lines into the diagram,
while there are $m$ solid lines and $n$ dotted lines out of the diagram (see Fig.~\ref{fig:equm}).
A solid line into the diagram is either connected to a solid line out of the diagram
(directly or through a square dot), or it is connected to a dotted line into the
diagram, and in the latter case, the momentum carried by the solid and dotted lines
must be the same. Without loss of generality, we assume that there are $x$ solid lines connected
to dotted lines into the diagram, and the momenta that they carry are
$\bold{p}_1$, $\bold{p}_2$, $\cdots$, and $\bold{p}_x$, respectively.
Therefore, $\left(a-x\right)$ solid (dotted) lines into the diagram
are connected to $\left(a-x\right)$ solid (dotted) lines out of the diagram.
And due to the momentum conservation, the outgoing lines (whether solid or dotted)
must carry the same momentum as the entering lines, which
are $\bold{p}_{x+1}$, $\bold{p}_{x+2}$, $\cdots$, and $\bold{p}_a$, respectively. Now
the left $\left(m-a+x\right)$ outgoing solid lines and
$\left(n-a+x\right)$ dotted lines must be connected to each other,
and each pair of solid and dotted lines carry the same momentum.
This is possible only if $m=n$. Moreover, $\rho^{(\alpha)}_{\bold{p}'_1\cdots\bold{p}'_m,
\bold{q}'_1\cdots\bold{q}'_n}$ is nonzero if and only if the set of momenta
carried by the outgoing solid lines is as same as that carried by
the outgoing dotted lines, i.e.
\begin{equation}\label{eq:phi:den:mo}
\left\{\bold{p}'_1,\bold{p}'_2,\cdots,\bold{p}'_m\right\}=
\left\{\bold{q}'_1,\bold{q}'_2,\cdots,\bold{q}'_n\right\}.
\end{equation}

Eq.~\eqref{eq:phi:den:mo} is called the equal-momentum condition.
As a consequence of this condition, $\hat{\rho}^I_{\bold{p}_1\cdots\bold{p}_a}(T)$
in Eq.~\eqref{eq:phi:den:rha} can only have diagonal terms. In the presence
of a white-noise field like $dW(x)$, the density matrix after a large period of evolution
is diagonal in the momentum space, for arbitrary initial momenta of particles.
The coherence between states of different momentum is fully lost during
the evolution.

The equal-momentum condition of $\rho^{(\bold{p}_1\cdots\bold{p}_a)}_{\bold{p}'_1\cdots
\bold{p}'_m, \bold{q}'_1\cdots\bold{q}'_n}$ indicates the equivalence
of density matrices between the interaction and Schr\"{o}dinger pictures.
In the Schr\"{o}dinger picture, the density matrix at $t=T$
is defined as $\hat{\rho}_{\alpha} = \text{E}\left(\hat{U}(T,-T)\ket{\alpha}\bra{\alpha}
\hat{U}^\dag(T,-T)\right)$. At first sight, $\hat{\rho}_{\alpha}$ is different from $\hat{\rho}_{\alpha}^I$
in Eq.~\eqref{eq:phi:den:rha} since $\hat{U}(T,-T) = e^{-iT\hat H_0}\hat{U}_I(T,-T)
e^{-iT\hat H_0}$. But in the momentum basis, $e^{-iT\hat H_0}$ is
a phase factor, and the factor before $\ket{\alpha}$ cancels that after $\bra{\alpha}$,
at the same time, the factor after $\bra{\bold{p}'_1\cdots
\bold{p}'_m}$ always cancels that before $\ket{\bold{q}'_1\cdots\bold{q}'_n}$
due to the equal-momentum condition. Therefore, we have $\hat{\rho}_{\alpha}=
\hat{\rho}^I_{\alpha}$ for arbitrary $\ket{\alpha}= \ket{\bold{p}_1\cdots\bold{p}_a}$.
From now on, we will not distinguish $\hat{\rho}_{\alpha}$ and $\hat{\rho}^I_{\alpha}$ anymore,
and call both of them the density matrix.

In Eq.~\eqref{eq:phi:dia:Lo}, we already showed that the $S$-matrix is Lorentz-invariant.
From it, we can easily derive the Lorentz invariance of $\rho^{(\alpha)}$.
Again, we use $\tilde{\bold{p}}=L\bold{p}$ to denote
the Lorentz transformation of momentum. By using Eq.~\eqref{eq:phi:dia:Lo} and~\eqref{eq:phi:den:rpp},
we obtain
\begin{equation}\label{eq:phi:den:Si}
\begin{split}
\rho^{(\bold{p}_1\cdots\bold{p}_a)}_{\bold{p}'_1\cdots\bold{p}'_m,
\bold{q}'_1\cdots\bold{q}'_n} = & \prod_{l=1}^m \sqrt{\frac{E_{\tilde{\bold{p}}'_l}}
{E_{{\bold{p}}'_l}}}\prod_{j=1}^n \sqrt{\frac{E_{\tilde{\bold{q}}'_j}}
{E_{{\bold{q}}'_j}}} \prod_{k=1}^a {\frac{E_{\tilde{\bold{p}}_k}}
{E_{{\bold{p}}_k}}}  \\ & \times \rho^{(\tilde{\bold{p}}_1
\cdots\tilde{\bold{p}}_a)}_{\tilde{\bold{p}}'_1\cdots\tilde{\bold{p}}'_m,
\tilde{\bold{q}}'_1\cdots\tilde{\bold{q}}'_n} .
\end{split}
\end{equation}
One can prove Eq.~\eqref{eq:phi:den:Si} by using the diagrammatic rules. First, $\tilde{E}_\bold{p}$
does not change under a Lorentz transformation, because
${E}_\bold{p}/T= \left(E_\bold{p}/\Delta^3\bold{p}\right)\left(\Delta^4 p/\pi\right)$
and both $E_\bold{p}/\Delta^3\bold{p}$ and $\Delta^4 p$ are scalars. As a consequence,
$Z$ must be a scalar under Lorentz transformation. Furthermore, the factors
of $\rho^{(\alpha)}$ include the Dirac-$\delta$ function and $1/\left(1+\tilde{E}_\bold{p}\right)$,
as shown in Fig.~\ref{fig:den}. The former transforms as Eq.~\eqref{eq:phi:den:Si} and
the latter is a scalar. Therefore, Eq.~\eqref{eq:phi:den:Si} stands for arbitrary $\left(\bold{p}_1,\cdots,
\bold{p}_a\right)$, $\left(\bold{p}'_1,\cdots,\bold{p}'_m\right)$ and $\left(\bold{q}'_1,\cdots,
\bold{q}'_n\right)$.

From the Lorentz invariance of $\rho^{(\alpha)}$, we can derive the Lorentz invariance
of density matrix. Under a Lorentz transformation, the free-particle state transforms as
\begin{equation}\label{eq:phi:den:Usv}
\ket{\tilde{\bold{p}}_1\cdots\tilde{\bold{p}}_a} = \prod_{k=1}^a \sqrt{\frac{E_{\bold{p}_k}}
{E_{\tilde{\bold{p}}_k}}} \hat{U}(L) \ket{{\bold{p}}_1\cdots{\bold{p}}_a} ,
\end{equation}
where $\hat{U}(L)$ is the unitary representation of the Lorentz transformation $L$.
By using Eq.~\eqref{eq:phi:den:Si} and~\eqref{eq:phi:den:Usv}, we obtain
\begin{equation}\label{eq:phi:den:out}
\hat{\rho}_{\tilde{\bold{p}}_1\cdots\tilde{\bold{p}}_a}(T) = \left(\prod_{k=1}^a {\frac{E_{\bold{p}_k}}
{E_{\tilde{\bold{p}}_k}}}\right)  \hat{U} \ \hat{\rho}_{{\bold{p}}_1\cdots{\bold{p}}_a}(T) \ \hat{U}^\dag .
\end{equation}
In a scattering experiment, the density matrix encodes the information of the probability distribution of
final outcomes. Eq.~\eqref{eq:phi:den:out} tells us that the outcome of
an experiment is independent of which reference frame we choose.

\subsection{Expressions of density matrix \label{sec:phi:speden}}

\begin{figure}[tbp]
\vspace{0.2cm}
\includegraphics[width=0.9\linewidth]{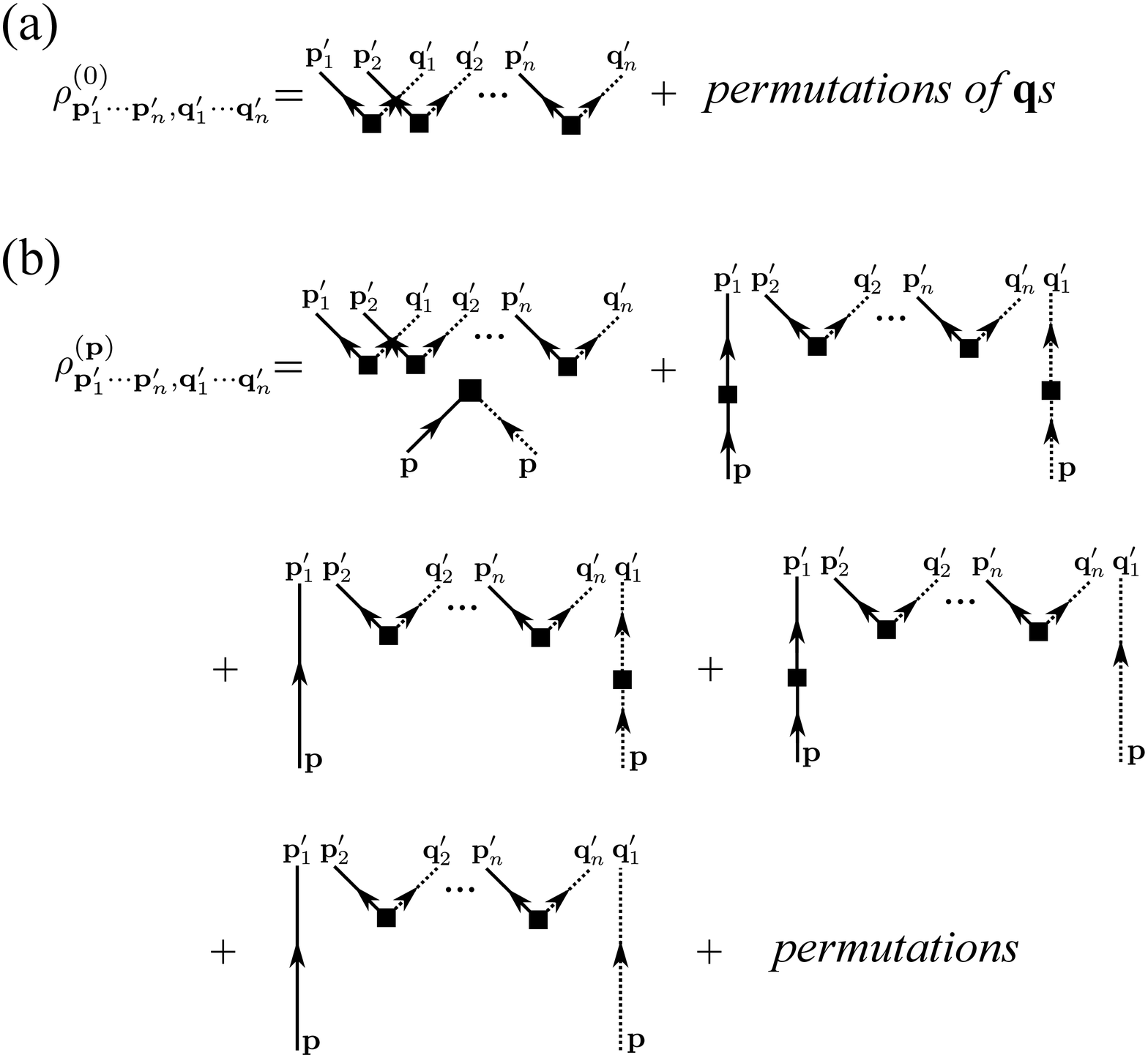}
\caption{Graphical expressions of (a) $\rho^{(0)}_{\bold{p}'_1\cdots\bold{p}'_n,\bold{q}'_1
\cdots\bold{q}'_n}$ and (b) $\rho^{(\bold{p})}_{\bold{p}'_1\cdots\bold{p}'_n,\bold{q}'_1
\cdots\bold{q}'_n}$. This figure is used in the calculation of $\hat \rho_0$ and
$\hat{\rho}_{\bold{p}}$.}\label{fig:rho0}
\end{figure}
Let us calculate the density matrix at the time $t=T$ for given initial states.
We first consider the vacuum state as the initial state.
Fig.~\ref{fig:rho0}(a) displays the diagrams for calculating $\rho^{(0)}_{\bold{p}'_1\cdots
\bold{p}'_n, \bold{q}'_1\cdots\bold{q}'_n}$. There are totally $n!$ diagrams,
corresponding to $n!$ permutations of $\bold{q}$s. Using the diagrammatic rules, we obtain
\begin{equation}
\begin{split}
& \rho^{(0)}_{\bold{p}'_1\cdots
\bold{p}'_n, \bold{q}'_1\cdots\bold{q}'_n} \\ &= \frac{1}{Z} \left(\prod_{j=1}^n 
\frac{1}{1+\tilde{E}_{\bold{p}'_j}}\right)  \sum_{\pi} \left[ \prod_{j=1}^n \delta^3\left(\bold{p}'_j-
\bold{q}'_{\pi_j}\right) \right], 
\end{split}
\end{equation}
where $\pi$ denotes the permutation. And the density matrix turns out to be
\begin{equation}\label{eq:phi:den:rho0T}
\begin{split}
\hat{\rho}_0(T)=& \frac{1}{Z}\sum_{n=0}^\infty \frac{1}{n!} \int \left(\prod_{j=1}^n \frac{d^3\bold{p}_j}
{1+\tilde{E}_{\bold{p}_j}} \right) \\
&\times \ket{\bold{p}_1\cdots\bold{p}_n} \bra{\bold{p}_1\cdots\bold{p}_n}.
\end{split}
\end{equation}

We check the trace of this density matrix. If
there are $m_1$ particles of momentum $\bold{p}_1$, $m_2$ particles of momentum $\bold{p}_2$,
etc., finally, there are $m_j$ particles of momentum $\bold{p}_j$ with
$\bold{p}_1 \neq \bold{p}_2 \neq \cdots \neq\bold{p}_j$,
the squared norm of the state vector $\ket{\bold{p}_1
\cdots\bold{p}_1\bold{p}_2\cdots\bold{p}_2\cdots \bold{p}_j\cdots\bold{p}_j}$ 
is $m_1!m_2!\cdots m_j!/ \left(\Delta^3\bold{p}\right)^{m_1+m_2+\cdots+m_j}$.
Using this result and
\begin{equation}
Z =  \displaystyle\prod_\bold{p} \frac{1+\tilde{E}_\bold{p}}{\tilde{E}_\bold{p}}=
\prod_{\bold{p}} \left( \sum_{m=0}^\infty \frac{1}{\left(1+\tilde{E}_\bold{p}\right)^m}\right),
\end{equation}
we verify $\text{Tr} \left[\hat{\rho}_0(T)\right]=1$. The trace of the density matrix
keeps unity, as it should be during a unitary evolution.

The long time limit of the density matrix~\eqref{eq:phi:den:rho0T} is trivial. As $T\to\infty$, 
we have $\tilde{E}_\bold{p} \to 0$ for arbitrary $\bold{p}$. By using the complete relation
of the Hilbert space, we find that
$\displaystyle\lim_{T\to\infty} \hat{\rho}_0(T) ={1}/{Z}$ is a constant.
A constant density matrix describes a state at infinite temperature.
Therefore, after infinitely long evolution the
universe is driven into an infinite-temperature state, even it contains no
particle initially.

On the other hand, as $T\to 0$, we have $\tilde{E}_\bold{p} \to \infty$ and then
$Z\to 1$. In Eq.~\eqref{eq:phi:den:rho0T}, the terms with $n>0$ all vanish
due to the factor $1/\left(1+\tilde{E}_\bold{p}\right)$.
$\hat{\rho}_0(T)$ is then equal to the initial density matrix. In general,
due to $E_\bold{p}\geq m$ with $m$ being the mass, we can say
${E}_\bold{p}\gg \lambda^2 T$ for arbitrary $\bold{p}$ once if $m\gg \lambda^2T$.
Therefore, as $T \ll m/\lambda^2$, there exists no significant excitation at each momentum.

For an intermediate $T$, Eq.~\eqref{eq:phi:den:rho0T} describes a state
with many-particle excitations. The density matrix is diagonal in the momentum basis.
The probability of observing $n$-particle excitations is nonzero for arbitrary $n>0$. A detailed
calculation of the probability needs the renormalization of $\lambda$, which will
be discussed in next subsection.

Let us move on and calculate the density matrix when
there exists initially a single particle of momentum $\bold{p}$.
The diagrams for calculating $\rho^{(\bold{p})}_{\bold{p}'_1\cdots
\bold{p}'_n,\bold{q}'_1\cdots\bold{q}'_n}$ are displayed in Fig.~\ref{fig:rho0}(b).
Using the diagrammatic rules, we find
\begin{equation}\label{eq:phi:spe:rhop}
\begin{split}
&\rho^{(\bold{p})}_{\bold{p}'_1\cdots
\bold{p}'_n,\bold{q}'_1\cdots\bold{q}'_n} \\ & =
\frac{\rho^{(0)}_{\bold{p}'_1\cdots
\bold{p}'_n,\bold{q}'_1\cdots\bold{q}'_n}}{1+\tilde{E}_\bold{p}}
\left[\delta^3(0)+\tilde{E}^2_\bold{p}\sum_{j=1}^n \delta^3
\left( \bold{p}'_j-\bold{p} \right)\right],
\end{split}
\end{equation}
where $\delta^3(0)=\braket{\bold{p}|\bold{p}}=1/\Delta^3\bold{p}$.
Substituting Eq.~\eqref{eq:phi:spe:rhop} into Eq.~\eqref{eq:phi:den:rha}, we obtain
\begin{equation}\label{eq:phi:spe:rhp}
\begin{split}
\hat{\rho}_\bold{p}(T)=&  \frac{1}{\Delta^3\bold{p}} \ 
\frac{ 1}{Z}\sum_{n=0}^\infty \frac{1}{n!} \int \left(\prod_{j=1}^n \frac{d^3\bold{p}_j}
{1+\tilde{E}_{\bold{p}_j}} \right)\\
&\times \frac{1+  \tilde{E}^2_\bold{p} \sum_{k=1}^n \delta^3\left(
\bold{p}_k-\bold{p}\right)\Delta^3\bold{p} }
{1+\tilde{E}_\bold{p}} \\ & \times 
\ket{\bold{p}_1\cdots\bold{p}_n} \bra{\bold{p}_1\cdots\bold{p}_n},
\end{split}
\end{equation}
where the factor $ {1}/{\Delta^3\bold{p}} $
comes from the squared norm of the initial state vector $\ket{\bold{p}}$.

Again, in the limit $T\to\infty $ the density matrix~\eqref{eq:phi:spe:rhp} goes to a constant, indicating
that the universe thermalizes to an infinite-temperature state. And in the
limit $T\to 0$, $\hat{\rho}_\bold{p}(T)$ equals the initial density matrix.
The Dirac-$\delta$ and Kronecker-$\delta$ functions are related to each other by $\delta^3\left(
\bold{p}_k-\bold{p}\right)\Delta^3 \bold{p}=\delta_{\bold{p}_k,\bold{p}} $. As a consequence,
$\sum_{k=1}^n \delta^3\left(
\bold{p}_k-\bold{p}\right)\Delta^3\bold{p}$ is the number of particles
of momentum $\bold{p}$ in the state vector $\ket{\bold{p}_1\cdots\bold{p}_n}$.
Once if $T\ll m/\lambda^2$, we must have $\tilde{E}_\bold{p}\gg 1$
for arbitrary $\bold{p}$, hence, the most significant term
in Eq.~\eqref{eq:phi:spe:rhp} is $\ket{\bold{p}}\bra{\bold{p}}$ because its
prefactor (after neglecting the constants)
is $\left(1+\tilde{E}_\bold{p}^2\right)/\left(1+\tilde{E}_\bold{p}\right)^2 \sim 1$
but the prefactors of other terms are at most $\sim 1/\tilde{E}_\bold{p}$.
Therefore, only the signature of a single momentum-$\bold{p}$ particle is significant as $T\ll m/\lambda^2$.

If we compare $\hat{\rho}_0$ in Eq.~\eqref{eq:phi:den:rho0T}
with $\hat{\rho}_\bold{p}$ in Eq.~\eqref{eq:phi:spe:rhp}, we find that
the excitations caused by the random field are independent of
the initial state. For both initial states (vacuum or a single particle),
the condition for observing an additional excitation of momentum $\bold{p}$
is $\tilde{E}_\bold{p} \sim 1$, i.e. $T \sim E_\bold{p} /\lambda^2$.
If a particle has a bigger energy, more time must be costed for it to be excited
by the random field.

In general, we can even choose the initial state to be a superposition
of basis vectors in momentum space. In this case, the problem is changed into
how the off-diagonal elements in the density matrix evolve.
By using the diagrammatic rules, we obtain
\begin{equation}\label{eq:phi:va:rpq}
\begin{split}
& \text{E}\left( \hat{U}_I (T,-T) \ket{\bold{p}} \bra{\bold{q}}\hat{U}_I^\dag (T,-T) \right) \\
&=  \frac{\delta^3\left(\bold{p}-\bold{q}\right)}{1+\tilde{E}_\bold{p}} \hat{\rho}_0(T) +
\frac{\tilde{E}_\bold{p}\tilde{E}_\bold{q}}{\left(1+\tilde{E}_\bold{p}\right)\left(1
+\tilde{E}_\bold{q}\right)} \hat{a}^\dag_\bold{p} \hat{\rho}_0(T) \hat{a}_\bold{q}.
\end{split}
\end{equation}
The off-diagonal element keeps invariant for $T\ll m/\lambda^2$
or $\tilde{E}_\bold{p}\gg 1$. Only at $T\sim m/\lambda^2$, the prefactor of
$\ket{\bold{p}}\bra{\bold{q}}$ expriences a significant reduction and the
multiple excitations become important. As $T\gg m/\lambda^2$, Eq.~\eqref{eq:phi:va:rpq}
is similar to $\hat{\rho}_0(T)$, indicating that the initial condition is
unimportant to the long time behavior of density matrix.

One can also choose an initial state with two or more
particles. The calculation is done by using the same diagrammatic technique.
The physical picture of excitations is similar, hence, we will not discuss them anymore.

\subsection{Renormalization of $\lambda$ \label{sec:phi:re}}

For the density matrices such as $\hat{\rho}_0(T)$ and $\hat{\rho}_\bold{p}(T)$,
we want to know the probability distribution of the total number of particles.
Before doing the calculation, we must solve the problem of ultraviolet divergence.
In the above discussions, we assume a discretized spacetime with the volume
of each element being $\Delta^4x $. Correspondingly, the total volume of
the four-momentum space is finite, being $(2\pi)^4/\Delta^4x$. In other words,
we manually assume an ultraviolet cutoff in the momentum space. But there is
no reason for us to believe that there does exist a cutoff in the physical world.
Therefore, we need to take $\Delta^4x \to 0$, or equivalently, integrate over the infinite momentum space.

Let us first consider the partition function $Z$, which reads
\begin{equation}\label{eq:phi:re:Z}
\begin{split}
Z= & \displaystyle\prod_\bold{p} \frac{1+\tilde{E}_\bold{p}}{\tilde{E}_\bold{p}} \\ = &
\ exp\left\{ \frac{V}{(2\pi)^3} \int d^3\bold{p} \ \ln \left(\frac{1}{\tilde{E}_\bold{p}}+1\right) \right\}.
\end{split}
\end{equation}
It is easy to see that the integral with respect to $\bold{p}$ diverges.
For large enough $\left|\bold{p}\right|$, we have $\tilde{E}_\bold{p} \gg 1$ and then
$\ln \left(1+{1}/{\tilde{E}_\bold{p}}\right) \approx 1/{\tilde{E}_\bold{p}}$. Since $\tilde{E}_\bold{p}$
increases only as $\left|\bold{p}\right|$, the integral $\displaystyle\int d^3\bold{p} \ 1/{\tilde{E}_\bold{p}}$
diverges badly.

As is well known, the ultraviolet divergence repeatedly appears
in conventional QFTs, and the orthodox treatment is to regularize the integral
by e.g., cutting off it at some maximum momentum, and then cancel
the divergence by renormalizing the parameters of the theory.
In a stochastic QFT, the divergence is of a different type.
Since $\ln Z$ diverges algebraically, $Z$ must diverge in an exponential way,
and then $1/Z$ goes to zero exponentially. Because $1/Z$ is a common factor that appears
in each element of the density matrix, all the elements
vanish exponentially in the ultraviolet limit. The essential reason for this
vanishing is that the dimension of Hilbert space
increases exponentially with the momentum space. As a consequence,
the probability of finding the random state vector within a finite-dimensional
subspace vanishes exponentially.

Even the divergence in a stochastic QFT is of a new type, we can still
cancel it by renormalizing the parameter $\lambda$. As the coupling strength
to a random field, $\lambda$ has not been met before in conventional QFTs. But
it is a piece of common sense to renormalize the parameters of a field theory.
There is then no reason for us to not renormalize $\lambda$.

We regularize the integral with respect to $\bold{p}$ by introducing
a momentum cutoff $\Lambda$. In detail, we do the integral in
Eq.~\eqref{eq:phi:re:Z} over $\left|\bold{p}\right|\leq \Lambda$ and find
\begin{equation}\label{eq:phi:re:Za}
\begin{split}
\ln Z = \frac{V}{2\pi^2} \int_0^\Lambda d\kappa  \ \kappa^2 
\ln \left(\frac{\lambda^2 T}{\sqrt{\kappa^2+m^2}}+1\right)  .
\end{split}
\end{equation}
For convenience, we first consider $m\gg \lambda^2 T$ (the renormalization
process is independent of this condition). Now Eq.~\eqref{eq:phi:re:Za}
becomes
\begin{equation}\label{eq:phi:re:Zcut}
\begin{split}
\ln Z \approx & \frac{V}{2\pi^2} \int_0^\Lambda d\kappa  \ \kappa^2 
\frac{\lambda^2 T}{\sqrt{\kappa^2+m^2}} \\
=& \frac{TV}{4\pi^2}m^2\left(\lambda^2\frac{\Lambda^2}{m^2}\right)\bigg[
\sqrt{1+\frac{1}{\left(\Lambda/m\right)^2}} \\ & - 
\frac{\ln\left( \sqrt{1+\left(\Lambda/m\right)^2}+{\Lambda}/{m}
\right)}{\left(\Lambda/m\right)^2}  \bigg].
\end{split}
\end{equation}
To get rid of the divergence, we redefine the coupling constant as $\lambda \to \lambda_p=\lambda \Lambda/m$.
Note that $\Lambda/m$ is dimensionless so that $\lambda_p$
has the same dimension as $\lambda$.
Following convention in QFTs, we call $\lambda$ the bare coupling but
$\lambda_p$ the physical coupling. Finally, we take the limit
$\Lambda/m \to \infty$. In terms of the physical coupling, the partition function
reads
\begin{equation}\label{eq:phi:re:fZ}
Z=exp\left\{ \frac{TV}{4\pi^2}m^2 \lambda^2_p\right\}.
\end{equation}
In fact, Eq.~\eqref{eq:phi:re:fZ} stands for arbitrary $m$ and $T$, but not only for $m\gg \lambda^2 T$.
To show it, we replace $\lambda$ (the bare coupling) in Eq.~\eqref{eq:phi:re:Za} 
by $\lambda_p m/\Lambda$. Taking $\Lambda/m\to \infty$, we obtain Eq.~\eqref{eq:phi:re:fZ} again.

Briefly speaking, the renormalization is to regulate the integral
with respect to $\bold{p}$ by setting a momentum cutoff $\Lambda$, replace the bare coupling $\lambda$
by the physical coupling $\lambda_p$, and then take the limit
$\Lambda/m\to\infty$. After this process, the probabilities become finite.

The expression of $Z$ in Eq.~\eqref{eq:phi:re:fZ}
is clearly Lorentz-invariant, because $TV$ is half of the total spacetime volume which keeps
invariant under Lorentz transformations. And $1/Z=e^{-{TV}m^2 \lambda^2_p/({4\pi^2})}$
is then a scalar factor existing in each density matrix element.

Let us study the probability distribution of the particle number.
Starting from the vacuum state, the density matrix becomes $\hat{\rho}_0(T)$ after an evolution
of period $2T$. From the density matrix, we can easily derive
the probability of the total number of particles being $n$. The probability is denoted by $P_0(n)$.
According to Eq.~\eqref{eq:phi:den:rho0T}, we have
\begin{equation}\label{eq:phi:re:P0}
\begin{split}
P_0(n) = & \frac{1}{Z} \frac{1}{n!} \int \left(\prod_{j=1}^n d^3\bold{p}_j\right)
\left(\prod_{l=1}^n \frac{1}{1+\tilde{E}_{\bold{p}_l}}\right) \\
&\times \text{Tr}\left(\ket{\bold{p}_1\cdots\bold{p}_n} \bra{\bold{p}_1\cdots\bold{p}_n}\right)
\\  = & \frac{1}{Z} \frac{1}{n!} \left(\frac{V}{2\pi^2}\int_0^\Lambda d\kappa \frac{\kappa^2}
{1+\sqrt{\kappa^2+m^2}/(\lambda^2T)}\right)^n,
\end{split}
\end{equation}
where we have used $\braket{\bold{p}_1\cdots\bold{p}_n|\bold{p}_1\cdots\bold{p}_n}=
\left[\delta^{3}(0)\right]^n = \left(V/\left(2\pi\right)^3\right)^n$ for $\bold{p}_1\neq\bold{p}_2\neq
\cdots\neq \bold{p}_n$, and we have regulated the integral
by setting an ultraviolet cutoff. The $\lambda$ in Eq.~\eqref{eq:phi:re:P0}
is the bare coupling. Replacing it by $\lambda_pm/\Lambda$ and taking
the limit $\Lambda/m\to\infty$, we obtain
\begin{equation}\label{eq:phi:re:P0f}
P_0(n) =
\frac{1}{n!}\left(\frac{TV}{4\pi^2}m^2 \lambda^2_p\right)^n 
exp\left\{ - \frac{TV}{4\pi^2}m^2 \lambda^2_p\right\}.
\end{equation}
Eq.~\eqref{eq:phi:re:P0f} is recognized as the probability function of Poisson distribution
with the parameter ${TV}m^2 \lambda^2_p/({4\pi^2})$.

According to the properties of Poisson distribution, as $T$ increases, the function $P_0(n)$
becomes flattened, and its peak moves towards the limit of infinite particle number.
Furthermore, the average of the particle number is ${TV}m^2 \lambda^2_p/({4\pi^2})$,
which increases linearly with time and is also proportional to the squared coupling strength
or the squared mass.

Similarly, we study the distribution of particle number for $\hat{\rho}_{\bold{p}}(T)$
when there is initially a single particle of momentum $\bold{p}$. The expression
of $\hat{\rho}_{\bold{p}}(T)$ is given in Eq.~\eqref{eq:phi:spe:rhp}. Noticing
$\text{Tr}\left[\hat{\rho}_{\bold{p}}(T)\right]=1/\Delta^3\bold{p}$,
we then use $\hat{\rho}_{\bold{p}}(T)\Delta^3\bold{p}$ to calculate the probability.
$\bold{p}$ is special in the momentum space,
because the probability of finding a particle of momentum $\bold{p}$ is $100\%$ at
the initial time. Therefore, we use $P_\bold{p}(j,n)$ to denote the probability of finding
$j$ particles of momentum $\bold{p}$ and $\left(n-j \right)$ particles of different momentum from $\bold{p}$.
If $\bold{p}_1$, $\cdots$, $\bold{p}_{n-j}$ are all different from
each other and also different from $\bold{p}$,
the vector $\ket{\bold{p}_1\cdots\bold{p}_{n-j}\bold{p}\cdots\bold{p} }$ with
$\left(n-j\right)$ $\bold{p}$s has the squared norm $\left[\delta^3(0) \right]^{n} j!$.
Furthermore, there are $\displaystyle{n!}/
\left[{j!\left(n-j\right)!}\right]$ different ways of choosing $j$ particles from
the total $n$ particles and letting them have the momentum $\bold{p}$. With these
considerations, we find the probability to be
\begin{equation}
\begin{split}
P_\bold{p}(j,n)= & \frac{1}{Z} \frac{1}{\left(n-j\right)!}\left(
\frac{V}{(2\pi)^3}\int d^3\bold{p}\frac{1}{1+\tilde{E}_\bold{p}}\right)^{n-j} \\ &
\times \frac{1+j\tilde{E}^2_\bold{p}}{\left(1+\tilde{E}_\bold{p}\right)^{j+1}}.
\end{split}
\end{equation}
Replacing $\lambda$ by $\lambda_p m/\Lambda$ and taking $\Lambda/m\to\infty$,
we obtain
\begin{equation}\label{eq:phi:re:Pjn}
\begin{split}
P_\bold{p}(j,n) = \left\{ \begin{array}{ccc}
P_0(n-j) &  & j=1,n\geq j \\
& & \\
0 &  & \text{otherwise.}
\end{array}\right.  
\end{split}
\end{equation}

Eq.~\eqref{eq:phi:re:Pjn} tells us that the initial particle is always
there with a fixed momentum. Besides it, more particles are excited
by the random field with time passing by. And the number of additionally excited particles
follows the same Poisson distribution as Eq.~\eqref{eq:phi:re:P0f}. The excitation
caused by the random field is independent of whether there is initially
a particle or not.

We do the calculation for the initial state being a superposition
of different $\bold{p}$ by using the result~\eqref{eq:phi:va:rpq}. We also
study the case in which there are more particles in the initial state.
The results are similar. We draw a conclusion that the excitation
caused by the random field is independent of the initial state.

Now the consequence of coupling to a white-noise field is clear.
The random field thermalizes the universe, increasing its temperature continuously
towards infinity by exciting particles. The number of excited particles
follows the Poisson distribution with an expectation value ${TV}m^2 \lambda^2_p/({4\pi^2})$.

\subsection{Revisit of the differential equations for state vector and
density matrix\label{sec:phi:revisit}}

In the renormalization, we replace the bare coupling by the physical one. It is then natural
to consider the influence of this replacement on the differential equations
for state vector (Eq.~\eqref{eq:phi:ca:sch}) and density matrix (Eq.~\eqref{eq:phi:ca:rho}).
In other words, we will carry the renormalization of $\lambda$ back to
the evolution equations, from which we derived the path-integral formalism.

The renormalization process requires a momentum cutoff. It is
therefore necessary to reexpress the evolution equations in the momentum
space. This can be done by expressing $\hat\phi(\bold{x})$ in terms of
$\hat{a}_\bold{p}$ and $\hat{a}_\bold{p}^\dag$.
With the cutoff $\Lambda$, Eq.~\eqref{eq:phi:ca:sch} becomes
\begin{equation}\label{eq:phi:ca:schrenorm}
\begin{split}
d\ket{\psi(t)} =  & -i dt \int\displaylimits_{\mathcal{B}(\Lambda)} d^3\bold{p} \ E_\bold{p}
\hat{a}^\dag_\bold{p} \hat{a}_\bold{p} \ket{\psi(t)}
+ i \frac{\lambda_p m}{\Lambda} \frac{1}{\sqrt{\left(2\pi\right)^3}}  \\ &  \times
 \int\displaylimits_{\mathcal{B}(\Lambda)} \frac{d^3\bold{p}}{\sqrt{2E_\bold{p}}}
\left[\hat{a}_\bold{p} d \omega(t,\bold{p}) +\hat{a}^\dag_\bold{p} d\omega^*(t,\bold{p})\right] \ket{\psi(t)} \\
& + \text{the second-order term},
\end{split}
\end{equation}
where $\mathcal{B}(\Lambda)$ denotes a ball of radius $\Lambda$ centered at the origin
in the momentum space. And $d \omega(t,\bold{p}) = 
\displaystyle\int_{R^3} dW(t,\bold{x}) e^{i\bold{p}\cdot \bold{x}}$
is a random number. Notice that $d\omega$ is defined as an integral
over the three-dimensional space, therefore, its variance is proportional to $dt$.
This explains why we use the symbol $d\omega$ instead of $\omega$.
We determine the distribution of $d\omega$ by using a similar analysis
as we did below Eq.~\eqref{eq:phi:dwpana}. For convenience, we introduce
\begin{equation}
\begin{split}
& dW^{(1)}_t(\bold{p}) = \frac{\text{Re}\left\{ d\omega(t,\bold{p})\right\}}{\sqrt{V/2}},
\\ & dW^{(2)}_t(\bold{p}) = \frac{\text{Im}\left\{ d\omega(t,\bold{p})\right\}}{\sqrt{V/2}}.
\end{split}
\end{equation}
It is easy to prove that $dW^{(1)}_t(\bold{p})$ and $dW^{(2)}_t(\bold{p})$ are two
independent random numbers with Gaussian distribution of variance $dt$.
Therefore, we can definitely see $dW^{(1)}_t(\bold{p})$ and $dW^{(2)}_t(\bold{p})$ as
the differentials of two independent Wiener processes, respectively. Furthermore, $\left\{ W^{(1,2)}_t(\bold{p}_1),
W^{(1,2)}_t(\bold{p}_2), \cdots | \bold{p}_i \neq - \bold{p}_j\right\}$ are a set
of independent Wiener processes. By using these Wiener processes and the
Majorana bosonic operators $\hat{\gamma}_\bold{p}^{(1)}=\left(\hat{a}_\bold{p}+
\hat{a}^\dag_\bold{p}\right)/\sqrt{2}$, $\hat{\gamma}_\bold{p}^{(2)}=\left(i\hat{a}_\bold{p}-i
\hat{a}^\dag_\bold{p}\right)/\sqrt{2}$, we rewrite Eq.~\eqref{eq:phi:ca:schrenorm} as
\begin{equation}\label{eq:phi:ca:schrenormF}
\begin{split}
d\ket{\psi(t)} =  & -i dt \int\displaylimits_{\mathcal{B}(\Lambda)} d^3\bold{p} \ E_\bold{p}
\hat{a}^\dag_\bold{p} \hat{a}_\bold{p} \ket{\psi(t)}
+ i \frac{\lambda_p m}{\Lambda} \sqrt{\frac{V}{\left(2\pi\right)^3}}  \\ &  
 \int\displaylimits_{\mathcal{B}(\Lambda)} \frac{d^3\bold{p}}{\sqrt{2E_\bold{p}}}
\left[\hat{\gamma}^{(1)}_\bold{p} d W^{(1)}_t(\bold{p}) +
\hat{\gamma}^{(2)}_\bold{p} d W^{(2)}_t(\bold{p}) \right] \ket{\psi(t)} \\
& + \text{the second-order term}.
\end{split}
\end{equation}

Eq.~\eqref{eq:phi:ca:schrenormF} is not Lorentz-invariant for a finite $\Lambda$.
But the Lorentz invariance is recovered after we take the limit $\Lambda\to\infty$.
Next we set the Hamiltonian term equal zero, and then study how the random term
works. Without the Hamiltonian term, Eq.~\eqref{eq:phi:ca:schrenormF} is strictly solvable.
The solution can be easily expressed in terms of a time-ordering operator. Even better,
if we use the Baker-Campbell-Hausdorff formula and neglect an unimportant
overall phase of the state vector, the solution becomes
\begin{equation}
\begin{split}
\ket{\psi(t)}= & \prod_{\left|\bold{p}\right|\leq \Lambda} exp\bigg\{ 
i \frac{\lambda_p m}{\Lambda} \sqrt{\frac{\left(2\pi\right)^3}{ 2 V E_\bold{p}}} 
\\ & \left[\hat{\gamma}^{(1)}_\bold{p} W^{(1)}_t(\bold{p}) +
\hat{\gamma}^{(2)}_\bold{p} W^{(2)}_t(\bold{p}) \right] \bigg\} \ket{\psi(0)}.
\end{split}
\end{equation}
The state vector at time $t$ depends on a set of Wiener processes, indicating
the stochasticity of evolution equation. The evolutions at different $\bold{p}$-modes
are independent to each other, coinciding with our previous analysis.
As $\Lambda\to\infty$, the contribution of each $\bold{p}$-mode becomes infinitesimal due to
the factor $1/\Lambda$ in the exponent. But the sum of excitations over the whole
momentum space is nonzero. This is seen by the analysis of density matrix below.

One can derive the density matrix from $\ket{\psi(t)}$. A more straightforward
way of calculating the particle-number generation rate is to carry out the renormalization
for the differential equation of density matrix. Setting a cutoff in Eq.~\eqref{eq:phi:ca:rho},
or starting from Eq.~\eqref{eq:phi:ca:schrenormF}, we obtain
\begin{equation}\label{eq:phi:ca:rhorenorm}
\begin{split}
\frac{d\hat \rho(t)}{dt} = & -i \big[\int\displaylimits_{\mathcal{B}(\Lambda)}
d^3\bold{p} \ E_\bold{p} \hat{a}^\dag_\bold{p} \hat{a}_\bold{p}, \hat \rho(t) \big] +
\frac{\lambda_p^2 m^2}{\Lambda^2} \int\displaylimits_{\mathcal{B}(\Lambda)}
\frac{d^3\bold{p}}{2E_\bold{p}} \\ & \bigg\{ \left[ 
\hat{a}_\bold{p}^\dag \hat\rho(t) \hat{a}_\bold{p}- \frac{1}{2} \hat{a}_\bold{p}
\hat{a}^\dag_{\bold{p}} \hat\rho(t) -\frac{1}{2}\hat\rho(t) \hat{a}_\bold{p}
\hat{a}^\dag_{\bold{p}}   \right] \\ & + \text{the other three terms}\bigg\}.
\end{split}
\end{equation}
Therefore, the generation rate at each momentum $\bold{p}$ is
\begin{equation}
\begin{split}
\frac{dn_\bold{p}}{dt} = \left\{ 
\begin{array}{ccc} \frac{\lambda^2_p m^2 V}{2\cdot (2\pi)^3 \Lambda^2 E_\bold{p}} & &
\left|\bold{p}\right| \leq \Lambda \\ & & \\
0 & & \left|\bold{p}\right| > \Lambda \end{array} \right.
\end{split}
\end{equation}
with $n_\bold{p}(t)=\text{Tr}\left[\hat\rho(t)\hat{a}_\bold{p}^\dag \hat{a}_\bold{p}\right]$.
As easily seen, the generation rate at arbitrary momentum vanishes in the limit $\Lambda\to\infty$.
But if we first sum up the generation rates over $\left|\bold{p}\right| \leq \Lambda$, and then
take the ultraviolet limit, we directly find
\begin{equation}
\frac{dN}{dt} = \frac{\lambda_p^2 m^2 V}{8\pi^2},
\end{equation}
where $N$ is the total number of particles and $V$ is the volume of space.
This generation rate is exactly consistent with our previous result that, the
number of particles generated from $-T$ to $T$ is ${TV}m^2 \lambda^2_p/({4\pi^2})$.
Therefore, the renormalization process can be consistently
carried back to the stochastic evolution equations.

\section{$\phi^4$ theory \label{sec:phi4} of random unitary evolution}

We successfully developed a stochastic QFT of neutral bosons
of spin zero. Our approach can be easily generalized to more complicated
QFTs coupled to an external random scalar field.
As an example, we explore a theory with particle-particle interaction.
The simplest Lorentz-invariant action with an interaction term reads
\begin{equation}\label{eq:4:action}
\begin{split}
I_4 = & \int d^4 x \left(-\frac{1}{2}\partial_\mu \phi \partial^\mu \phi - \frac{1}{2}
m^2 \phi^2 -\frac{g}{4!}\phi^4 \right) \\ & + \lambda \int d W(x) \ \phi(x),
\end{split}
\end{equation}
where $g$ and $\lambda$ are the strength of interaction and
coupling, respectively. As $\lambda=0$, Eq.~\eqref{eq:4:action} is the action of $\phi^4$-theory
which is a textbook example for studying the interacting QFTs.
For a generic $\lambda$, the model~\eqref{eq:4:action} describes bosons which
have a $\delta$-function type of interaction between each other and at the same time, are driven
by an external random field. Next, we quantize this field theory by using the aforementioned
techniques.

\subsection{Quantization and diagrammatic rules for $S$-matrix \label{sec:4:S}}

Since $\phi^4(x)$ can be seen as a potential term, there is
nothing new in the canonical quantization. We follow the procedure introduced
in Sec.~\ref{sec:phi:quan}. In the Schr\"{o}dinger picture, the
equations of motions for the state vector and density matrix are
as same as Eq.~\eqref{eq:phi:ca:sch} and~\eqref{eq:phi:ca:rho}, respectively,
except that $\hat H_0$ in Eq.~\eqref{eq:phi:ca:sch} or~\eqref{eq:phi:ca:rho}
is replaced by the interacting Hamiltonian:
$\hat H=\hat H_0+\displaystyle\frac{g}{4!} \displaystyle\int d^3\bold{x} \ \hat{\phi}^4(\bold{x})$.

The $S$-matrix is calculated by using the path integral approach.
We repeat the procedure in Sec.~\ref{sec:phi:scatter}
and~\ref{sec:phi:dia}. The result is expressed as
\begin{equation}\label{eq:4:S:Sba}
\begin{split}
S_{\beta,\alpha}^g=  \frac{ \displaystyle\int D\phi 
\bra{\beta}e^{iT\hat H_0} \ket{\phi_T} \bra{\phi_{-T}}e^{iT\hat H_0} \ket{\alpha} e^{iI_4} }
 {\displaystyle\int D\phi
\bra{0}e^{iT\hat H_0} \ket{\phi_T}  \bra{\phi_{-T}}e^{iT\hat H_0} \ket{0} e^{iI_0} },
\end{split}
\end{equation}
where we use $S^g$ to denote the $S$-matrix in the presence of interaction.
$S^g_{\beta,\alpha}$ is distinguished from $S_{\beta,\alpha}$
which denotes the $S$-matrix as $g=0$. Eq.~\eqref{eq:4:S:Sba}
is almost as same as Eq.~\eqref{eq:phi:path:Sba}
except that $I_W$ in the latter is replaced by $I_4$ in the former. The path-integral formula stands
for the stochastic $\phi^4$-theory.

Again, we choose $\ket{\alpha}$ and $\ket{\beta}$ to be
the basis vectors in the momentum space. The inner products
$\bra{\beta}e^{iT\hat H_0} \ket{\phi_T}$ and $\bra{\phi_{-T}}e^{iT\hat H_0} \ket{\alpha} $
are then given by Eq.~\eqref{eq:phi:path:phit}.
They contribute an additional $i\eta$-term to the action.
And $e^{i I_4}$ can be expanded into a series, reading
\begin{equation}\label{eq:4:S:eI4}
e^{i I_4} = e^{iI_W} \sum_{k=0}^\infty \frac{1}{k!} \left(\frac{-ig}{4!} \int d^4 x \ \phi^4(x) \right)^k.
\end{equation}
Substituting Eq.~\eqref{eq:4:S:eI4} into Eq.~\eqref{eq:4:S:Sba}, we find
$S_{\beta,\alpha}^g$ to be the path integral of
$e^{iI_W^{(\eta)}}$ multiplied by a polynomial of $\phi$, just like $S_{\beta,\alpha}$.
The path integral with respect to $\phi$ can be transformed into the functional derivative
of $e^{iI_W^{(J)}}$. We then express the $S$-matrix as
\begin{widetext}
\begin{equation}\label{eq:4:S:sexp}
\begin{split}
& S^g_{\bold{p}'_1\cdots\bold{p}'_m,
\bold{p}_1\cdots\bold{p}_n}=
(-i)^{n+m} \int \left(\prod_{j=1}^n d^3 \bold{x}_j\right) \left(\prod_{l=1}^m d^3 \bold{x}'_l\right) 
\prod_{j=1}^n \left(\sqrt{\frac{2E_{\bold{p}_j}}{(2\pi)^3}} 
e^{i \left( T E_{\bold{p}_j}+ \bold{p}_j \cdot \bold{x}_j \right)} \right)
\prod_{l=1}^m \left(\sqrt{\frac{2E_{\bold{p}'_l}}{(2\pi)^3}}
e^{i\left(T E_{\bold{p}'_l}-\bold{p}'_l \cdot \bold{x}'_l\right)} \right) 
 \\ & \times \sum_{k=0}^\infty \frac{\left(-i\right)^{4k} \left(-ig\right)^k}
 { 4!^k \ k! }\int \left(\prod_{s=1}^k 
 d^4 y_s\right) 
 \left. \frac{\displaystyle\delta^{4k+n+m} }{\delta J^4(y_1) \cdots \delta J^4(y_k) \delta 
J(-T,\bold{x}_1) \cdots \delta J(-T,\bold{x}_n)\delta J(T,\bold{x}'_1)\cdots
\delta J(T,\bold{x}'_m)}  e^{i\mathcal{F}\left[J\right]} \right|_{J=0} \\ &
\ \ \ \ \ \ \ \ \ \ \ \ \ \ \ \ \ \ \ \ \  \ -\cdots .
\end{split}
\end{equation}
\end{widetext}
Here we only display part of $S^g$. The functional derivative of $e^{i\mathcal{F}\left[J\right]}$
is calculated by grouping $J$s in the denominator, with each group
consisting of either a pair of $J$s or a single $J$. The omitted terms in Eq.~\eqref{eq:4:S:sexp}
forbid the pairing of $J(-T,\bold{x}_i)$ with $J(-T,\bold{x}_j)$ 
or the pairing of $J(-T,\bold{x}'_i)$ with $J(-T,\bold{x}'_j)$ for arbitrary $i$ and $j$.

\begin{figure}[tbp]
\vspace{0.2cm}
\includegraphics[width=0.9\linewidth]{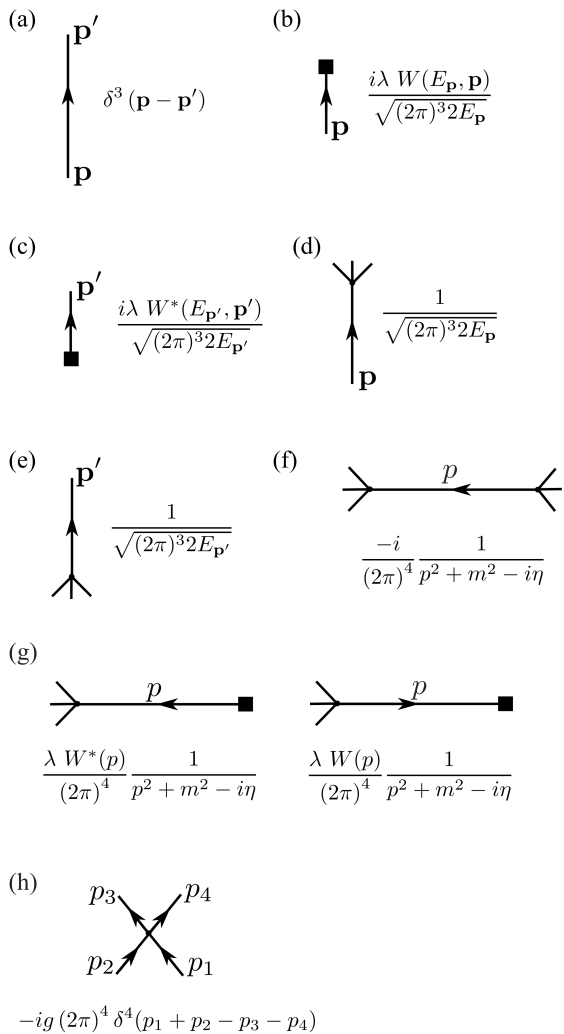}
\caption{Diagrammatic rules for the calculation of $S$-matrix in
the stochastic $\phi^4$-theory.}\label{fig:Sp}
\end{figure}
We still use diagrams to keep track of all the ways of grouping $J$s.
The diagrammatic rules for calculating $S$-matrix are the combination of
rules for free stochastic QFT and rules for $\phi^4$-theory,
but with some new additions (see Fig.~\ref{fig:Sp}).
In each diagram, there are vertices representing the interaction
between particles, square dots representing the scattering of a particle
on random field, and solid lines representing the propagator of a particle.
More specifically:

\noindent (a) For each isolated line carrying an arrow pointed upwards,
include a Dirac $\delta$-function.

\noindent (b) For each line of momentum $\bold{p}$ running into a
square dot, include a factor $i\lambda W(E_\bold{p},\bold{p})/\sqrt{(2\pi)^32E_\bold{p}}$.

\noindent (c) For each line of momentum $\bold{p}'$ running out of a
square dot, include a factor $i\lambda W^*(E_{\bold{p}'},\bold{p}')/\sqrt{(2\pi)^32E_{\bold{p}'}}$.

\noindent (d) For each line of momentum $\bold{p}$ running into a
vertex, include a factor $1/\sqrt{(2\pi)^32E_\bold{p}}$.

\noindent (e) For each line of momentum $\bold{p}'$ running out of a
vertex, include a factor $1/\sqrt{(2\pi)^32E_{\bold{p}'}}$.

\noindent (f) For each internal line carrying a four-momentum $p$ running
from one vertex to another vertex, include a factor $-i/\left[\left(2\pi\right)^4\left(
p^2+m^2-i\eta\right)\right]$.

\noindent (g) For each internal line carrying a four-momentum $p$ running from
a square dot to a vertex, include a factor $\lambda W^*\left(p\right)/\left[\left(2\pi\right)^4\left(
p^2+m^2-i\eta\right)\right]$. For such a line running from a vertex to a square dot,
include a factor $\lambda W\left(p\right)/\left[\left(2\pi\right)^4\left(
p^2+m^2-i\eta\right)\right]$. Notice that the direction of arrow matters.

\noindent (h) For each vertex, include a factor $-ig(2\pi)^4 \delta^4\left(p_1+p_2-p_3-p_4\right)$
where $p_1$ and $p_2$ are the four-momenta entering the vertex, and
$p_3$ and $p_4$ are the four-momenta leaving the vertex. This $\delta$-function
ensures that the four-momenta is conserved at each vertex.

\noindent (i) For each diagram, include a factor $S_{0,0}$. This factor
comes from the coupling to random field, which appears in both the noninteracting
and interacting stochastic QFTs.

\noindent (k) For each diagram with $k$ vertices, include a combinatoric factor
$\mathcal{C}/\left(4!^k k!\right)$, where $\mathcal{C}$ denotes the
number of different ways of grouping $J$s that result in the same diagram.

\begin{figure}[tbp]
\vspace{0.7cm}
\includegraphics[width=0.9\linewidth]{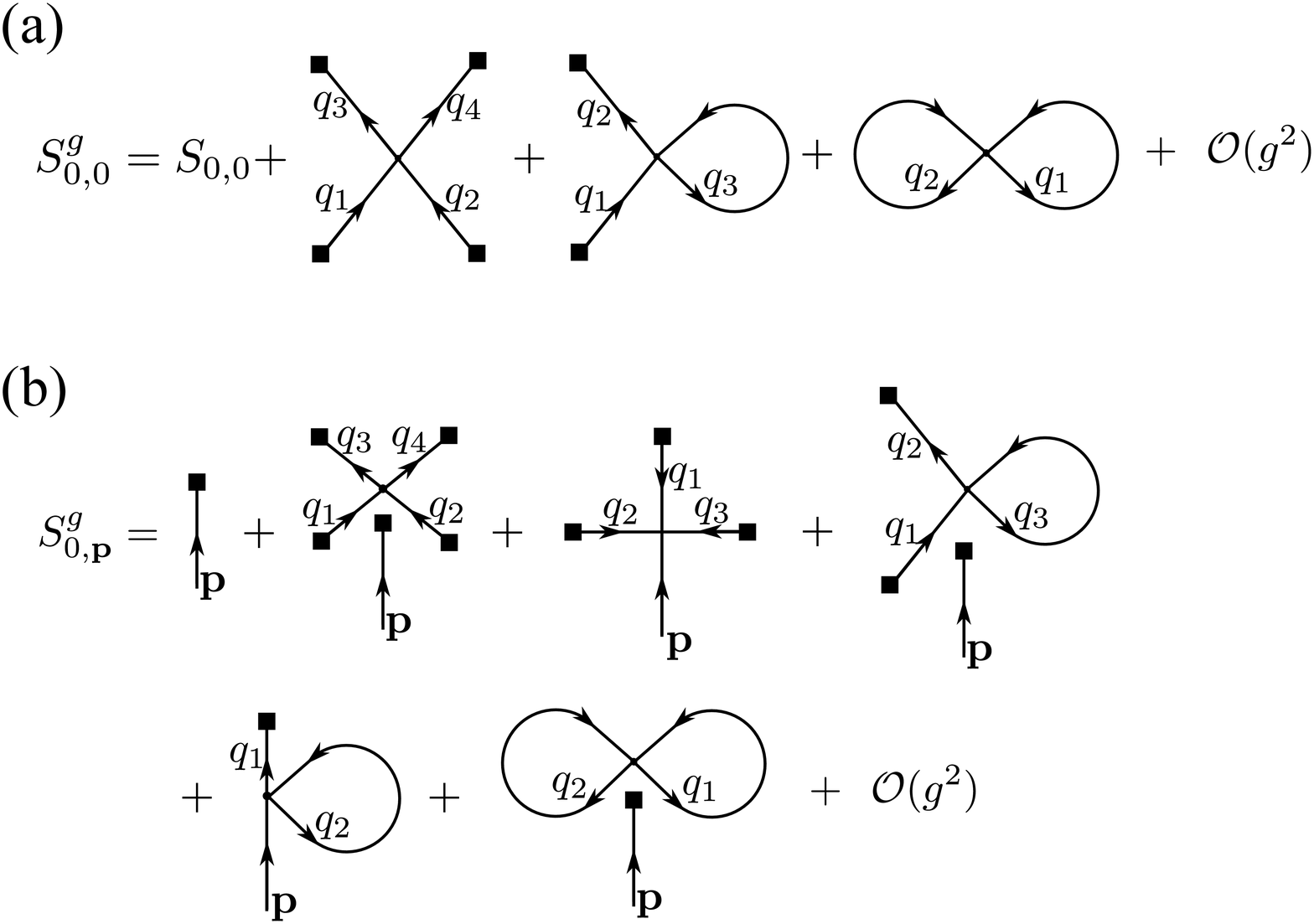}
\caption{Graphical expressions of (a) $S^g_{0,0}$ and (b) $S^g_{0,\bold{p}}$.
This figure explains how to calculate the $S$-matrix of stochastic $\phi^4$-theory
by using the diagrammatic technique.}\label{fig:Sg0gp}
\end{figure}
Finally, we integrate the product of these factors over
all the four-momenta carried by internal lines, and then obtain the $S$-matrix.
As examples, we display the diagrams for calculating $S^g_{0,0}$ and
$S^g_{0,\bold{p}}$ in Fig.~\ref{fig:Sg0gp}(a) and~(b), respectively.
We consider both the connected and disconnected diagrams.
Note that there are infinite diagrams in the presence of interaction.
We only show diagrams to the order $g$.

By using the diagrammatic rules, we obtain
\begin{widetext}
\begin{equation}\label{eq:4:S:sg00e}
\begin{split}
S^g_{0,0} = & S_{0,0} \bigg\{ 1 +\frac{-ig\left(2\pi\right)^4}{4!} \int \left( \prod_{l=1}^4 d^4q_l \right)
\delta^4\left(q_1+q_2-q_3-q_4\right) \frac{\lambda^4 W^*(q_1)W^*(q_2)
W(q_3)W(q_4)}{\left(2\pi\right)^{16}} 
\left( \prod_{j=1}^4 \frac{1}{q_j^2+m^2-i\eta}\right) \\ & + 
\frac{-ig\left(2\pi\right)^4}{4} \int \left( \prod_{l=1}^3 d^4q_l \right)
\delta^4\left(q_1-q_2\right) \frac{\lambda^2 W^*(q_1)W(q_2)}{\left(2\pi\right)^{8}} 
\frac{-i}{(2\pi)^4}
\left( \prod_{j=1}^3 \frac{1}{q_j^2+m^2-i\eta}\right) \\ & +
\frac{-ig\left(2\pi\right)^4}{8} \int d^4q_1 d^4q_2 \ 
\delta^4\left(0\right)  \frac{\left(-i\right)^2}{(2\pi)^8}
\left( \prod_{j=1}^2 \frac{1}{q_j^2+m^2-i\eta}\right) 
\bigg\} \\ & + \mathcal{O}(g^2),
\end{split}
\end{equation}
\end{widetext}
where $\delta^4(0)=1/\Delta^4 p = 2TV/(2\pi)^4$. Similarly, we can obtain
the expression of $S^g_{0,\bold{p}}$.
In Eq.~\eqref{eq:4:S:sg00e}, $S^g_{0,0}$ is an integral over the four-momentum space.
Some of the integration variables are eliminated by the $\delta$-functions
associated to vertices. But in a diagram consisting of square dots, there
are usually integration variables left even after we take all the $\delta$-functions into account.
This is different from the noninteracting theory in which the expression of $S$-matrix
contains no integral, and is also different from
the conventional $\phi^4$-theory in which there are no square dots in a Feynman diagram.
Because $W(q)$ is a random number, we cannot integrate out $q$.
We have to be satisfied by expressing $S^g$ as an integral of $W(q)$,
since $S^g$ is a random number.

By looking carefully at the diagrammatic rules, we find $S^g$ to be Lorentz invariant, that is $S^g$
satisfies Eq.~\eqref{eq:phi:dia:Lo}. The Lorentz invariance can be proved
by checking the factors of $S$-matrix one by another. The factors contributed by
internal lines or vertices do not change under the Lorentz transformations,
because they are $\delta^4(p)$, $1/\left(p^2+m^2-i\eta\right)$ or $W(p)$, being
all scalars. The factors contributed by external lines are $\delta^3\left(\bold{p}-\bold{p}'\right)$,
$W\left(E_\bold{p},\bold{p}\right)/\sqrt{E_\bold{p}}$ or $1/ \sqrt{E_\bold{p}}$,
which do transform in the way shown by Eq.~\eqref{eq:phi:dia:Lo}.
The path integral approach to stochastic QFTs naturally results in a Lorentz-invariant scattering matrix.

\subsection{Diagrammatic rules for density matrix \label{sec:4:den}}

Using the $S$-matrix, we easily obtain the state vector
after a random unitary evolution of period $2T$ for arbitrary
initial state. The final state is a random vector in the Hilbert space.
To study its properties, we calculate the corresponding density matrix.
We follow the procedure introduced in Sec.~\ref{sec:phi:den}.
To be distinguished from $\rho$ of noninteracting theory, $\rho^{g}$
is employed to denote the density-matrix element of an interacting theory.

$\rho^{g\left(\bold{p}_1\cdots\bold{p}_a\right)}_{\bold{p}'_1\cdots\bold{p}'_m,\bold{q}'_1
\cdots\bold{q}'_n}$ is the product of $S^g_{\bold{p}'_1\cdots\bold{p}'_m,
\bold{p}_1\cdots\bold{p}_a}$ and $S^{g*}_{\bold{q}'_1\cdots\bold{q}'_n,
\bold{p}_1\cdots\bold{p}_a}$ averaged over the
configurations of random field. The factors of $S$-matrix are displayed
in Fig.~\ref{fig:Sp}, which include the random numbers $W(p)$
and $W^*(p)$ with the latter being equal to $W(-p)$.
To calculate $\rho^{g}$, we need to know the expectation of
$\left| S_{0,0}\right|^2$ multiplied by a sequence of $W(p)$.
Here, the four-momentum $p$ can be either on-shell (see Fig.~\ref{fig:Sp}(b,c))
or off-shell (see Fig.~\ref{fig:Sp}(g)). We cannot directly utilize
Eq.~\eqref{eq:phi:den:SV} which gives the expectation value as there are only
on-shell four momenta. Instead, we redo the calculation and consider
generic four-momenta. The generator is redefined as
$\text{E} \left(\left| S_{0,0}\right|^2 exp\left\{ \int d^4{p} \ \chi_p
W(p)  \right\}\right)$ where $\chi_p$ is an arbitrary real function of $p$.
In Sec.~\ref{sec:phi:sin}, we already proved that $\left\{ \left. \text{Re}W(p), \text{Im}W(p) \right| 
p^0>0 \right\}$ is a set of independent Gaussian random numbers of
zero mean and variance $\sigma_p^2=(2\pi)^4/\left(2\Delta^4 p\right)$.
And the expression of $\left| S_{0,0}\right|^2$ can be found in Eq.~\eqref{eq:phi:sin:S2}.
We then obtain
\begin{equation}\label{eq:4:S:Ege}
\begin{split}
& \text{E} \left(\left| S_{0,0}\right|^2 exp\left\{ \int d^4{p} \ \chi_p
W(p)  \right\}\right) \\
= & \int \prod_{{p}^0 > 0} \left(
\frac{d \text{Re}W(p) \ d\text{Im}W(p) }{2\pi\sigma^2_p}\right)
exp\left\{ -\sum_{p^0>0} \frac{ \left| W(p) \right|^2}{2\sigma^2_{p}}\right\} \\
& \times exp \bigg\{ -\frac{\lambda^2}{\left(2\pi\right)^3}
\sum_{p^0>0} \Delta^4 {p} \left| W(p) \right|^2 \delta(p^2+m^2) \\ &
+ \sum_{p^0>0} \Delta^4 {p} \left[ \text{Re}W(p) \left( \chi_p +\chi_{-p} \right)
+ i \text{Im}W(p) \left(\chi_p -\chi_{-p} \right)\right]\bigg\} \\ 
= & \frac{1}{Z} \ exp \left\{ \sum_{p^0>0} \Delta^4{p} \ \frac{\left(2\pi\right)^4 \chi_p
\chi_{-p} }{1+2\pi \lambda^2 \delta\left(p^2+m^2\right)}\right\}.
\end{split}
\end{equation}
A useful relation between Dirac and Kronecker $\delta$-functions
is $\delta\left(p^2+m^2\right)=\delta_{\left|p_0\right|,
E_\bold{p}}/ \left(2\left|p^0\right| \Delta p^0\right)$.
According to it, we have $2\pi \lambda^2 \delta\left(p^2+m^2\right)= \delta_{p_0,
E_\bold{p}}/\tilde{E}_\bold{p}$ for $p^0>0$.

By calculating the functional derivative of Eq.~\eqref{eq:4:S:Ege} with respect to $\chi_p$
at $\chi_p=0$, we find
\begin{equation}\label{eq:4:S:Epa}
\begin{split}
& \text{E} \left[ \left|S_{0,0}\right|^2 W(p_1) W(p_2)\cdots W(p_{2n})\right] \\
= & \frac{1}{Z} \bigg\{ e(p_1,p_2) \ e(p_3,p_4) \ \cdots \ e(p_{2n-1},p_{2n}) \\ & +
e(p_1,p_3) \ e(p_2,p_4) \ \cdots \ e(p_{2n-1},p_{2n}) \\ & 
+ \text{the} \ \text{other} \ \text{ways} \ \text{of} \ \text{pairing} \ p\text{s}\bigg\},
\end{split}
\end{equation}
where
\begin{equation}
e(p,q)= \left(2\pi\right)^4 \delta^4\left(p+q\right) \left(1- \frac{
\delta_{\left|p^0\right|,E_\bold{p} }}{ 1+\tilde{E}_{\bold{p}} }\right).
\end{equation}
It is easy to see that Eq.~\eqref{eq:4:S:Epa} reduces to Eq.~\eqref{eq:phi:den:SV}
if we choose the four-momenta to be on-shell.
In Eq.~\eqref{eq:4:S:Epa}, we need to consider all the different ways of
pairing $W(p)$s in the set $\left\{ W(p_1), W(p_2),\cdots, W(p_{2n}) \right\}$.
This is reminiscent of the Wick's theorem. Again, we use diagrams to keep track of the ways
of pairing. $W(p)$ is represented by a square dot in the
Feynman diagram. If we pair $W(p_i)$ with $W(p_j)$, we then
merge the corresponding two square dots into a single one.
The resulting diagram is a paired-diagram.

\begin{figure}[tbp]
\vspace{0.5cm}
\includegraphics[width=0.9\linewidth]{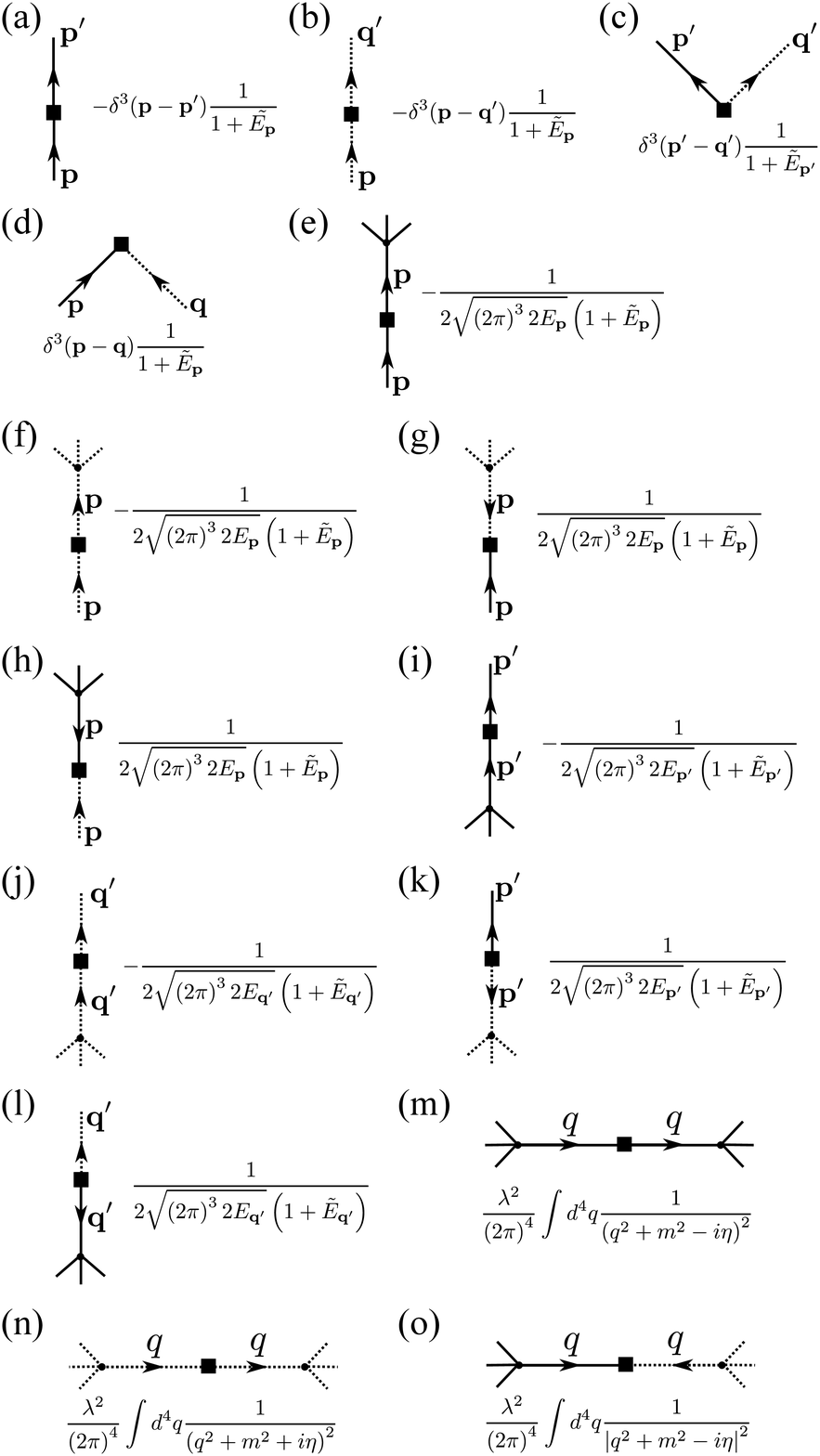}
\caption{Diagrammatic rules for the calculation of density matrix in the
stochastic $\phi^4$-theory.}\label{fig:4den}
\end{figure}
A paired-diagram is the combination of two Feynman diagrams
with one in solid line representing $S^g$ and the other in dotted line representing
$S^{g*}$. Fig.~\ref{fig:4den} displays the components of a paired diagram
that contain square dots. The rules for calculating $\rho^g$
are summarized as follows. Some of them have been given in Sec.~\ref{sec:phi:den},
but the others are new addtitions.

\noindent (a) For a square dot with one entering solid line of momentum $\bold{p}$
and one leaving solid line of momentum $\bold{p}'$,
include the factor $-\delta^3\left(\bold{p}-\bold{p}'\right)/\left(1+\tilde{E}_\bold{p}\right)$.

\noindent (b) For a square dot with one entering dotted line of momentum $\bold{p}$
and one leaving dotted line of momentum $\bold{q}'$,
include the factor $-\delta^3\left(\bold{p}-\bold{q}'\right)/\left(1+\tilde{E}_\bold{p}\right)$.

\noindent (c) For a square dot with one leaving solid line of momentum $\bold{p}'$
and one leaving dotted line of momentum $\bold{q}'$,
include the factor $\delta^3\left(\bold{p}'-\bold{q}'\right)/\left(1+\tilde{E}_{\bold{p}'}\right)$.

\noindent (d) For a square dot with one entering solid line of momentum $\bold{p}$
and one entering dotted line of momentum $\bold{q}$,
include the factor $\delta^3\left(\bold{p}-\bold{q}\right)/\left(1+\tilde{E}_\bold{p}\right)$.

\noindent (e) For a square dot with one entering solid line of momentum $\bold{p}$
and one solid line leaving towards a vertex,
include the factor $-1/\left[2\sqrt{(2\pi)^32E_\bold{p}}\left(1+\tilde{E}_\bold{p}\right)\right]$.

\noindent (f) For a square dot with one entering dotted line of momentum $\bold{p}$
and one dotted line leaving towards a vertex,
include the factor $-1/\left[2\sqrt{(2\pi)^32E_\bold{p}}\left(1+\tilde{E}_\bold{p}\right)\right]$.

\noindent (g) For a square dot with one entering solid line of momentum $\bold{p}$
and one dotted line entering from a vertex,
include the factor $1/\left[2\sqrt{(2\pi)^32E_\bold{p}}\left(1+\tilde{E}_\bold{p}\right)\right]$.

\noindent (h) For a square dot with one entering dotted line of momentum $\bold{p}$
and one solid line entering from a vertex,
include the factor $1/\left[2\sqrt{(2\pi)^32E_\bold{p}}\left(1+\tilde{E}_\bold{p}\right)\right]$.

\noindent (i) For a square dot with one leaving solid line of momentum $\bold{p}'$
and one solid line entering from a vertex,
include the factor $-1/\left[2\sqrt{(2\pi)^32E_{\bold{p}'}}\left(1+\tilde{E}_{\bold{p}'}\right)\right]$.

\noindent (j) For a square dot with one leaving dotted line of momentum $\bold{q}'$
and one dotted line entering from a vertex,
include the factor $-1/\left[2\sqrt{(2\pi)^32E_{\bold{q}'}}\left(1+\tilde{E}_{\bold{q}'}\right)\right]$.

\noindent (k) For a square dot with one leaving solid line of momentum $\bold{p}'$
and one dotted line leaving towards a vertex,
include the factor $1/\left[2\sqrt{(2\pi)^32E_{\bold{p}'}}\left(1+\tilde{E}_{\bold{p}'}\right)\right]$.

\noindent (l) For a square dot with one leaving dotted line of momentum $\bold{q}'$
and one solid line leaving towards a vertex,
include the factor $1/\left[2\sqrt{(2\pi)^32E_{\bold{q}'}}\left(1+\tilde{E}_{\bold{q}'}\right)\right]$.

\noindent (m) For a square dot through which one solid line between two vertices is going,
include the factor
\begin{equation}
\begin{split}
& \frac{\lambda^2}{\left(2\pi\right)^4} \int d^4 q \ \frac{1}{\left(q^2+m^2-i\eta\right)^2}
\left(1- \frac{\delta_{\left|q^0\right|, E_\bold{q}}}{1+\tilde{E}_\bold{q}}\right)
\\ & = \frac{\lambda^2}{\left(2\pi\right)^4} \int d^4 q \ \frac{1}{\left(q^2+m^2-i\eta\right)^2},
\end{split}
\end{equation}
where we have used the fact that the on-shell momenta occupy a
negligible fraction of the four-momentum space in the limit $\Delta^4 p \to 0$.

\noindent (n) For a square dot through which one dotted line between two vertices is going,
include the factor $\displaystyle\frac{\lambda^2}{\left(2\pi\right)^4} 
\displaystyle\int d^4 q \displaystyle\frac{1}{\left(q^2+m^2+i\eta\right)^2}$.

\noindent (o) For a square dot with one dotted and one solid lines
from vertices entering it, include the factor $\displaystyle\frac{\lambda^2}{\left(2\pi\right)^4} 
\displaystyle\int d^4 q \displaystyle\frac{1}{\left|q^2+m^2-i\eta\right|^2}$.

\noindent (p) For each solid line that is not connected to a square dot,
or for each vertex connecting to four solid lines,
include the factor of the corresponding Feynman diagram.
For each dotted line that is not connected to a square dot,
or for each vertex connecting to four dotted lines,
include the complex conjugate of the Feynman-diagram factor.

\noindent (q) For each paired-diagram, include a factor $1/Z$.

\noindent (r) For each paired-diagram that is the combination
of a solid-line diagram with $k_1$ vertices and a dotted-line diagram
with $k_2$ vertices, include a combinatoric factor
$\mathcal{C}\mathcal{C}_1\mathcal{C}_2/\left(4!^{k_1+k_2} k_1!k_2!\right)$,
where $\mathcal{C}_1$ and $\mathcal{C}_2$ are the combinatoric numbers
of the solid-line and dotted-line diagrams, respectively, and $\mathcal{C}$
denotes the number of ways of pairing square dots that result in the same
paired-diagram.

The diagrammatic rules for calculating $\rho^g$ are much more complicated
than those for calculating $\rho$. But
it is not difficult to see the Lorentz invariance of $\rho^g$. The factors of $\rho^g$
include $1/\left(1+\tilde{E}_\bold{p}\right)$, $1/\left(q^2+m^2\pm i\eta\right)^2$
or $1/\left|q^2+m^2 - i\eta\right|^2$ which are all scalars, and also
$\delta^3\left(\bold{p}-\bold{p}'\right)$ and $1/\sqrt{E_\bold{p}}$ if there
exist external lines in the paired-diagram. Under a Lorentz transformation, $\rho^g$ transforms as
Eq.~\eqref{eq:phi:den:Si}, and then the density matrix transforms as
Eq.~\eqref{eq:phi:den:out}. The Lorentz invariance of density matrix
guarantees that the outcomes of experiment do not depend on our choice of reference frame.

In Sec.~\ref{sec:phi:den}, we proved an equal-momentum condition
for the density-matrix element. There exist similar
conditions for $\rho^{g\left(\bold{p}_1\cdots\bold{p}_b\right)}_{\bold{p}'_1\cdots\bold{p}'_m,\bold{q}'_1
\cdots\bold{q}'_n}\neq 0$. Each paired-diagram of $\rho^{g\left(\bold{p}_1
\cdots\bold{p}_b\right)}_{\bold{p}'_1\cdots\bold{p}'_m,\bold{q}'_1\cdots\bold{q}'_n}$
has $b$ solid lines and $b$ dotted lines running into the diagram,
and also $m$ solid lines and $n$ dotted lines running out of the diagram.
We do not care whether the paired-diagram is a connected or disconnected diagram.
Two lines (solid or dotted) meet at a square dot, while four lines meet at a vertex.
We assume that there are totally $s$ square dots and
$v$ vertices in the diagram. Furthermore, we assume that $u$ external lines running into the diagram are
directly connected to external lines running out of the diagram, and there are $r$ internal lines.
It is clear to see $2s+4v= 2b+m+n+2r-2u$. Hence, $m+n$ must be even, which
is one of the conditions of $\rho^g \neq 0$.

\begin{figure}[tbp]
\vspace{0.5cm}
\includegraphics[width=0.9\linewidth]{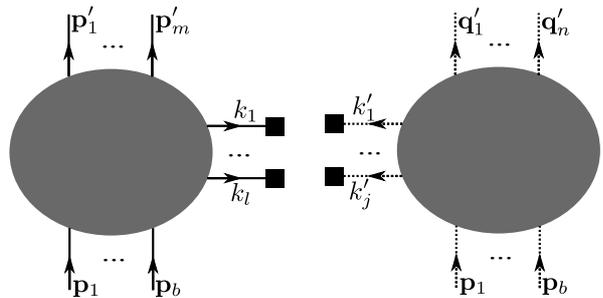}
\caption{Graphical expressions of $S^g_{\bold{p}'_1\cdots\bold{p}'_m,\bold{p}_1\cdots\bold{p}_b}$
and $S^{g*}_{\bold{q}'_1\cdots\bold{q}'_n,\bold{p}_1\cdots\bold{p}_b}$. This
figure is used to prove the equal-momentum condition
of density matrix in the stochastic $\phi^4$-theory.}\label{fig:4dpro}
\end{figure}
Fig.~\ref{fig:4dpro} shows the Feynman diagrams of
$S^g_{\bold{p}'_1\cdots\bold{p}'_m,\bold{p}_1\cdots\bold{p}_b}$
(the left panel) and $S^{g*}_{\bold{q}'_1\cdots\bold{q}'_n,\bold{p}_1\cdots\bold{p}_b}$ (the right panel).
All the external lines and square dots are displayed. But the vertices and internal lines between vertices
are hidden in the shaded regions. Without loss of generality, we assume that
the arrows always point to square dot. Indeed, we can reverse the direction
of an arrow by simply changing the sign of four-momentum carried by the line.
At each vertex, the total four-momentum entering it must equal the
total four-momentum leaving it. The four-momentum conservation requires
\begin{equation}\label{eq:4:den:em}
\begin{split}
& p_1 + \cdots + p_b =  p'_1 +\cdots + p'_m + k_1 +\cdots k_l, \\
& p_1 + \cdots + p_b =  q'_1 +\cdots + q'_n + k'_1 +\cdots k'_j,
\end{split}
\end{equation}
where $p_i = \left(E_{\bold{p}_i}, \bold{p}_i\right)$,
$p'_i = \left(E_{\bold{p}'_i}, \bold{p}'_i\right)$ and
$q'_i = \left(E_{\bold{q}'_i}, \bold{q}'_i\right)$ are the on-shell four-momenta of
external lines. If all the square dots in Fig.~\ref{fig:4dpro} can be paired with each other,
we then obtain a paired-diagram of
$\rho^{g\left(\bold{p}_1\cdots\bold{p}_b\right)}_{\bold{p}'_1\cdots\bold{p}'_m,\bold{q}'_1
\cdots\bold{q}'_n}$. To pair a square dot connected to solid (dotted) line
with a square dot connected to solid (dotted) line, the momenta carried by the lines
must sum to zero. To pair a square dot connected to solid line
with a square dot connected to dotted line, the momenta carried by the lines
must equal each other. Therefore, we have $k_1+\cdots+k_l = k'_1+\cdots
+ k'_j$. By comparing it with Eq.~\eqref{eq:4:den:em}, we immediately find
\begin{equation}\label{eq:4:den:pq}
p'_1 +\cdots +p'_m = q'_1 +\cdots q'_n.
\end{equation}
This is the equal-momentum condition for the density matrix of an interacting
stochastic QFT. $\rho^{g\left(\bold{p}_1\cdots\bold{p}_b\right)}_{\bold{p}'_1
\cdots\bold{p}'_m,\bold{q}'_1 \cdots\bold{q}'_n}$ is nonzero if and only if $n+m$
is even and Eq.~\eqref{eq:4:den:pq} stands.

A direct consequence of Eq.~\eqref{eq:4:den:pq} is that the density
matrices in the Schr\"{o}dinger and interaction pictures equal each other,
just like in the noninteracting theory. This
is an important result, because we are usually interested in
the density matrix in the Schr\"{o}dinger picture but what we obtain from the paired-diagramms
is the density matrix in the interaction picture. Since they are equivalent
to each other, we do not need to change pictures.

\subsection{Two-particle collision\label{sec:4:renor}}

As an example, let us see how to use the stochastic $\phi^4$-theory
to study the collision of two particles with initial momentum $\bold{p}_1$
and $\bold{p}_2$, respectively. In the conventional $\phi^4$-theory,
one calculates the $S$-matrix and interprets it as the probability amplitude.
In the stochastic $\phi^4$-theory, the $S$-matrix becomes a matrix of random numbers,
but the density matrix is still deterministic. The density matrix encodes
the information of the final quantum state after collision. Its diagonal elements are
usually interpreted as the probabilities of experimental outcomes.
In the stochastic $\phi^4$-theory, the density matrix of final state,
denoted by $\hat{\rho}^g_{\bold{p}_1\bold{p}_2}$,
can be obtained by using the diagrammatic technique.

We have learned from the noninteracting stochastic QFT that,
if we want to obtain some convergent results in the limit $\Delta^4x\to 0$,
a renormalization of $\lambda$ is necessary. Fortunately, in the presence of a $\phi^4$-interaction,
this renormalization procedure keeps the same.
Let us see what comes out in the calculation of $\rho^g$.
The renormalization of $\lambda$ is necessary when we calculate the factors of $\rho^g$
that contain $\lambda$. These factors are shown in Fig.~\ref{fig:4den}.
We first look at the factors in Fig.~\ref{fig:4den}(m), Fig.~\ref{fig:4den}(n) and Fig.~\ref{fig:4den}(o),
which are all represented by a square dot through which an internal line between
two vertices goes. In these factors, there exists an integral with respect to four-momentum,
e.g. $\displaystyle\int d^4 q \ 1/\left(q^2+m^2-i\eta\right)^2$.
We integrate out $q^0$ by using the residue theorem, and then regularize the integral
with respect to $\bold{q}$ by setting a momentum cutoff. In terms of
the physical coupling, the result is
\begin{equation}
\begin{split}
& \frac{\lambda^2}{\left(2\pi\right)^4} \int d^4 q \frac{1}{\left(q^2+m^2\mp i\eta\right)^2} \\ &
=\pm \frac{i  \lambda^2}{32\pi^3} \int d^3\bold{q}\frac{1}{E^3_\bold{q}} \\ &
=\pm \frac{i\lambda_p^2}{8\pi^2} \frac{m^2}{\Lambda^2} \left( \ln\left[ \frac{\Lambda}{m}+
\sqrt{\left(\frac{\Lambda}{m}\right)^2+1}\right] - \frac{\Lambda/m}{\sqrt{\left(\Lambda/m\right)^2
+1}}\right) \\ & \overset{\Lambda\to \infty}{\longrightarrow} 0.
\end{split}
\end{equation}
After renormalization, Fig.~\ref{fig:4den}(m) or Fig.~\ref{fig:4den}(n) have no contribution to $\rho^g$.
Similarly, we find the factor of Fig.~\ref{fig:4den}(o) to be
\begin{equation}
\begin{split}
& \frac{\lambda^2}{\left(2\pi\right)^4} \int d^4 q \frac{1}{\left|q^2+m^2- i\eta\right|^2} \\ &
=\frac{\lambda^2}{\left(2\pi\right)^4}\frac{1}{2} \int d^4 q\bigg\{
\frac{1}{\left(q^2+m^2- i\eta\right)^2} + \frac{1}{\left(q^2+m^2+ i\eta\right)^2} \\
& -\left(\frac{2i\eta}{\left(q^2+m^2\right)^2+\eta^2}\right)^2 \bigg\} \\
& =  \frac{\lambda^2}{\left(2\pi\right)^4} 2\pi^2 \int d^4 q \ \delta(q^2+m^2)\delta(q^2+m^2)\\ &
= \frac{\lambda_p^2T}{16\pi^3} \frac{m^2}{\Lambda^2} \int d^3\bold{q} \frac{1}{E^2_\bold{q}}
\overset{\Lambda\to\infty}{\longrightarrow} 0 .
\end{split}
\end{equation}
The contribution of Fig.~\ref{fig:4den}(o) also vanishes after renormalization.
Therefore, in the calculation of $\rho^g$, we do not need to consider
diagrams in which the internal lines are connected to square dots.

\begin{figure}[tbp]
\vspace{0.5cm}
\includegraphics[width=0.9\linewidth]{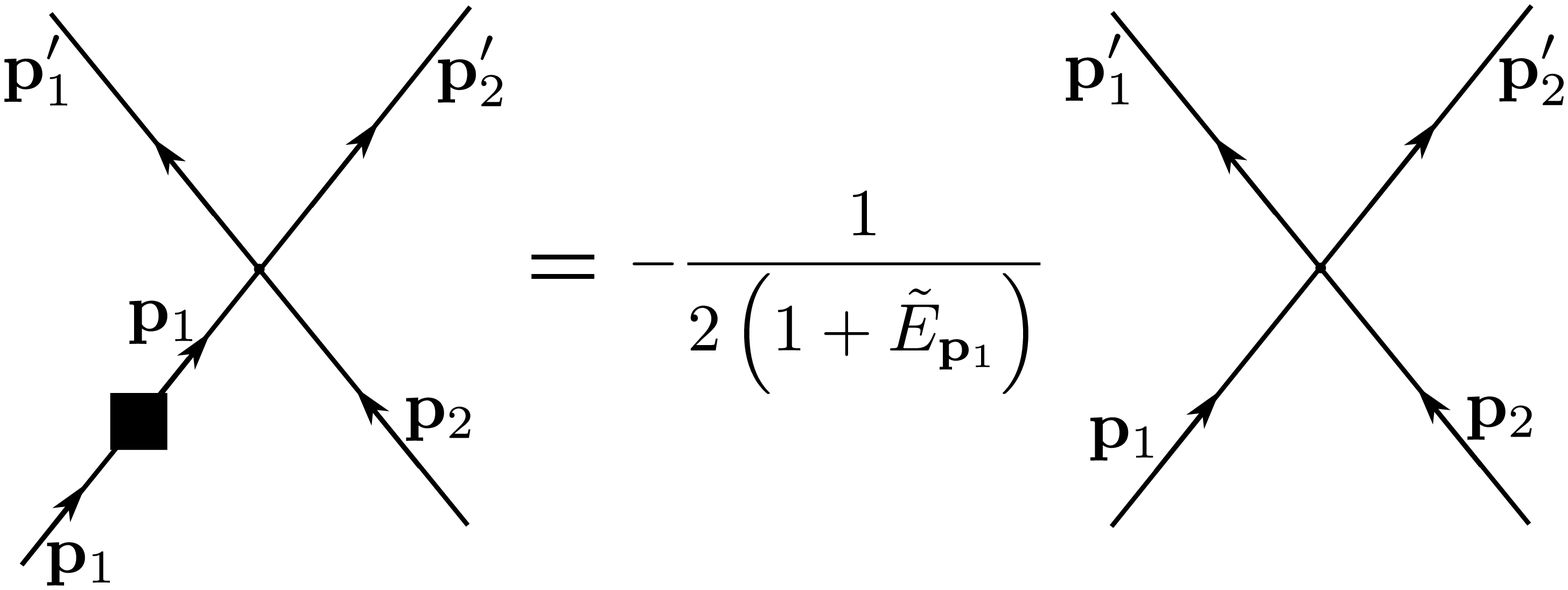}
\caption{Each diagram with an external line carrying square dot can
be expressed as the product of a diagram with bare external lines
and a vanishing factor.}\label{fig:4resq}
\end{figure}
Next, let us study the factors in Figs.~\ref{fig:4den}(e-l) which are represented
by an external line carrying square dot that runs into or out of a vertex.
For each of these diagrams, there exists a corresponding diagram with bare external lines.
Compared with the latter, the diagram with square dot includes an additional factor
$1/\left(1+\tilde{E}_{\bold{p}_1}\right)$ (see Fig.~\ref{fig:4resq}).
But the factor $1/\left(1+\tilde{E}_{\bold{p}_1}\right)$
vanishes after we replace the bare coupling by the physical one
and take $\Lambda\to\infty$. Therefore, the contribution
of Figs.~\ref{fig:4den}(e-l) vanishes after renormalization, compared to the
corresponding diagrams with bare external lines.
For the same reason, the contribution of Fig.~\ref{fig:4den}(a) and Fig.~\ref{fig:4den}(b) vanishes.

Now, only Fig.~\ref{fig:4den}(c) and Fig.~\ref{fig:4den}(d) are left.
In both of them, there exists a factor $1/\left(1+\tilde{E}_{\bold{p}}\right)$.
In Fig.~\ref{fig:4den}(c), $\bold{p}$ denotes the final momentum, but
in Fig.~\ref{fig:4den}(d), it denotes the initial momentum. This is
a key difference, because we must integrate out the final momentum
when calculating the probability distribution of experimental outcomes,
but we do not integrate with respect to initial momentum. As a consequence,
Fig.~\ref{fig:4den}(d) can be neglected, since
the factor $1/\left(1+\tilde{E}_{\bold{p}}\right)$ vanishes after renormalization.
Conversely, Fig.~\ref{fig:4den}(c) must
be kept in the calculation of $\rho^g$, because the volume of momentum
space diverges as $\Lambda\to\infty$ so that the integral of $1/\left(1+\tilde{E}_{\bold{p}}\right)$
is possibly finite.

The above analysis leads to a great reduction in the number of possible diagrams.
Now we only need to consider the paired-diagram which can be
divided into a diagram with no square dot plus a sequence of Fig.~\ref{fig:4den}(c).
A paired-diagram with no square dot consists of two isolated Feynman diagrams
with one in solid line and the other in dotted line. Such a diagram has the
properties of Feynman diagram in the conventional $\phi^4$-theory.
For examples, the number of entering external lines
equals the number of leaving external lines (the number of particles is conserved
in the conventional theory), and the total energy or momentum carried by the entering external lines
equal those carried by the leaving external lines, respectively.

\begin{figure}[tbp]
\vspace{0.5cm}
\includegraphics[width=0.9\linewidth]{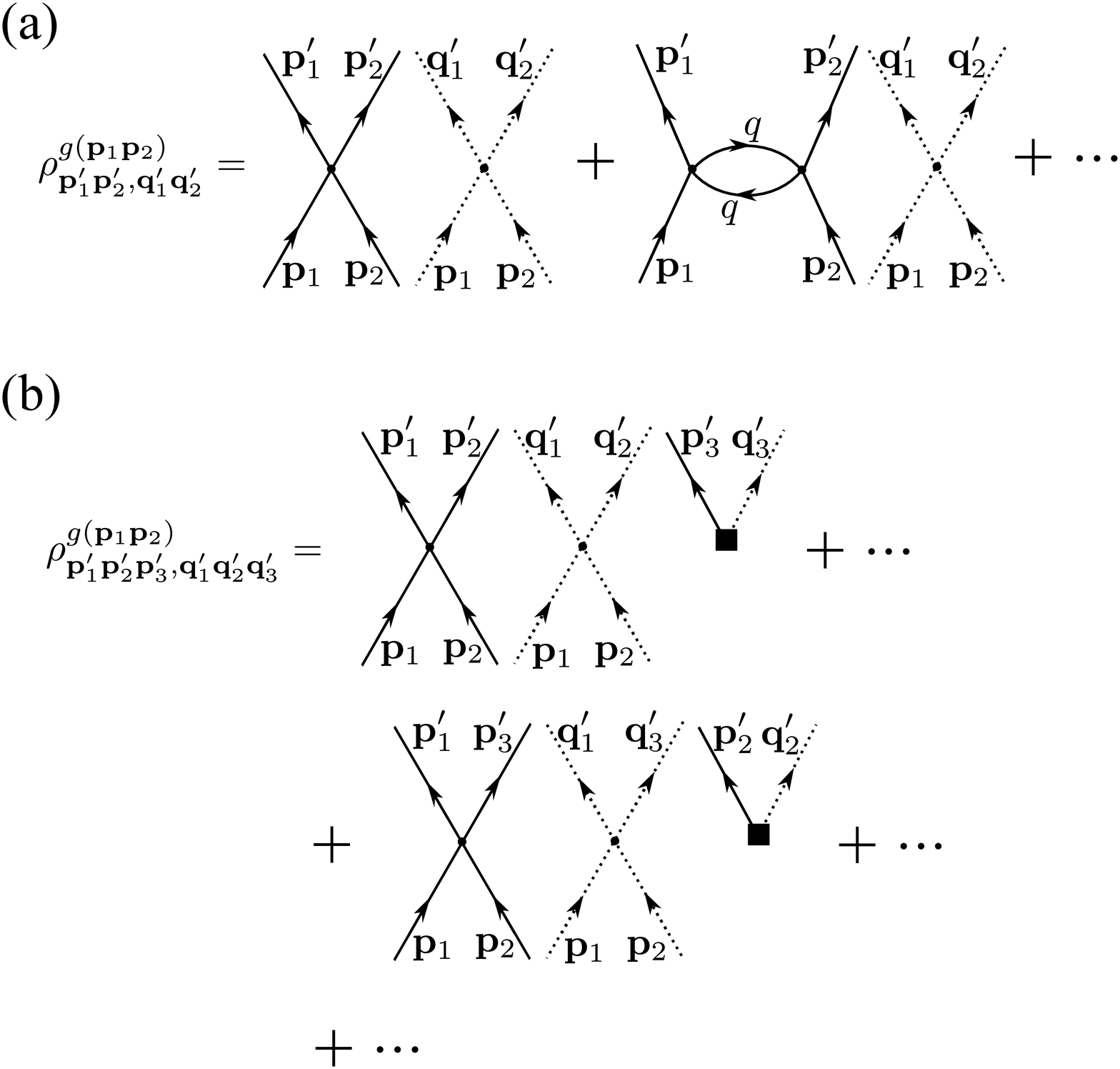}
\caption{Graphical expressions of (a)
$\rho^{g(\bold{p}_1\bold{p}_2)}_{\bold{p}'_1\bold{p}'_2,\bold{q}'_1\bold{q}'_2}$
and (b) $\rho^{g(\bold{p}_1\bold{p}_2)}_{\bold{p}'_1\bold{p}'_2\bold{p}'_3,
\bold{q}'_1\bold{q}'_2\bold{q}'_3}$.}\label{fig:4coll}
\end{figure}
In the problem of two-particle collision, we calculate the density-matrix elements
$\rho^{g(\bold{p}_1\bold{p}_2)}_{\bold{p}'_1\cdots\bold{p}'_m,\bold{q}'_1\cdots\bold{q}'_n}$.
Fig.~\ref{fig:4coll}(a) displays some paired-diagrams of
$\rho^{g(\bold{p}_1\bold{p}_2)}_{\bold{p}'_1\bold{p}'_2,\bold{q}'_1\bold{q}'_2}$.
It is clear that no diagram contains square dots.
Because in a diagram of $\rho^{g(\bold{p}_1\bold{p}_2)}_{\bold{p}'_1
\bold{p}'_2,\bold{q}'_1\bold{q}'_2}$,
the total numbers of entering and leaving external lines equal each other.
But if the diagram contains square dots (a sequence of Fig.~\ref{fig:4den}(c)),
there must be more leaving external lines than entering ones.
In the calculation of $\rho^{g(\bold{p}_1\bold{p}_2)}_{\bold{p}'_1\bold{p}'_2,\bold{q}'_1\bold{q}'_2}$,
each paired-diagram can be separated into one solid-line and one dotted-line
Feynman diagrams, the sum of paired-diagrams is then equal to
the product of sum of solid-line Feynman diagrams and sum of dotted-line Feynman diagrams.
And there exists a factor $1/Z$ for each paired-diagram. We then obtain
\begin{equation}\label{eq:4:re:rg2}
\rho^{g(\bold{p}_1\bold{p}_2)}_{\bold{p}'_1\bold{p}'_2,\bold{q}'_1\bold{q}'_2}=
\frac{1}{Z} \left.S^g_{\bold{p}'_1\bold{p}'_2,\bold{p}_1\bold{p}_2}\right|_{\lambda=0}
\left.S^{g*}_{\bold{q}'_1\bold{q}'_2,\bold{p}_1\bold{p}_2}\right|_{\lambda=0},
\end{equation}
where $ \left.S^g\right|_{\lambda=0}$ is the $S$-matrix at ${\lambda=0}$
which equals the $S$-matrix of conventional $\phi^4$-theory. As $\lambda=0$, we have $Z=1$
and then Eq.~\eqref{eq:4:re:rg2} reduces to the density-matrix
of conventional $\phi^4$-theory. In the two-particle block,
the only difference between the matrix elements of stochastic and conventional
$\phi^4$-theories is the factor $1/Z$.

The Feynman diagram is a part of the paired-diagram. If the Feynman diagram
contains loops, then the corresponding integral with respect to four-momentum
diverges in the limit $\Lambda\to\infty$,
and we have to renormalize the parameters $g$ or $m$ for obtaining
a convergent result. Fortunately, in each paired-diagram, the Feynman diagram is separated
from diagrams like Fig.~\ref{fig:4den}(c).
The former contains no square dot, and then we do not need to renormalize $\lambda$
when calculating them. At the same time, the latter contains no
loop, and then in the calculation we do not need to renormalize parameters other than $\lambda$.
In the stochastic $\phi^4$-theory, the renormalization of $\lambda$ is independent of the
renormalization of other parameters. Since the conventional $\phi^4$-theory
is renormalizable, the stochastic $\phi^4$-theory must be also renormalizable.
The paired-diagrams always give finite results after
we properly renormalize $\lambda$ together with other parameters.

Since $1/Z=exp\left\{-TVm^2\lambda^2_p/\left(4\pi^2\right)\right\}$,
Eq.~\eqref{eq:4:re:rg2} tells us that the density matrix elements in the two-particle block
decay exponentially with time increasing, and all the elements decay at
the same rate. To keep the trace of density matrix invariant, the elements
in the $n$-particle block with $n\neq 2$ must increase with time. Indeed,
the density matrix is block-diagonal, that is
$\rho^{g(\bold{p}_1\bold{p}_2)}_{\bold{p}'_1\cdots\bold{p}'_m,\bold{q}'_1\cdots\bold{q}'_n}$
is nonzero only if $m=n \geq 2$. Fig.~\ref{fig:4coll}(b) displays the paired-diagrams
for calculating $\rho^{g(\bold{p}_1\bold{p}_2)}_{\bold{p}'_1\bold{p}'_2\bold{p}'_3,
\bold{q}'_1\bold{q}'_2\bold{q}'_3}$. By some analysis, we find
\begin{equation}\label{eq:4:two:333}
\begin{split}
& \rho^{g(\bold{p}_1\bold{p}_2)}_{\bold{p}'_1\bold{p}'_2\bold{p}'_3,
\bold{q}'_1\bold{q}'_2\bold{q}'_3} = \frac{1}{Z} \frac{\delta^3\left(\bold{p}'_3-\bold{q}'_3\right)}
{1+\tilde{E}_{\bold{p}'_3}} \left.\rho^{g(\bold{p}_1\bold{p}_2)}_{\bold{p}'_1\bold{p}'_2,
\bold{q}'_1\bold{q}'_2}\right|_{\lambda=0} \\ & +  \frac{1}{Z} \frac{\delta^3\left(\bold{p}'_2-\bold{q}'_2\right)}
{1+\tilde{E}_{\bold{p}'_2}} \left.\rho^{g(\bold{p}_1\bold{p}_2)}_{\bold{p}'_1\bold{p}'_3,
\bold{q}'_1\bold{q}'_3}\right|_{\lambda=0} + \text{the other 7 terms}.
\end{split}
\end{equation}
In Eq.~\eqref{eq:4:two:333}, the omitted terms are obtained by using various permutations
of $\bold{p}'$s or $\bold{q}'$s in the first term. In a similar way, we can obtain
$\rho^{g(\bold{p}_1\bold{p}_2)}_{\bold{p}'_1\cdots\bold{p}'_n,\bold{q}'_1\cdots\bold{q}'_n}$
for $n>3$. Finally, the density matrix is found to be
\begin{equation}\label{eq:4:re:fin}
\begin{split}
\hat{\rho}^g_{\bold{p}_1\bold{p}_2}= & \frac{1}{Z}
\sum_{n=0}^\infty \frac{1}{n!} \int \left(\prod_{j=1}^n \frac{d^3\bold{p}_j}
{1+\tilde{E}_{\bold{p}_j}}\right)\\ & \times \hat{a}^\dag_{\bold{p}_1}\cdots \hat{a}^\dag_{\bold{p}_n}
\left. \hat{\rho}^g_{\bold{p}_1\bold{p}_2}\right|_{\lambda=0} \hat{a}_{\bold{p}_n}
\cdots \hat{a}_{\bold{p}_1},
\end{split}
\end{equation}
where $\left. \hat{\rho}^g_{\bold{p}_1\bold{p}_2}\right|_{\lambda=0}$
is the density matrix at $\lambda=0$ which is also the density matrix
of conventional $\phi^4$-theory.

It is interesting to compare Eq.~\eqref{eq:4:re:fin}
with Eq.~\eqref{eq:phi:den:rho0T}. The latter is the density matrix of noninteracting
stochastic theory as the initial state is a vacuum.
If we replace $\left. \hat{\rho}^g_{\bold{p}_1\bold{p}_2}\right|_{\lambda=0}$
by $\ket{0}\bra{0}$ (the vacuum density matrix),
Eq.~\eqref{eq:4:re:fin} then becomes Eq.~\eqref{eq:phi:den:rho0T}.
Therefore, the random field in the $\phi^4$-theory plays the same role
as what it plays in the noninteracting theory. In the presence of interaction, two particles
collide with each other and scatter, as if the random field does not exist.
At the same time, the random field excites particles
from the vacuum as if the interaction does not exist. The scattering
and excitation processes are independent of each other.

The picture of two-particle collision is now clear. Driven by the random field, the
universe as a whole is thermalized, with new particles excited from the vacuum.
The number of additional excitations has the Poisson distribution. 
Two original particles are scattered in a way described by the conventional
$\phi^4$-theory. But their signals are gradually covered by the background excitations.

Finally, it is worth mentioning the difference between the prediction of
our stochastic $\phi^4$-theory and wave function collapse (state-vector reduction).
It is clear that the theory~\eqref{eq:4:action} cannot explain the wave function collapse.
After two particles collide with each other, according to Eq.~\eqref{eq:4:re:rg2},
the off-diagonal elements of density matrix (e.g.
$\bra{\bold{p}'_1\bold{p}'_2}\hat{\rho}^g_{\bold{p}_1\bold{p}_2}\ket{\bold{q}'_1\bold{q}'_2}$)
do decay exponentially. This seemingly suggests the superposition
of different eigenstates of momentum be suppressed.
However, the decay is indeed the result of more particles being excited out of the vacuum.
The evidence is that the diagonal elements in the two-particle block also
decay at the same rate. Therefore, what is suppressed is not the superposition of
$\ket{\bold{p}'_1\bold{p}'_2}$ and $\ket{\bold{q}'_1\bold{q}'_2}$,
but is the probability of final state staying in the two-particle sector.
It is better to say that the random field causes thermalization,
rather than wave function collapse.

\section{Summary \label{sec:summary}}

In summary, we develop an action formulation of stochastic QFTs which
describe random unitary evolutions of state vector in the many-body Hilbert space.
The theory is determined by an action which is a functional of random field.
The symmetry of the theory is a statistical symmetry. The probability distribution
of action keeps invariant under symmetry transformations. A significant
advantage of action formulation is that one can easily write down an action with
the wanted symmetries. For examples, in the action~\eqref{eq:ha:lag} of harmonic
oscillator, the time translational symmetry is preserved by coupling the coordinate
of particle to a Wiener process. In the actions~\eqref{eq:phi:ra:IW} and~\eqref{eq:4:action}
of scalar bosons, both the spacetime translation and Lorentz symmetries are
preserved by coupling scalar field to a scalar random field. The scalar random field
is defined by partitioning the spacetime into a set of infinitesimal elements of volume $d^4x$
and then assigning to each element an independent Gaussian random number (denoted
by $dW(x)$) of zero mean and variance $d^4x$. $dW(x)$ can be also seen as
the product of a white-noise field and $d^4x$.

The canonical quantization of a random action results in a SDE of state vector
which has the same statistical symmetry as the action. On the other hand, the dynamical equation
of density matrix which is a Lindblad equation, has explicit symmetries corresponding
to the statistical ones of action. More important, the $S$-matrix and density matrix
which can be calculated by using the path integral approach, have also the symmetries
of action. For a stochastic QFT, the diagrammatic technique of calculation is able to guarantee
the Lorentz invariance of both $S$-matrix (see Eq.~\eqref{eq:phi:dia:Lo}) and density
matrix (see Eq.~\eqref{eq:phi:den:out}). And the Lorentz invariance is robust even
after we consider the interaction between particles, which shows the power
of our formulation. In general, by coupling $dW(x)$ to a product of quantum fields
that transforms as a scalar, we always obtain a stochastic QFT in which the Lorentz
invariance of $S$-matrix and density matrix is preserved in the diagrammatic calculations.

The stochastic QFT in the action formulation is a natural generalization of conventional
QFT. It describes identical particles and the interaction between particles can be considered
easily. As in conventional QFTs, we develop the path integral approach and diagrammatic
technique for calculating the $S$-matrix. The diagrammatic rules for free field and
$\phi^4$-theory are summarized in Figs.~\ref{fig:FM} and~\ref{fig:Sp}, respectively.
Distinguished from the diagrams of QFT, the Feynman diagrams of stochastic theory contain square dots
which represent the random field and each square dot is connected to a single propagator line.
With the help of diagrams, we obtain an exact expression of $S$-matrix in the absence
of interaction, and then the final quantum state after scattering
for an arbitrary initial state (see Eqs.~\eqref{eq:phi:sin:U0} and~\eqref{eq:phi:sin:Up}).
The $S$-matrix of interacting theory can be also calculated in a systematic way.
Furthermore, by using the probability distribution of random field, we develop the diagrammatic
technique for calculating the density matrix of final state.
The density matrix is graphically represented by paired-diagrams. Each paired-diagram
consists of one Feynman diagram in solid line and one in dotted line with the latter representing
the complex conjugate of $S$-matrix. Each square dot in a paired-diagram is connected to
two propagator lines. With the help of paired-diagrams, we obtain the density matrix in
the absence of interaction (see Eqs~\eqref{eq:phi:den:rho0T} and~\eqref{eq:phi:spe:rhp}).
In the presence of interaction, we prove a relation between the density matrices
of stochastic QFT and conventional QFT (see Eq.~\eqref{eq:4:re:fin}).
Our approach of calculating $S$-matrix and density matrix avoid
solving the stochastic differential equations, and its generalization to
more complicated stochastic QFTs is straightforward.

The continuous translation and Lorentz symmetries require an infinite momentum
space. For a model with finite bare coupling (denoted by $\lambda$) between random and
quantum fields, our formalism leads to an ultraviolet divergence which has the similar
origin as the ultraviolet divergence in QFTs. The divergence can be canceled by
the renormalization of $\lambda$. We first regulate the integral over
momentum space by setting an ultraviolet cutoff $\Lambda$, replace the bare
coupling by the physical coupling $\lambda_p=\lambda \Lambda/m$, and finally
take $\Lambda\to\infty$. In terms of physical coupling, the $S$-matrix and density
matrix become finite. The renormalization process indeed suppresses the bare coupling strength
to infinitesimal, which explains why the divergence vanishes. The renormalization
of $\lambda$ is independent of the renormalization of other parameters in the $\phi^4$-theory.
We prove that the stochastic $\phi^4$-theory is a renormalizable theory. In general,
coupling $dW(x)$ linearly to the field of a renormalizable QFT always results in
a renormalizable stochastic QFT.

The explicit time translational symmetry is broken by the random field.
The energy and particle numbers are then not conserved. The random field continuously excites
particles out of the vacuum with the total number of excited particles following the Poisson
distribution. The expectation value of particle number is $TVm^2\lambda_p^2/\left(4\pi^2\right)$
where $T$ is the half of driving time, $V$ is the total volume of space, and $m$ is the
mass of particle. As a consequence, the universe is heated up, evolving
towards an infinite-temperature state. In the presence of interaction, the collision
between particles is not affected by random field. But the signals of colliding
particles are gradually covered by the background excitations caused by random field.

Finally, we would like to mention the difference between our models and the relativistic
spontaneous collapse models. Our action formulation results in a linear stochastic differential equation of state vector.
And in this paper, we only consider a linear coupling between random and quantum
fields. Due to the lack of nonlinearity, the models~\eqref{eq:phi:ra:IW} and~\eqref{eq:4:action}
cannot explain the collapse of wave function, instead, they predict a heating effect
of background spacetime. The lack of nonlinearity is a problem in the application
of our approach to describing wave-function collapse. Even there are no linear collapse models
up to now, but no fundamental laws forbid the existence of such a model.
A study of more complicated quantum fields, interactions or even gravity fields
may tell us whether the action formulation of stochastic QFTs can explain wave-function collapse.

\section*{Acknowledgement}
Pei Wang is supported by NSFC under Grant Nos.~11774315 and~11835011,
and by the Junior Associates program of the Abdus Salam International Center for Theoretical Physics.

\end{document}